\documentclass[11pt,a4paper]{article}
 
\usepackage{amsmath,amsthm,amssymb}
\usepackage{graphics,graphicx}
\usepackage{epsfig}
\usepackage{multicol}
\usepackage{color}
\usepackage{cite} %
\makeatletter
\@addtoreset{equation}{section}
\makeatother

\begin{document}

\begin{center}
\LARGE{\bf Modelling Cosmic Springs with Finsler and Generalised Finsler Geometries}
\end{center}

\begin{center}
\large{\bf Matthew J. Lake} ${}^{a,b,c,d}$\footnote{matthewjlake@narit.or.th}
\end{center}
\begin{center}
\emph{ ${}^a$National Astronomical Research Institute of Thailand, \\ 260 Moo 4, T. Donkaew,  A. Maerim, Chiang Mai 50180, Thailand \\}
\emph{ ${}^b$ Department of Physics, Faculty of Science, Chiang Mai University, \\ 239 Huaykaew Road, T. Suthep, A. Muang, Chiang Mai 50200, Thailand \\}
\emph{ ${}^c$ School of Physics, Sun Yat-Sen University, \\ Guangzhou 510275, People’s Republic of China \\}
\emph{ ${}^d$ Department of Physics, Babe\c s-Bolyai University, \\ Mihail Kog\u alniceanu Street 1, 400084 Cluj-Napoca, Romania \\}
\vspace{0.1cm}
\end{center}


\begin{abstract}
We show that the equations of motion governing the dynamics of strings in a compact internal space can be written as dispersion relations, with a local speed that depends on the velocity and curvature of the string in the large dimensions. 
From a $(3+1)$-dimensional perspective these can be viewed as dispersion relations for waves propagating in the string interior and are analogous to those for current-carrying topological defects. 
This allows us to construct a unified framework with which to study and interpret the internal structure of various field-theoretic and fundamental string species, in a simple physically intuitive coordinate system, without the need for dimensional reduction or approximate effective actions. 
This, in turn, allows us to identify the precise conditions under which higher-dimensional strings and current-carrying defects are observationally indistinguishable, for macroscopic observers.
Our approach naturally incorporates the description of so-called `cosmic springs', whose dynamics are expressed in terms of an effective Finsler geometry, for circular loops, or generalised Finsler geometry, for non-circular configurations. 
This demonstrates the importance of these novel geometric structures and their utility in modelling complex physical phenomena in cosmology and astrophysics. 
\end{abstract}

{\bf Keywords}: superconducting cosmic string; cosmic spring; Finsler geometry; higher-dimensional windings; mathematical modelling

{\it Symmetry, MDPI special issue, Mathematical Modelling of Physical Systems, \\ P. Nicolini, ed., 2022.}

\newpage
\tableofcontents

\section{Introduction} \label{Sect1}
Cosmic strings are line-like concentrations of mass-energy, thought to have formed during symmetry-breaking phase transitions in the early Universe \cite{NO,ViSh00,hep-ph/9411342,hep-th/0508135v2,Preskill,Lake:2011aa}.
Though the string width is determined by the inverse of the symmetry-breaking energy scale, this is small compared to cosmological distances, and they may be approximated as one-dimensional objects for many purposes \cite{Go71,And02}. 
Such strings may have been produced at the epoch of electro-weak symmetry-breaking \cite{Na77,Vachaspati:1992uz,James:1992zp}, or the GUT scale, and their formation is also a generic feature of the phase transitions predicted by many extensions of the Standard Model \cite{Jeannerot:2003qv,Rocher:2004rc,Sakellariadou:2004rf}.
\\ \indent
In field theory they are a type of topological defect, analogous to the magnetic flux tubes and other types of vortex configuration created in condensed matter systems \cite{Abrikosov1,Zurek1,Zurek2,Zurek3,Bowick1,Williams1,Hendry1,Chuang1,Annett1}. 
They are produced via the Kibble-Zurek mechanism \cite{ICTP/75/5} when the vacuum manifold of the fields ($\mathcal{M}$) possesses a nontrivial first homotopy group, e.g., $\pi_1(\mathcal{M})=\mathbb{Z}$, in which each nonzero integer corresponds to a possible winding number of the vortex-string cross-section. 
In recent years, the formation of `cosmic', i.e. horizon-sized, fundamental strings ($F$-strings) and one-dimensional $D$-branes ($D$-strings) has also been extensively studied in string theory \cite{Polchinski_Intro,0911.1345v3,CMP_FD1,CMP_FD2,0811.1277v1,hep-th/0505050v1,astro-ph/0410073v2,Sakellariadou:2009,Rajantie:2007,Copeland:2005}, particularly in the context of brane inflation \cite{Cline1,Gasperini1,Tye1,Carroll:TASI,Quevedo:Lectures,Danielsson1}, in which such defects can be copiously produced \cite{Sarangi1,Jones_etal1,Pogosian_Obs1}.
\\ \indent
However, regardless of the precise details of individual models, the main phenomenological and observationally relevant parameters that characterise the string are its linear mass density, or energy per unit length,
often denoted as $\mu$ or $T$, and tension, which is often denoted, somewhat confusingly, by the letters $U$ or $\mathcal{T}$.
\footnote{Different conventions are prevalent in different areas of research, e.g., $\mu$ is the preferred notation for the linear mass density of defect strings, whereas $\mathcal{T}$ is more commonly used for $F$-strings in string theory. In this work, we employ various notations, in keeping with the conventions of the relevant research fields, but clearly define our notation at the beginning of the relevant section(s) of the paper.} 
For vacuum strings in field theory and the fundamental strings of string theory, $\mu = -U$, but for current-carrying strings the mass density and tension may differ so that $\mu \neq -U$ \cite{Witten:1985,Peter92}. 
The relationship between the macroscopic string dynamics and the current at a given point on the string, at a given time, is complicated, since the local tension is determined by the local flux density. 
This can vary as a function of both position and time, giving rise to so-called `cosmic springs', i.e., strings with variable effective local tension. 
\\ \indent
In this paper, we study the complex interplay between the spring-like properties of current-carrying strings and their macroscopic evolution. 
We begin by demonstrating the equivalence, under dimensional reduction, between current-carrying defect strings and wound-strings in geometries with compact extra dimensions, providing a unified framework for modelling a variety of cosmic string species. 
However, the demonstration of equivalence between wound-strings and superconducting defect strings, under dimensional reduction, is not new and was first shown by Nielsen \cite{Ni79}. 
We then show how, by rewriting the equations of motion (EOM) for the current as dispersion relations, the correct EOM for the macroscopic evolution of the string can be obtained without dimensional reduction and without the need to construct an approximate effective action. 
\\ \indent
When the EOM for the fluxes on the effective string world sheet in $(3+1)$ dimensions are written as dispersion relations, the local speed of propagation depends on the velocity and curvature of the string in the noncompact space. 
For circular string loops, this implies that the macroscopic dynamics can be modelled in terms of an effective Finsler metric, but for noncircular configurations a generalised Finsler geometry is required. 
This method provides a new set of tools, and a new geometric language, with which to describe the dynamics of cosmic springs, i.e., strings with variable effective tension due to the presence of fluxes within the string core.
\\ \indent
The structure of this paper is as follows. 
In Sec. \ref{Sect2}, we review the necessary background for our model of $F$-strings with higher-dimensional windings, including the space-time embedding, EOM, and physical observables of the string. 
Sec. \ref{Sect2.1.1} covers the background geometry, embedding, string action and covariant EOM, while in Sec. \ref{Sect2.1.2} we determine general expressions for the constants of motion and the effective pressures and shears, from a $(3+1)$-dimensional perspective. 
In Sec. \ref{Sect2.1.3}, we demonstrate the one-to-one equivalence of the Euler-Lagrange equations and the conservation equations for the space-time energy-momentum tensor in a simple, physically intuitive coordinate system. 
We adopt these coordinates, and the associated gauge conditions for the string, throughout the rest of the paper. 
The wound-string model parameters, which determine the physically observable properties such as the effective linear mass density and tension, are defined in Sec. \ref{Sect2.1.4}. 
Sections \ref{Sect2.2}$-$\ref{Sect2.4} then deal with long strings, circular loops and arbitrary planar loops, respectively. 
Each subsection is further divided into two. 
The part first deals with unwound strings whereas the second considers the effects of higher-dimensional windings on the chosen ansatz for the configuration of the string in the macroscopic dimensions. 
Hence, in each case, we first review the string dynamics in the absence of high-dimensional effects, before considering the generalisation to strings with nontrivial embeddings in the internal space. 
We explicitly show that the conservation of momentum in the extra dimensions is equivalent to the conservation of current from a $(3+1)$-dimensional perspective and that the string EOM take the form of dispersion relations for waves `inside' the string. 
A brief summary of these results is given in Sec. \ref{Sect2.5}. 
\\ \indent
In Sec. \ref{Sect3}, we review the necessary background for our model of current-carrying topological defect strings. 
Though many superconducting string models exist in the literature, their basic phenomenology is similar \cite{ViSh00}, and can be well approximated by considering the simplest field theory in which vortex-line solutions exist, namely, the Abelian-Higgs model. 
In Sec. \ref{Sect3.1}, we review the dynamics of the Abelian-Higgs fields and couple them to an `external' source of charge which is capable of generating current when confined to the string core. 
The simplest vortex-line solution known as the Nielsen-Olesen string, which does not contain a superconducting current, is also reviewed. 
Section \ref{Sect3.2} considers the standard approach to modelling both non-current-carrying and current-carrying strings. 
In this approach, the effective action for a zero-width and {\it non-superconducting} string is first constructed, by integrating over the `internal' degrees of freedom that describe the microscopic structure of the string core in the Nielsen-Olesen solution. 
It is straightforward to show that this leads directly to the Nambu-Goto action for the $F$-string. 
Current-carrying strings are then modelled by adding world-sheet fluxes to the effective action. 
\\ \indent
To the best of our knowledge no detailed study has been made, in the existing literature, of the evolution of the internal structure of a finite-width string, as it undergoes dynamical evolution as a solitonic field state in the background space-time. 
The first goal of this work is to construct such a model, via an appropriate choice of ansatz for the fundamental field variables. 
This allows us to explicitly demonstrate the interplay between the dynamical evolution of the `microscopic' degrees of freedom (i.e., field states), that characterise the string interior, and the macroscopic variables that describe the dynamics of the string as an extended solitonic object. 
Our second goal is to show that the evolution of these internal degrees of freedom is completely analogous to the evolution of $F$-string windings in the compact `internal' space predicted by string theory. 
We stress that this program goes beyond the simple process of dimensional reduction. 
We do not dimensionally reduce either the $F$-string windings or the internal degrees of freedom of the finite-width core of the defect string. 
Instead, both microscopic and macroscopic, `internal' and `external' evolution are considered together. 
\\ \indent
In the case of long, straight strings, we find that the two levels of description can easily be reconciled within the usual pseudo-Riemannian geometry of special and general relativity. 
(For simplicity, we consider a $(3+1)$-dimensional Minkowski or warped Minkowski background as the geometry of the large dimensions, throughout this work.) 
However, for circular defect-string loops, the evolution of the fundamental field variables takes place in an {\it effective} Finsler geometry, in which the metric depends both on the space-time coordinates, $x$, and on their first derivatives with respect to a time-like internal parameter, $\partial_{\tau}x$. 
In the absence of circular symmetry, the effective metric also depends on the first derivatives of the coordinates with respect to a space-like internal variable, $\partial_{\sigma}x$, corresponding to a form of generalised Finsler geometry. 
The connection to string theory is then made by associating $\tau$ and $\sigma$ with the time-like and space-like world-sheet parameters. 
Hence, long strings are modelled in Minkowski space, circular loops are modelled using an effective Finsler geometry, and arbitrary planar loops are modelled using an effective generalised Finsler geometry. 
These cases are dealt with in Secs. \ref{Sect5.1}-\ref{Sect5.3}, respectively, and a short summary of the defect string model is given in Sec. \ref{Sect5.4}. 
Our Conclusions, and a brief discussion of prospects for future work, are given in Sec. \ref{Sect6}. 

\section{Fundamental strings with higher-dimensional windings} \label{Sect2}
%
In \cite{Ni79} it was shown that the higher-dimensional dynamics of a fundamental string embedded in $M^4 \times S^1$ can be reinterpreted from a four-dimensional perspective as an effective world-sheet current.
There follows a formal equivalence between superconducting strings (of zero width) in $M^4$ and Nambu-Goto strings \cite{Go71,Na77} in $M^4 \times S^1$ under dimensional reduction. In \cite{YaLa15}, it was shown that the equations of motion (EOM) for $F$-strings in $M^4 \times S^1$ admit critical solutions in which the effective four-dimensional tension vanishes locally everywhere on the string. 
For the critical case, the string may adopt arbitrary static configurations in the large dimensions. 
Phenomenologically, this is consistent with similar results for current-carrying strings \cite{NiOl87,CoHiTu87,BlOlVi01} via the general correspondence demonstrated in \cite{Ni79}. 
\\ \indent
In \cite{LaYo12}, specific higher-dimensional $F$-string configurations were investigated together with corresponding configurations of superconducting defect strings with finite width. 
For long strings and circular loops, the embedding of the $F$-string in the higher-dimensional space was found to be formally analogous to the phase structure of superconducting Nielsen-Olesen strings \cite{NO} at critical coupling. 
A key feature of this correspondence is that the radius of the compact space $R$ and the string width in the field-theoretic model $r_{c}$ play equivalent roles in determining the constants of motion. 
This suggests that the radius of the windings in the higher-dimensional space provides an effective thickness for the Nambu-Goto string in four dimensions. 
\\ \indent
Based on these suggestive results, we conjecture that $F$-strings, which (by definition) have no internal structure, can mimic the internal structure of current carrying defect strings via their embedding in the `internal' higher-dimensional space required by string theory, in which three spatial dimensions are extended and $n$ are compactified. 
As a qualitative statement, it is unsurprising that this may \emph{sometimes} be true and, since the number of string models available both in string theory and in various field theories is huge, it is too vague to be of much use in distinguishing string species (or in determining when different species may be indistinguishable) via current or future observations. 
However, it is clear that it would be useful to be able to state under precisely which circumstances a given defect string will be successfully mimicked by the embedding of a higher-dimensional $F$-string. 
\\ \indent
Since cosmic superstring phenomenology offers a probe of high energy physics in the early universe in the presence of extra dimensions, the detection of higher-dimensional signatures from strings would be a key piece of evidence in favour of string theory. 
It is therefore vital for us to understand what distinguishable higher-dimensional signatures may actually exist. 
In this section of the paper, we extend the results for $F$-strings presented in  \cite{LaYo12} to the case of arbitrary planar loops. 
Our purpose is to work towards developing a general formalism through which the EOM governing higher-dimensional $F$-string dynamics may be interpreted as dispersion relations for waves `in' the effective four-dimensional width of the string. 
These can then be compared with dispersion relations which govern the flow of current (and the evolution of genuine internal structure) in field-theoretic strings, and the conditions under which two string species give rise to identical phenomenology may be determined. 
\\ \indent
Any equivalence should also include the string constants of motion, which should be formally analogous under a one-to-one correspondence between string theory and field-theoretic parameters, as in \cite{LaYo12}. The finite-width field-theoretic counterparts to the higher-dimensional $F$-string configurations considered here are presented in Sec. \ref{Sect3}. 
Though, for the sake of brevity, we restrict our present analysis to planar loop configurations, we expect no significant obstacles in extending this treatment to arbitrary $(3+1)$-dimensional configurations. 
Finally, we note that the results presented here also hold for strings wrapping $S^1$ sub-cycles of constant radius in \emph{any} compactified geometry. 
The only caveat required is that the solutions must be dynamically stabilised, where necessary, if topological stability is not guaranteed (as it is for $M^4 \times S^1$ compactification). 
This is the case with the solutions obtained in \cite{BlIg05,LaYo12,LaWa10}, which are valid for strings wrapping great circles in the $S^3$ manifold which regularises the conifold tip of the Klebanov-Strassler geometry \cite{KlSt00}. 

\subsection{The background geometry, equations of motion, and physical observables for wound-strings}\label{Sect2.1}
In this section we review the background geometry, EOM, and physical observables of the wound $F$-string model. 

\subsubsection{The background geometry, string action and the Euler-Lagrange equations for wound-strings} \label{Sect2.1.1}
We use the metric signature $(+---)$ and consider a background space-time with the line element 
\begin{eqnarray} \label{Met_2.1}
{\rm d}s^2 = g_{IJ} {\rm d}x^{I}{\rm d}x^{J} = a^2\eta_{\mu\nu}{\rm d}x^{\mu}{\rm d}x^{\nu} - R^2{\rm d}\varphi^2,
\end{eqnarray}
where $\mu,\nu \in \left\{0,1,2,3\right\}$, $\eta_{\mu\nu}$ is the Minkowski metric, $\varphi \in [0,2\pi)$ is an angular coordinate in the internal space and $R$ is a (constant) radius.
\footnote{The assumption of constant radius is made for simplicity and this condition should be loosened in a more thorough treatment. Interestingly, this could lead to the description of so-called `lumpy' cosmic strings, from a $(3+1)$-dimensional perspective \cite{Lake:2015gra}, but we leave this study to a future work.} 
The phenomenological `warp factor', $a \in (0,1]$, accounts for the fact that the internal dimensions may be flux-compactified, as expected in string theory. 
In this case, $a^2<1$ represents the back reaction of the fluxes on the large dimensions \cite{Denef:2007pq}.  
\\ \indent
Note that we need not literally assume an $M^4 \times S^1$ compactification and, in principle, the internal space may be far more complicated. 
Equation (\ref{Met_2.1}) represents the effective metric `seen' by any string wrapping an $S^1$ sub-cycle of constant radius in an (almost) arbitrary Calabi-Yau manifold. 
The internal space is almost arbitrary, since the only condition we impose is that an $S^1$ sub-cycle of constant radius exists. 
As the string is one-dimensional, it necessarily wraps (topologically) $S^1$ sub-cycles in the any compact space, though these need not be of uniform shape or length. 
For simplicity, we assume windings of constant radius, though it would be interesting to consider the more general case in which the effective compactification scale $R^{{\rm eff}}(\tau,\sigma)$ is a function of the world-sheet coordinates (c.f. \cite{Jacobi}). 
For a constant compactification radius, the part of the background metric seen by the string is of the form given in Eq. (\ref{Met_2.1}), regardless of its macroscopic structure \cite{YaLa15}. 
\footnote{For example, in \cite{LaYo12,LaWa10}, strings wrapping $S^1$ sub-cycles of constant radius at the tip of the Klebanov-Strassler geometry \cite{KlSt00} were considered. Here the target space manifold is $M^4 \times S^3$ and the line element is given by ${\rm d}s^2 = a^2\eta_{\mu\nu}{\rm d}x^{\mu}{\rm d}x^{\nu} - R^2{\rm d}\Omega_3^2$, where ${\rm d}\Omega_3^2$ is the line-element on the unit three-sphere. In Hopf coordinates \cite{MaWo_Hopf}, this is given by ${\rm d}\Omega_3^2 = {\rm d}\psi^2 + \sin^2(\psi){\rm d}\chi^2 + \cos^2(\psi){\rm d}\varphi^2$ where $\psi \in [0,\pi)$ is the polar angle and $\chi,\varphi \in [0,2\pi)$ are the two azimuthal angles. Taking  $\psi(\tau,\sigma)$, $\chi(\tau,\sigma)$ and $\varphi(\tau,\sigma)$ as embedding coordinates for the string, it is clear that the value of $\psi$ controls the effective radius of the windings, which may vary as a function of both $\tau$ and $\sigma$. For $\psi={\rm const.}$, the winding radius is also constant and for $\psi=0$ it takes the maximum value $R$, the radius of the $S^3$. Similar arguments hold true for more complicated manifolds, as long as they contain at least one $S^1$ submanifold of constant radius. The advantage of using the effective metric (\ref{Met_2.1}) is that we do not need to make any further assumptions regarding the general structure of the internal space. Specifically, we do not assume that the string windings will (or will not) be topologically stabilised.}
\\ \indent
In the absence of additional fluxes, the basic string action is the Nambu-Goto action \cite{Go71,Na77} 
\begin{eqnarray} \label{Act_2.1}
S = -\mathcal{T}\int {\rm d}^2\zeta \sqrt{-\gamma}, 
\end{eqnarray}
where $\gamma$ is the the determinant of the induced metric on the world-sheet
\begin{eqnarray} \label{IndMet_2.1}
\gamma_{ab}(X) = g_{IJ}\left(X\right) \partial_{a} X^{I} \partial_{b} X^{J},
\end{eqnarray}
with $I,J \in \left\{0,1,2,3, \dots , d\right\}$, $a,b \in \left\{0,1\right\}$, $\partial_{a}X^{I} = \partial X^{I}/\partial \zeta^{a}$, and $\zeta^{0}=\tau,\zeta^{1}=\sigma$, in an arbitrary number of space-time dimensions, $D = 1 + d$. 
The intrinsic string tension is given by
\begin{eqnarray} \label{FundTens_2.1}
\mathcal{T} = \frac{1}{2\pi \alpha'},
\end{eqnarray}
where $\alpha'$ is the Regge slope parameter, which is related to the (fundamental) string scale via
\begin{eqnarray} \label{StrLen_2.1}
l_{\rm st} = \sqrt{\alpha'}.
\end{eqnarray}
In the formulae above, and throughout the rest of this paper, we work in natural units, $\hbar = 1$, $c = 1$. 
Variation of the action with respect to the embedding coordinates gives the canonical Euler-Lagrange equations, plus a boundary term
\begin{eqnarray} \label{BoundTerm_2.1}
\left[\mathcal{P}^{\sigma}{}_I(\tau,\sigma) \delta X^I  \right]_{0}^{\sigma_f}  = 0,
\end{eqnarray} 
where
\begin{eqnarray} \label{CanMom_2.1}
\mathcal{P}^{a}{}_I(\tau,\sigma) = \frac{\partial(\mathcal{L}\sqrt{-\gamma})}{\partial (\partial_a X^I)}
\end{eqnarray}
is the canonical momentum of $X^{I}$ with respect to $\zeta^{a}$. 
To satisfy the boundary term, we may impose Dirichlet, Neumann or periodic boundary conditions 
\begin{eqnarray} \label{Dir_2.1} 
X^{I}(\tau,0) = {\rm const.}, \ \ \ X^{I}(\tau,\sigma_f) = {\rm const.},
\end{eqnarray}
\begin{eqnarray} \label{Neu_2.1}
\mathcal{P}^{\sigma}{}_I(\tau,0) = \mathcal{P}^{\sigma}{}_I(\tau,\sigma_f) = 0,
\end{eqnarray}
\begin{eqnarray} \label{Per_2.1}
X^{I}(\tau,\sigma) = X^{I}(\tau,\sigma + m\sigma_f), \ \ \ m \in \mathbb{Z},
\end{eqnarray}
where the imposition of Neumann boundary conditions implies that the string end points move at the speed of light \cite{Zwi09}.
Using the identity $\delta(-\gamma) = (-\gamma)\gamma^{ab}\delta\gamma_{ab}$, the canonical EOM may also be rewritten in the form \cite{And02},
\begin{eqnarray} \label{AltEL_2.1}
\frac{\partial}{\partial \zeta^{a}}\left(\sqrt{-\gamma}\gamma^{ab}g_{IJ}(X)\partial_{b}X^{J}\right) - \frac{1}{2}\sqrt{-\gamma}\gamma^{cd}\frac{\partial g_{KL}(X)}{\partial X^{I}}\partial_{c}X^{K}\partial_{d}X^{L}= 0.
\end{eqnarray}

\subsubsection{The space-time energy-momentum tensor, constants of motion, and effective pressures and shears for wound-strings} \label{Sect2.1.2}
The space-time energy-momentum tensor is defined by varying the action with respect to the metric $g_{IJ}(x)$
\begin{eqnarray} \label{T_mu_nu*}
\delta S = \int T^{IJ} \delta g_{IJ} \sqrt{-g}{\rm d}^dx,
\end{eqnarray}
where from here on we use $x^{I}$ to refer to space-time coordinates and $X^{I}$ to refer to the embedding coordinates of the string. 
For the Nambu-Goto action, this gives \cite{And02,ViSh00}
\begin{eqnarray} \label{T_mu_nu}
T^{IJ} = \frac{1}{\sqrt{-g}}\mathcal{T}\int \sqrt{-\gamma} \gamma^{ab}\partial_{a}X^{I}\partial_{b}X^{J} \delta^D(x-X){\rm d}\tau {\rm d}\sigma.
\end{eqnarray}
In the static gauge, $X^{0} = t = \zeta \tau$, where $\zeta$ is constant with dimensions of length, this gives
\begin{eqnarray} \label{StatGaugeEMT}
T^{IJ} = \frac{1}{\sqrt{-g}}\zeta^{-1}\mathcal{T}\int \sqrt{-\gamma} \gamma^{ab}\partial_{a}X^{I}\partial_{b}X^{J} \delta^d(x-X){\rm d}\sigma.
\end{eqnarray}
The conserved currents associated with the space-time symmetries are
\begin{eqnarray} \label{}
J^{(I)A} = K^{(I)}{}_{B}T^{AB}
\end{eqnarray}
where $K^{(I)}{}_{B}$ are the Killing vectors, so that the constants of motion are given by
\begin{eqnarray} \label{CoM_2.2}
\Pi^{I} =  \int K^{(I)}{}_{B}T^{0B} \sqrt{-g} {\rm d}^{d}x = \mathcal{T}\int \sqrt{-\gamma} \gamma^{\tau b}K^{(I)}{}_{B}\partial_{b}X^{B} {\rm d}\sigma.
\end{eqnarray}
We also define the covariant components
\begin{eqnarray} \label{4MomCov_2.2}
\Pi_{I} =  \int K_{(I)B}T^{0B} \sqrt{-g} {\rm d}^{d}x = \mathcal{T}\int \sqrt{-\gamma} \gamma^{\tau b}g_{IJ}(X)K^{(J)}{}_{B}\partial_{b}X^{B} {\rm d}\sigma,
\end{eqnarray}
though these are not conserved if the index $I$ refers to a non-Cartesian coordinate. 
In general, indices in the space-time energy-momentum tensor are raised and lowered using the background metric
\begin{eqnarray} \label{RaiLow_2.2}
T^{I}{}_{J} = g_{JK}(x)T^{IK},
\end{eqnarray}
but the integral over ${\rm d}^dx$ together with the factor of $\delta^d(x-X)$ in the expressions above, fixes $x = X$. 
\\ \indent 
Using the letters $E$, $P$ and $l$ to denote energy, linear momentum and angular momentum, respectively, we then define the warp factor-averaged values of the constants of motion via
\begin{eqnarray} \label{En_2.2}
E = \sqrt{P^{0}P_{0}}, \ \ \ \mathcal{P}^{I} = \pm \sqrt{-P^{I}P_{(I)}}, \ \ \ \Lambda^{J} = \pm \sqrt{-l^{J}l_{(J)}(t_i)},
\end{eqnarray}
where the space-time index $I$ labels a Cartesian coordinate and $J$ an angular coordinate in the background space. 
The presence of of brackets around a repeated index implies that no summation is implied. 
For unwarped background geometries, these recover the standard expressions when $a^2=1$. 
By convention, we take the signs of $\mathcal{P}^{I}$ and $\Lambda^{J}$ to match those of $P^{I}$ and $l^{J}$, and $t_i$ denotes the initial time at which we consider the state of the system. 
Such definitions are problematic unless the metric $g_{IJ}(X)$ is independent of $\sigma$. 
However, this can be achieved by adopting a gauge in which $\sigma$ is identified directly with a background space-time coordinate, such that $X^{I} \propto \sigma$ for some $I$, and the implications of this for the string EOM and conservation equations are considered in Sec. \ref{Sect2.1.3}. 
It is then straightforward to define the warp factor-averaged total linear and angular momentum
\begin{eqnarray} \label{TotLinMon_2.2}
\mathcal{P} = \pm \sqrt{-P^{I}P_{I}}, \ \ \ \Lambda = \pm \sqrt{-l^{J}l_{J}(t_i)},
\end{eqnarray}
and the total $4$-momentum
\begin{eqnarray} \label{Tot4Mom_2.2}
\Pi = +\sqrt{\Pi^{I}\Pi_{I}} = +\sqrt{E^2 - \mathcal{P}^2 - \Lambda^2}.
\end{eqnarray}
Finally, we define the contravariant and covariant integrated pressures and shears via 
\begin{eqnarray} \label{PressCont_2.2}
T^{K} = \int T^{KK} \sqrt{-g} {\rm d}^dx, \ \ \ T_{K} = \int T^{(K)}{}_{K}(t_i) \sqrt{-g} {\rm d}^dx,
\end{eqnarray}
for $K \neq 0$, and
\begin{eqnarray} \label{ShearCont_2.2}
S^{KL} = \int T^{KL}  \sqrt{-g} {\rm d}^dx, \ \ \  S^{K}{}_{L} = \int T^{K}{}_{L}(t_i)  \sqrt{-g} {\rm d}^dx,
\end{eqnarray}
for $K,L \neq 0$, $K \neq L$, so that the warp-averaged values are
\begin{eqnarray} \label{TotPress_2.2}
\mathcal{T}^{K} = \pm \sqrt{-T^{K}T_{(K)}(t_i)}, \ \ \ \Sigma^{KL} = \pm \sqrt{S^{K}{}_{(L)}S_{(K)}{}^{L}(t_i)}.
\end{eqnarray}
\\ \indent 
The magnitudes of the total integrated pressures and shears are given by $\mathcal{T} = +\sqrt{-T^{K}T_{K}}$ and $\Sigma = +\sqrt{S^{K}{}_{L}S_{K}{}^{L}}$, where summation is now permitted over repeated indices. In addition to giving warp factor averaged values for the conserved quantities in warped geometries, the definitions above are also convenient in the sense that the $\mathcal{T}^{K}$ always have canonical units of tension/pressure $[E][l]^{-1} = [l]^{-2}$, whereas neither $T^{K}$ nor $T_{K}$ do, individually, unless $X^{K}$ represents a Cartesian coordinate. 

\subsubsection{The one-to-one equivalence of the EOM and conservation equations in a physically intuitive coordinate system} \label{Sect2.1.3}
The conservation of the space-time energy-momentum tensor may be expressed as \cite{Dir75}
\begin{eqnarray} \label{ConsLawAlt_2.3}
\nabla_{J}T^{J}{}_{I}\sqrt{-g} = \frac{\partial}{\partial x^{J}}\left(T^{J}{}_{I}\sqrt{-g}\right)  - \frac{1}{2} \frac{\partial g_{AB}(x)}{\partial x^{I}}T^{AB}\sqrt{-g} = 0.
\end{eqnarray}
Substituting the energy-momentum tensor for the Nambu-Goto string gives
\begin{eqnarray} \label{SubsEMT1_2.3}
&{}& \frac{\partial}{\partial x^{J}}\left[\int {\rm d}\tau {\rm d}\sigma \delta^{D}(x-X)\sqrt{-\gamma}\gamma^{ab}g_{IK}(x)\partial_{a}X^{J}\partial_{b}X^{K}\right] 
\nonumber\\
&-& \frac{1}{2}\frac{\partial g_{AB}(x)}{\partial x^{I}}\int {\rm d}\tau {\rm d}\sigma \delta^{D}(x-X)\sqrt{-\gamma}\gamma^{cd}\partial_{c}X^{A}\partial_{d}X^{B} = 0.
\end{eqnarray}
Restricting ourselves to embeddings in which the world-sheet coordinates can be directly identified with two background space-time coordinates 
\begin{eqnarray} \label{Emb/2.3}
X^{0} = x^{0} = \zeta \tau, \ \ \ X^{1} = x^{0} = \chi \sigma,
\end{eqnarray}
where $\zeta,\chi={\rm const.}$, so that
\begin{eqnarray} \label{IntMeas_2.3}
{\rm d}\tau {\rm d}\sigma = \zeta^{-1}\chi^{-1}{\rm d}X^{0} {\rm d}X^{1}, 
\end{eqnarray}
and integrating over ${\rm d}^{d}x$, gives
\begin{eqnarray} \label{SubsEMT3_2.3}
&{}& \frac{\partial}{\partial x^{0}} \left[g_{IK}(X)\sqrt{-\gamma}\gamma^{ab}\partial_{a}x^{0}\partial_{b}X^{K}\right] 
+ \frac{\partial}{\partial x^{1}} \left[g_{IK}(X)\sqrt{-\gamma}\gamma^{ab}\partial_{a}x^{1}\partial_{b}X^{K}\right]
\nonumber\\
&+& \int {\rm d}^{D-2}x\left[\frac{\partial \delta^{D-2}(x-X)}{\partial x^{M}} g_{IK}(x) + \frac{\partial g_{IK}(x)}{\partial x^{M}} \delta^{D-2}(x-X) \right]\sqrt{-\gamma}\gamma^{ab}\partial_{a}X^{M}\partial_{b}X^{K}
\nonumber\\
&-& \frac{1}{2} \frac{\partial g_{AB}(X)}{\partial X^{I}}\sqrt{-\gamma}\gamma^{cd}\partial_{c}X^{A}\partial_{d}X^{B} = 0, 
\end{eqnarray}
for $M \notin \left\{0,1\right\}$. Using $\partial_{a}x^{0} \propto \delta^{0}{}_{a}$ and $\partial_{a}x^{1} \propto \delta^{1}{}_{a}$, together with the identity \cite{MaWo_DiracDelta},
\begin{eqnarray} \label{Ident_2.3}
\int {\rm d}x f(x) \frac{{\rm d} \delta(x)}{{\rm d}x} = -\int {\rm d}x \delta(x) \frac{{\rm d} f(x)}{{\rm d}x},
\end{eqnarray}
this becomes
\begin{eqnarray} \label{SubsEMT4_2.3}
&{}& \partial_{a}x^{0}\frac{\partial}{\partial x^{0}} \left[g_{IK}(X)\sqrt{-\gamma}\gamma^{ab}\partial_{b}X^{K}\right] + \partial_{a}x^{1}\frac{\partial}{\partial x^{1}} \left[g_{IK}(X)\sqrt{-\gamma}\gamma^{ab}\partial_{b}X^{K}\right]
\nonumber\\
&+& \left\{-\left[\frac{\partial g_{IK}(X)}{\partial X^{M}}\right]_{\partial X^{M}} + \left[\frac{\partial g_{IK}(X)}{\partial X^{M}}\right]_{\partial X^{M}} \right\} \sqrt{-\gamma}\gamma^{ab}\partial_{a}X^{M}\partial_{b}X^{K}
\nonumber\\
&-& \frac{1}{2} \frac{\partial g_{AB}(X)}{\partial X^{I}}\sqrt{-\gamma}\gamma^{cd}\partial_{c}X^{A}\partial_{d}X^{B} = 0. 
\end{eqnarray}
Canceling the boundary terms and rewriting the summation over world-sheet indices in a compact form yields the Euler-Lagrange equations obtained using the metric variation identity, Eq. (\ref{AltEL_2.1}). Hence, in this simple gauge, there is one-to-one correspondence between the string EOM and the conservation equations for the space-time energy-momentum tensor. For a given index $I$, the corresponding equations are identical.

\subsubsection{Definition of the higher-dimensional wound-string model parameters} \label{Sect2.1.4}
Let $l(\sigma)$ denote the length of string in the interval $[0,\sigma]$ at time $t$. In a $D=1+d$ dimensional space-time
\begin{eqnarray} \label{dl_2.4}
{\rm d}l^2 = -g_{IJ}(X)\partial_{\sigma}X^{I}\partial_{\sigma}X^{J}{\rm d}\sigma^2,
\end{eqnarray}
where $I,J \in \left\{1,2,3 . . . d\right\}$ denote only spatial coordinates. 
This may be split into the sum of a three-dimensional and an extra-dimensional part
\begin{eqnarray} \label{dlsum_2.4}
{\rm d}l^2 = {\rm d}l_{3}^2 + {\rm d}l_{d-3}^2,
\end{eqnarray}
where
\begin{eqnarray} \label{dl3D_2.4}
{\rm d}l_{3}^2 = -g_{kl}(X)\partial_{\sigma}X^{k}\partial_{\sigma}X^{l}{\rm d}\sigma^2 
\end{eqnarray}
for $k,l \in \left\{1,2,3\right\}$, and 
\begin{eqnarray} \label{dlExD_2.4}
{\rm d}l_{d-3}^2 = -g_{mn}(X)\partial_{\sigma}X^{m}\partial_{\sigma}X^{n}{\rm d}\sigma^2 
\end{eqnarray}
for $m,n \in \left\{4,5,6, \dots , d\right\}$. 
The (physical) perpendicular velocity of the string is defined in the static gauge in terms of $l(\sigma)$ \cite{Zwi09}
\begin{eqnarray} \label{vperp_2.4}
v_{\perp}^2 = g_{IJ}(X)\frac{\partial X^{I}}{\partial t}\frac{\partial X^{J}}{\partial t} - \left(g_{IJ}(X)\frac{\partial X^{I}}{\partial l}\frac{\partial X^{J}}{\partial l}\right)^2, \ I,J \in \left\{1,2,3 . . . d\right\},
\end{eqnarray}
so that in warped backgrounds the action may be written as
\begin{eqnarray} \label{ActWarp_2.4}
S = -\mathcal{T}a\int {\rm d}t {\rm d}\sigma \left(\frac{{\rm d}l}{{\rm d}\sigma}\right)\sqrt{1-\frac{v_{\perp}^2}{a^2}}.
\end{eqnarray}
This reduces to the standard form given in  \cite{Zwi09} for $a =1$. 
Thus, in warped geometries, the velocity of open string end points obeying Neumann boundary conditions is given by $v_{\perp}^2(t,0) = v_{\perp}^2(t,2\pi) = a^2$.
\\ \indent
We note that it is not permissible to substitute the ansatz (or partial ansatz) at the level of the action before varying to obtain the EOM. 
The EOM used throughout the rest of this paper are those obtained from the covariant Euler-Lagrange equations for the string, Eq. (\ref{AltEL_2.1}). 
However, this representation of the action is often useful, as it provides physical insight into the motion of the string. 
\\ \indent
In the study of higher-dimensional strings, it is also useful to define the parameter $\Omega^{2}(t) \in [0,1]$, which represents the (time-dependent) fraction of the string lying in the large spatial dimensions, via
\begin{eqnarray} \label{Omega_2.4}
\Omega^{-2}(t) = \frac{1}{2\pi}\int_{0}^{2\pi}\omega^{-2}(t,\sigma){\rm d}\sigma,
\end{eqnarray}
where $\omega^{2}(t,\sigma) \in [0,1]$ represents the instantaneous local value, at a given point $\sigma$ and at time $t$. In terms of our previously defined notation, this is given as
\begin{eqnarray} \label{omega_2.4}
\omega^{-2}(t,\sigma) = 
\left(\frac{{\rm d} l}{{\rm d} \sigma}\right)^2  \left(\frac{{\rm d} l_{3}}{{\rm d} \sigma}\right)^{-2} = \frac{g_{kl}(X)\partial_{\sigma}X^k\partial_{\sigma}X^l + g_{mn}(X)\partial_{\sigma}X^m \partial_{\sigma}X^n}{g_{kl}\partial_{\sigma}X^k\partial_{\sigma}x^l}.
\end{eqnarray}

\subsection{Long $F$-strings} \label{Sect2.2}
In this section, we consider long straight strings, both with and without higher-dimensional windings. Long strings with higher-dimensional windings are considered in detail in Sec. \ref{Sect2.2.2}, so that it is instructive to first review the treatment of long straight strings without windings, a brief analysis of which is presented in Sec. \ref{Sect2.2.1}.

\subsubsection{Long $F$-strings in $(3+1)$ dimensions} \label{Sect2.2.1}
Let us adopt Cartesian coordinates for the background metric
\begin{eqnarray} \label{Met_3.1}
{\rm d}s^2 = a^2({\rm d}t^2-{\rm d}x^2-{\rm d}y^2-{\rm d}z^2) - R^2{\rm d}\varphi^2,
\end{eqnarray}
and take the embedding
\begin{eqnarray} \label{Emb_3.1}
X^I(\tau,\sigma) \equiv X^I(t,\sigma) \equiv X^I(t,z) =\left(t(\tau)=\zeta \tau, x=0, y=0, z(\sigma)=(2\pi)^{-1}\Delta \sigma, \varphi=0 \right).
\end{eqnarray}
Physically, this represents either a string of length $\Delta$, lying along the $z$-axis, whose end points are fixed by Dirichlet boundary conditions, or a finite section of (formally infinite) string, obeying periodic boundary conditions. 
The action may be written as
\begin{eqnarray} \label{Act_3.1}
S = -\mathcal{T}a^2\Delta \int  {\rm d}t,
\end{eqnarray}
which is equivalent to the form given in Eq. (\ref{ActWarp_2.4}), with
\begin{eqnarray} \label{dl_3.1}
\frac{dl}{d\sigma}  = (2\pi)^{-1}a\Delta, \ \ \ \frac{v_{\perp}^2}{a^2}=0. 
\end{eqnarray}
The EOM are
\begin{subequations}
\begin{align}
\partial_{0}(T^{0}{}_{0}\sqrt{-g}) + \partial_{z}(T^{z}{}_{0}\sqrt{-g}) &= 0, \label{EOM_3.1a}\\
\partial_{0}(T^{0}{}_{z}\sqrt{-g}) + \partial_{z}(T^{z}{}_{z}\sqrt{-g}) &=  0. \label{EOM_3.1b}
\end{align}
\end{subequations}
From here on, we use the following notation:
\begin{eqnarray} \label{EMT_dens}
T^{I}{}_{J} \sqrt{-g} = \int \mathcal{T}^{I}{}_{J} \sqrt{-g} \delta^{d}(x-X){\rm d}\sigma
\end{eqnarray}
where $\mathcal{T}^{I}{}_{J}\sqrt{-g}$ denotes the energy-momentum tensor \emph{density}. Thus, $\mathcal{T}^{I}{}_{J}\sqrt{-g}$ has the same dimensions as $T^{I}{}_{J}\sqrt{-g}$, but must be multiplied by an appropriate delta function and integrated over ${\rm d}\sigma$ to give the true energy-momentum tensor of the string. 
For the sake of notational simplicity, we quote only the relevant components of $\mathcal{T}^{I}{}_{J}\sqrt{-g}$ for the various systems considered in the remainder of this work. 
For the long straight string, the components of $\mathcal{T}^{I}{}_{J}\sqrt{-g}$ appearing in the EOM are given by
\begin{subequations}
\begin{align}
\mathcal{T}^{0}{}_{0}\sqrt{-g} = \mathcal{T} a^2 (2\pi)^{-1}\Delta, \ \ \ \mathcal{T}^{z}{}_{0}\sqrt{-g} = 0, \label{EMT_3.1a}\\
\mathcal{T}^{0}{}_{z}\sqrt{-g} = 0, \ \ \ \mathcal{T}^{z}{}_{z}\sqrt{-g} = \mathcal{T} a^2 (2\pi)^{-1}\Delta. \label{EMT_3.1b}
\end{align}
\end{subequations}
Clearly, the EOM are trivially satisfied. 
The only nonzero constants of motion are the energy and the integrated pressure in the $z$-direction,
\begin{eqnarray} \label{E_3.1}
E = \mathcal{T} a \Delta, \ \ \ \mathcal{T}^{z} = -\mathcal{T} a \Delta.
\end{eqnarray}
These are equal in magnitude but opposite in sign, as for vacuum strings in field-theoretic models \cite{ViSh00}.

\subsubsection{Long $F$-strings with higher-dimensional windings} \label{Sect2.2.2}
Let us again adopt Cartesian coordinates for the background space and consider the new wound-string embedding
\begin{eqnarray} 
X^I(\tau,\sigma) \equiv X^I(t,\sigma) \equiv X^I(t,z) =\left(t(\tau)=\zeta \tau, x=0, y=0, z(\sigma)=(2\pi)^{-1}\Delta \sigma, \varphi(t,\sigma) \equiv \varphi(t,z) \right).
\nonumber
\end{eqnarray}
\begin{eqnarray} \label{Emb_3.2}
{}
\end{eqnarray}
Physically, this represents a finite section of an infinite string, so that we must impose periodic boundary conditions on $\varphi(t,\sigma)$. 
To consider a section of open string, obeying Neumann boundary conditions, we would need to adopt an embedding for the $z$-coordinate of the form $z(t,\sigma) = (2\pi)^{-1}\Delta\sigma + g(t)$. 
This corresponds to a comoving coordinate system which `tracks' the propagation of the higher-dimensional windings from a four-dimensional perspective (i.e. along the $z$-axis). 
In this case, the analysis is a straightforward generalization of that presented below but, from an  astrophysical/cosmological perspective, long strings are greater interest than open ones. 
We therefore restrict our current analysis to finite sections of long string. 
\\ \indent
For the embedding (\ref{Emb_3.2}), the action may be written as
\begin{eqnarray} \label{Act_3.2}
S = -\mathcal{T} \int {\rm d}t {\rm d}\sigma \sqrt{a^2((2\pi)^{-2}a^2\Delta^2+R^2(\partial_{\sigma}\varphi)^2)-a^2R^2 (2\pi)^{-2}\Delta^2\dot{\varphi}^2},
\end{eqnarray}
which is equivalent to the static gauge form (\ref{ActWarp_2.4}) with
\begin{eqnarray} \label{dl_3.2}
\frac{dl}{d\sigma} = \sqrt{(2\pi)^{-2}a^2\Delta^2 + R^2(\partial_{\sigma}\varphi^2)}, \ \ \ 
\frac{v_{\perp}^2}{a^2} = \frac{\Delta^2R^2\dot{\varphi}^2}{a^2\Delta^2 + (2\pi)^2R^2(\partial_{\sigma}\varphi^2)}.
\end{eqnarray}
The EOM are
\begin{subequations} 
\begin{align}
\partial_{0}(T^{0}{}_{0}\sqrt{-g}) + \partial_{z}(T^{z}{}_{0}\sqrt{-g}) &= 0, \label{EOM_3.2a}\\
\partial_{0}(T^{0}{}_{z}\sqrt{-g}) + \partial_{z}(T^{z}{}_{z}\sqrt{-g}) &=  0, \label{EOM_3.2b}\\
\partial_{0}(T^{0}{}_{\varphi}\sqrt{-g}) + \partial_{z}(T^{z}{}_{\varphi}\sqrt{-g}) &= 0.\label{EOM_3.2c}
\end{align}
The EOM in $\phi$ may be written explicitly as
\end{subequations}
\begin{eqnarray} \label{phiEOMExp_3.2}
(a^2 + R^2(\varphi'^2-\dot{\varphi}^2))(\phi'' - \ddot{\varphi}) + a^2R^2(2\dot{\varphi}\varphi'\dot{\varphi}' - \varphi'^2\varphi'' - \dot{\varphi}^2\ddot{\varphi}) = 0,
\end{eqnarray}
where a dash represents differentiation with respect to $z$. 
This is satisfied for any function obeying
\begin{eqnarray} \label{phi(t,z)EOM_3.2}
\dot{\varphi}^2 - \varphi'^2 = 0, \ \ \ \dot{\varphi} = \pm \varphi'
\end{eqnarray}
which automatically implies 
\begin{eqnarray} \label{phi(t,z)EOM*_3.2} 
\ddot{\varphi} - \varphi'' = 0.
\end{eqnarray}
However, Eq. (\ref{phi(t,z)EOM_3.2}) is a stronger condition than Eq. (\ref{phi(t,z)EOM*_3.2}), since the latter admits superpositions of both left and right-movers, whereas the former does not. 
Equation (\ref{phi(t,z)EOM_3.2}) is satisfied for any function of the form
\begin{eqnarray} \label{phi(t,z)_3.2}
\varphi(t,z) = \varphi(k_z z + \omega_z t) \equiv \varphi(t,\sigma) = \varphi(m\sigma + \omega_z t),
\end{eqnarray}
where
\begin{eqnarray} \label{DispRel_3.2*} 
\omega_z^2 = k_z^2, \ \  \omega_z = \pm k_z
\end{eqnarray}
and $m \in \mathbb{Z}$. Defining
\begin{eqnarray} \label{WavNo_3.2}
k_z = 2\pi/\lambda_z, 
\end{eqnarray}
and imposing periodic boundary conditions on $\varphi(t,z)$, 
\begin{eqnarray} \label{phi(t,z)BC_3.2}
\varphi(k_z z + \omega_z t) =  \varphi(k_z (z + \Delta) + \omega_z t) \equiv \varphi(m\sigma + \omega_z t) =  \varphi(m(\sigma + 2\pi) + \omega_z t),
\end{eqnarray}
it follows that
\begin{eqnarray} \label{WavLen_3.2}
\Delta = m\lambda_{z}.
\end{eqnarray}
\indent
Different plane wave modes in the Fourier expansion of $\varphi(z \pm t)$ correspond to different frequencies $\omega_j$ and wavelengths $\lambda_j$, such that $\omega_j = \pm k_j = \pm 2\pi/\lambda_j = \pm 2\pi m_j/ \Delta$, for some $m_j \in \mathbb{N}$. 
But, by writing $\varphi(z \pm t)$ in the form (\ref{phi(t,z)_3.2}), which implies $\partial \varphi/\partial t = \omega_z {\rm d}\varphi/{\rm d}u$, $\partial \varphi/\partial z = k_z {\rm d}\varphi/{\rm d}u$,  $\partial \varphi/\partial \sigma = m {\rm d}\varphi/{\rm d}u$, where $u = k_z z + \omega_z t = m\sigma + \omega_z t$, we select a single mode as being characteristic of the wave form. 
The natural choice is the mode with the highest amplitude which, for nonlinear functions of $u$, gives the approximate wavelength of any fluctuation in current density, as seen form a four-dimensional perspective. 
If $\varphi$ contains linear terms in $z$ and $t$, it is most natural to use the associated $k_z$ and $\omega_z$ to give the characteristic wavelength and frequency. 
Linear terms give rise to a uniform current and any additional nonlinear terms describe local fluctuations in current density around the mean value. 
In this case, the integer $m$ may be interpreted as the number of windings present in a four-dimensional string section of length $\Delta$. 
\\ \indent
Though we may express any functions of $t$ and $z$ as functions of $t$ and $\sigma$, or vice-versa, from here on we choose the latter, as this makes the connection between the long string EOM and their generalisation to loops in Sec. \ref{Sect2.3} more explicit. 
The functions $\omega^{2}(t,\sigma)$ and $\Omega^{2}(t)$ are given as
\begin{eqnarray} \label{omega_3.2}
\omega^{-2}(t,\sigma) = 1 + \frac{(2\pi)^2R^2}{a^2\Delta^2}(\partial_{\sigma}\varphi)^2, \ \ \ 
\Omega^{-2}(t) = 1 + \frac{2\pi R^2}{a^2\Delta^2}\int_{0}^{2\pi}(\partial_{\sigma}\varphi)^2{\rm d}\sigma,
\end{eqnarray}
and, after substituting for $\dot{\varphi}$, the components of $\mathcal{T}^{I}{}_{J}\sqrt{-g}$ appearing in the EOM may be written as
\begin{subequations} \label{+++}
\begin{align}
\mathcal{T}^{0}{}_{0}\sqrt{-g} &= \mathcal{T} a^2 (2\pi)^{-1}\Delta \omega^{-2}, \ \ \ \ \ \ \ \ \ \ \ \ \ \mathcal{T}^{z}{}_{0}\sqrt{-g} = \mp \mathcal{T} a^2 (2\pi)^{-1}\Delta \left(\frac{1-\omega^{2}}{\omega^{2}}\right), \label{EMT_3.2a*}\\
\mathcal{T}^{0}{}_{z}\sqrt{-g} &= \pm \mathcal{T} a^2 (2\pi)^{-1}\Delta \left(\frac{1-\omega^{2}}{\omega^{2}}\right), \ \mathcal{T}^{z}{}_{z}\sqrt{-g} = \mathcal{T} a^2 (2\pi)^{-1}\Delta \left(\frac{2\omega^{2}-1}{\omega^{2}}\right), \label{EMT_3.2b*}\\
\mathcal{T}^{0}{}_{\varphi}\sqrt{-g} &= \pm \mathcal{T} a (2\pi)^{-1}\Delta R \frac{\sqrt{1-\omega^{2}}}{\omega}, \ \ \ \mathcal{T}^{z}{}_{\varphi}\sqrt{-g} = \mp  \mathcal{T} a (2\pi)^{-1}\Delta R \frac{\sqrt{1-\omega^{2}}}{\omega}. \label{EMT_3.2c*}
\end{align}
\end{subequations}
Following \cite{Ni79}, we define the components of the current density as
\begin{eqnarray} \label{J_compts}
\mathcal{J}^{0}\sqrt{-g}  = \frac{e}{2\pi}\dot{\varphi}, \ \ \ \mathcal{J}^{z}\sqrt{-g}  = \frac{e}{2\pi}\varphi'.
\end{eqnarray}
By analogy with Eq. (\ref{EMT_dens}) the components of the effective world-sheet current, from a $(3+1)$-dimensional perspective, are given by
\begin{eqnarray} \label{J_dens}
J^{\mu} \sqrt{-g} = \int \mathcal{J}^{\mu} \sqrt{-g} \delta^{d}(x-X){\rm d}\sigma
\end{eqnarray}
but, for the sake of notational simplicity, we quote only the relevant values of $\mathcal{J}^{\mu}$ from here on. 
Using the EOM in $\varphi$, we have
\begin{eqnarray} \label{J_reln}
\mathcal{J}^{0}\sqrt{-g}  = \pm \frac{e}{2\pi}\varphi' =  -\mathcal{J}^{z}\sqrt{-g}  = \mp \frac{e}{2\pi}\dot{\varphi},
\end{eqnarray}
which may be written in terms of the higher-dimensional string variables as
\begin{eqnarray} \label{J_omega*}
\mathcal{J}^{0}\sqrt{-g}  = \frac{e}{2\pi \Delta R^2  \mathcal{T}}\mathcal{T}^{0}{}_{\varphi}\sqrt{-g} = -\mathcal{J}^{z}\sqrt{-g}  = \frac{e}{2\pi \Delta R^2  \mathcal{T}}\mathcal{T}^{z}{}_{\varphi}\sqrt{-g}.
\end{eqnarray}
Equation (\ref{J_omega*}) makes the link between the components of the effective $(3+1)$-dimensional current and the higher-dimensional components of the energy-momentum tensor explicit. 
Specifically, we see that the $(3+1)$-dimensional current density, at a given point on the string, is proportional to the momentum density in the compact direction at that point. 
Hence, the conserved current is proportional to the conserved momentum in the internal space. 
From Eqs. (\ref{J_dens})-(\ref{J_reln}), it follows that $\partial_{\mu}(J^{\mu}\sqrt{-g})=0$ and the current conservation equation is equivalent to Eq. (\ref{EMT_3.2c*}).
\\ \indent
The EOM form a self-consistent set, since each equation is equivalent to
\begin{eqnarray} \label{omega(t,sig)EOM_3.2}
\dot{\omega}^2 = \frac{(2\pi)^2}{\Delta^2}(\partial_{\sigma}\omega)^2, \ \ \ \dot{\omega} = \pm  \frac{2\pi}{\Delta}\partial_{\sigma}\omega^2,
\end{eqnarray}
which is in turn equivalent to Eq. (\ref{phi(t,z)EOM_3.2}). 
Substituting for  $\dot{\varphi}$ in Eq. (\ref{dl_3.2}), the perpendicular velocity and string length may be written in terms of the higher-dimensional string variables as
\begin{eqnarray} \label{vperp_3.2}
\frac{{\rm d}l}{{\rm d}\sigma} = (2\pi)^{-1}a\Delta \omega^{-1},  \ \ \ \frac{v_{\perp}^2}{a^2} = (1-\omega^{2}). 
\end{eqnarray}
The net number of windings in the compact space, which are distributed along the $z$-direction, is given by
\begin{eqnarray} \label{n_z_3.2}
n_z = \frac{1}{2\pi}\int_{0}^{2\pi}\partial_{\sigma}\varphi \, {\rm d}\sigma,
\end{eqnarray}
and for convenience we may split $\varphi$ into linear and nonlinear parts
\begin{eqnarray} \label{phi_NL_3.2}
\varphi(t,\sigma) = m\sigma + \omega_z t + \varphi_{\rm NL}(m\sigma + \omega_z t ).
\end{eqnarray}
However, by the periodic boundary conditions, $\varphi_{\rm NL}$ makes no contribution to the net winding number and
\begin{eqnarray} \label{n_z_3.2}
n_z = m.
\end{eqnarray}
In terms of integrals over $\varphi_{\rm NL}$, the constants of motion are
\begin{eqnarray} \label{E2_3.2}
E = \mathcal{T} a \Delta\left[1 + \frac{2\pi R^2}{a^2\lambda_z^2}\int_{0}^{2\pi}\left(1 + \frac{\partial_{\sigma}\varphi_{\rm NL}}{n_z}\right)^2 {\rm d}\sigma \right],
\end{eqnarray}
\begin{eqnarray} \label{P^z2_3.2}
\mathcal{P}^{z} = \pm 2\pi \mathcal{T} \times \frac{n_z R^2}{a\lambda_z} \int_{0}^{2\pi}\left(1 + \frac{\partial_{\sigma}\varphi_{\rm NL}}{n_z}\right)^2 {\rm d}\sigma,
\end{eqnarray}
\begin{eqnarray} \label{Lam^phi2_3.2}
\Lambda^{\varphi} = \pm 2\pi \mathcal{T} a R n_z, 
\end{eqnarray}
\begin{eqnarray} \label{Pi2_3.2}
\Pi = \mathcal{T} a \Delta\sqrt{1 + \frac{4\pi n_z^2 R^2}{a^2\Delta^2}\int_{0}^{2\pi}\left(1 + \frac{\partial_{\sigma}\varphi_{\rm NL}}{n_z}\right)^2 {\rm d}\sigma -  \frac{(2\pi)^2R^2n_z^2}{a^2\Delta^2}}.
\end{eqnarray}
The integrated tensions are
\begin{eqnarray} \label{T^z1_3.2}
\mathcal{T}^{z} = -\mathcal{T} a \Delta\left[1 - \frac{2\pi R^2}{a^2\lambda_z^2}\int_{0}^{2\pi}\left(1 + \frac{\partial_{\sigma}\varphi_{\rm NL}}{n_z}\right)^2 {\rm d}\sigma \right],
\end{eqnarray}
\begin{eqnarray} \label{T^phi_3.2}
\mathcal{T}^{\varphi} = 0,
\end{eqnarray}
since $\mathcal{T}^{\varphi\varphi}\sqrt{-g}=0$ by the EOM for $\varphi$, and the only nonzero component of the shear, which is between the $z$- and $\varphi$-directions, is 
\begin{eqnarray} \label{Sig^zphi_3.2}
\Sigma^{z\varphi} = -\frac{a}{R}\Lambda^{\varphi}.
\end{eqnarray}
\indent
In the linear case ($\varphi_{\rm NL}=0$), these may be written in a compact form as
\begin{eqnarray} \label{E2_3.2}
E = \mathcal{T} a \Delta\left(1 + \frac{(2\pi)^2R^2}{a^2\lambda_z^2}\right) =  \mathcal{T} a \Delta \Omega^{-2},
\end{eqnarray}
\begin{eqnarray} \label{P^z3_3.2}
\mathcal{P}^{z} = \pm \mathcal{T} n_z \times \frac{(2\pi)^2 R^2}{a\lambda_z} =  \pm \mathcal{T} a \Delta \left(\frac{1-\Omega^2}{\Omega^2}\right),
\end{eqnarray}
\begin{eqnarray} \label{Lam^phi4_3.2*}
\Lambda^{\varphi} = \pm 2\pi \mathcal{T} a R n_z = \pm \mathcal{T} a\Delta \frac{ \sqrt{1-\Omega^{2}}}{\Omega},
\end{eqnarray}
\begin{eqnarray} \label{Pi3_3.2}
\Pi = \mathcal{T} a \Delta\sqrt{1 + \frac{(2\pi)^2 R^2}{a^2\lambda_z^2}} = \mathcal{T} a \Delta \Omega^{-1},
\end{eqnarray}
\begin{eqnarray} \label{T^z3_3.2}
\mathcal{T}^{z} = -\mathcal{T} a \Delta\left(1- \frac{(2\pi)^2R^2}{a^2\lambda_z^2}\right) =  -\mathcal{T} a \Delta \left(\frac{2\Omega^{2}-1}{\Omega^{2}}\right).
\end{eqnarray}
Clearly, it is possible for the string to be effectively tensionless (everywhere) when the distance between windings is equal to the circumference of extra dimension
\begin{eqnarray} \label{NoTens1_3.2}
a^2\lambda_z^2 = (2\pi)^2R^2,
\end{eqnarray}
or equivalently when 
\begin{eqnarray} \label{NoTensLin2_3.2}
\Omega^2 = \omega^2 = 1/2, \ \forall t,\sigma.
\end{eqnarray}
Under these conditions we also have that
\begin{eqnarray}  \label{UniformRotn1}
\omega_{z}^2 \equiv \omega_{\varphi}^2 = \frac{v_{\varphi}^2}{R^2} = \frac{a^2}{R^2}, 
\end{eqnarray}
where $\omega_{\varphi}$ and $v_{\varphi}$ denote the angular and linear velocity of the string in the compact space. 
\\ \indent
In the classical theory, the four-dimensional string tension reverses sign and becomes an effective pressure for $\omega^2 < 1/2$. 
This leads to qualitative differences in the macroscopic (i.e. $(3+1)$-dimensional) string dynamics. 
For example, circular string loops with initial winding configurations such that $a\lambda_z(t_i) < 2\pi R$ oscillate between two critical radii, $\rho_{c1} = \rho(t_i)$ and $\rho_{c2} >  \rho(t_i)$, implying that loops initially expand, rather than contract, after chopping off from the long string network \cite{LaYo12,LaWa10}. 
This is also shown explicitly, in the classical theory of circular loops, in Sec. \ref{Sect2.3}. However, due to quantum effects, superconducting defect strings carrying either bosons or fermions are expected have critical currents of order 
\begin{eqnarray} \label{J_max}
J_{\rm max} \simeq \frac{e}{2\pi R}, 
\end{eqnarray}
where $e$ is the coupling (i.e. charge) \cite{ViSh00}. 
In field theory, an important generalisation of the original (vacuum) cosmic string models was given by Witten \cite{Witten:1985}, who proposed that ordinary cosmic strings may carry electric currents, thus behaving like superconducting wires. 
The charge carriers may either be bosons, in which case a charged Higgs field with a nonzero vacuum expectation value in the core of the string is required, or fermions, which are trapped as zero modes along the string. 
For bosonic currents, exceeding the critical threshold implies electric field strengths large enough to induce pair production via the Schwinger process whereas, for trapped fermions, this marks the critical point at which it becomes energetically favourable for particles to leave the string \cite{ViSh00}. 
\\ \indent
Although Nielsen \cite{Ni79} established a formal correspondence between superconducting strings and strings compactified on higher-dimensional cycles at the classical level, it is reasonable to consider the quantum bound (\ref{J_max}) in relation to this model. 
For the long straight string, this may be rewritten in terms of the higher-dimensional variable $\lambda_z$ as
\begin{eqnarray} \label{thresh}
a\lambda_{z}^{\rm min} = 2\pi R.
\end{eqnarray}
In the defect string picture, this is equivalent to requiring the distance between neighbouring twists in the magnetic field lines to be greater than the circumference of the string core \cite{LaYo12}. 
In the wound string picture, it implies that neighbouring windings must be separated by a distance larger than the compactification scale. 
Though a full quantum mechanical treatment of the wound-string model is needed to confirm this result, we note that it is physically intuitive and is analogous to a standard result for nonrelativistic strings of finite width, i.e., that one cannot put `twists' in a length of rope over distances significantly shorter than its width. 
It is also interesting to note that the maximum current allowed by quantum mechanics is exactly that which leads to the tensionless condition, so that strings with positive effective pressure, or `repulsive rods', are unable to form.
\\ \indent
For the treatment of nonlinear windings, it is useful to introduce the parameter $\alpha \in [0,1]$, defined via
\begin{eqnarray} \label{alpha_3.2*}
\alpha^{-2} = \frac{1}{2\pi n_z^2} \int_{0}^{2\pi}(\partial_{\sigma}\varphi)^2{\rm d}\sigma = \frac{1}{2\pi} \int_{0}^{2\pi}\left(1 + \frac{\partial_{\sigma}\varphi_{\rm NL}}{n_z}\right)^2{\rm d}\sigma.
\end{eqnarray}
We define $\alpha$ as the inverse of the positive square root of (\ref{alpha_3.2*}) so that $\alpha = 1$ when $\varphi_{\rm NL} = 0$ and $\alpha < 1$ when $\varphi_{\rm NL} \neq 0$.
It is then straightforward to show that in general
\begin{eqnarray} \label{E4_3.2}  
E = \mathcal{T} a \Delta\left(1 + \frac{(2\pi)^2R^2}{a^2\alpha^2\lambda_z^2}\right) = \mathcal{T} a \Delta \Omega^{-2},
\end{eqnarray}
\begin{eqnarray} \label{P^z4_3.2}
\mathcal{P}^{z} = \pm \mathcal{T} \frac{n_z}{\alpha} \times \frac{(2\pi)^2 R^2}{a\alpha\lambda_z} = \pm \mathcal{T} a \Delta \left(\frac{1-\Omega^{2}}{\Omega^{2}}\right),
\end{eqnarray}
\begin{eqnarray} \label{Lam^phi4_3.2}
\Lambda^{\varphi} = \pm 2\pi \mathcal{T} a R n_z = \pm \mathcal{T} a\Delta \frac{ \sqrt{1-\Omega^{2}}}{\Omega}\alpha
\end{eqnarray}
\begin{eqnarray} \label{Pi4_3.2}
\Pi = \mathcal{T} a \Delta\sqrt{1 + \frac{(2\pi)^2 R^2}{a^2\lambda_z^2}(2\alpha^{-2} - 1)} =  \mathcal{T} a \Delta \Omega^{-1}\sqrt{2-\alpha^2 - (1-\alpha^2)\Omega^2},
\end{eqnarray}
\begin{eqnarray} \label{T^z_3.2}
\mathcal{T}^{z} = -\mathcal{T} a \Delta\left(1 - \frac{(2\pi)^2R^2}{a^2\alpha^2\lambda_z^2}\right) = -\mathcal{T} a\Delta \left(\frac{2\Omega^{2}-1}{\Omega^{2}}\right).
\end{eqnarray}
Thus, the definition (\ref{alpha_3.2*}) allows us to express the physical parameters of the string in a compact form in terms of $\Omega(t)$ and $\alpha$, even for strings with complicated nonlinear windings. 
\\ \indent
In the nonlinear case - that is, for a string with linear extension in the $z$-direction of the Minkowski space-time but with nonlinear windings propagating along its $(3+1)$-dimensional length - it is not possible for the string to be everywhere tensionless at any moment in time. However, it is still possible for the average net tension in the $z$-direction to be zero and this occurs when
\begin{eqnarray} \label{NoTensNL1_3.2}
a^2\alpha^2\lambda_z^2 = (2\pi)^2R^2,
\end{eqnarray}
or equivalently for 
\begin{eqnarray} \label{NoTensNL2_3.2}
\Omega^2 = 1/2, \ \forall t, \ (\omega^2 \neq 1/2, \ \forall t,\sigma).
\end{eqnarray}
Note that although $\Omega^2$, as we have defined it for a general string configuration, is a function of time, it is clear that for the embedding considered here it is a constant, even in the nonlinear case:
\begin{eqnarray} \label{}
\Omega^{2} = \left(1 + \frac{(2\pi)^2R^2}{a^2\alpha^2\lambda_z^2}\right)^{-1} = {\rm const.}
\end{eqnarray}
\indent
The linear case corresponds to $\alpha = 1$ and $\alpha < 1$ quantifies the degree of nonlinearity in the windings. 
Specifically, we may interpret $a\alpha \lambda_z$ as the mean value of the local wavelength of the windings (in the warped geometry), averaged across the entire length of the string and $n_z/\alpha$ as the average value of the effective `local'  winding number at a given point $z$. 
To see this, we now define the local values of the various model parameters - that is, their effective values at a given point on the string, labelled by $\sigma$ (or $z$), at a given time $t$. 
Thus, the  local string length in $(3+1)$-dimensions and the local winding number are defined as 
\begin{eqnarray} \label{l_z^eff_3.2}
(l_z^{\rm eff})^2 =  (2\pi)^{2}(\partial_{\sigma}z)^2 = \Delta^2, \ \ \ (n_z^{\rm eff})^2 =  \frac{\Delta^2}{(2\pi)^2}\varphi'^2  = (\partial_{\sigma}\varphi)^2.
\end{eqnarray}
We define the effective wavelength, wavenumber, angular frequency and velocity of the string in the $z$-direction via
\begin{eqnarray} \label{omega_z^eff_3.2}
(\omega_z^{\rm eff})^2 = (v_z^{\rm eff})^2(k_z^{\rm eff})^2, \ \ \ (k_z^{\rm eff})^2 = \frac{(2\pi)^2}{(\lambda_z^{\rm eff})^2} = \frac{(2\pi)^2}{\Delta}(\partial_{\sigma}\varphi)^2.
\end{eqnarray}
The instantaneous angular and linear velocities in the $\varphi$-direction are then related via
\begin{eqnarray} \label{omega_z^eff*_3.2}
(\omega_z^{\rm eff})^2 = (\omega_{\varphi}^{\rm eff})^2 = \frac{(v_{\varphi}^{\rm eff})^2}{R^2} = \dot{\varphi}^2,
\end{eqnarray}
since the angular frequency associated with the movement of the windings in the large dimensions must be equal to the angular velocity of the string in the compact space.
\\ \indent
In our new notation, the EOM for $\varphi$ is equivalent to the {\it dispersion relation}
\begin{eqnarray} \label{DispRel_3.2}
(\omega_z^{\rm eff})^2 = (k_z^{\rm eff})^2, \ \ \  \omega_z^{\rm eff} = \pm k_z^{\rm eff},
\end{eqnarray}
and
\begin{eqnarray} \label{v_z^eff_3.2}
(v_z^{\rm eff})^2 = 1, \ \forall t,\sigma.
\end{eqnarray}
By combining the various relations above, it follows that
\begin{eqnarray} 
(l_z^{\rm eff})^2 = (n_{z}^{\rm eff})^2(\lambda_{z}^{\rm eff})^2, \ 
(v_{\varphi}^{\rm eff})^2(\lambda_z^{\rm eff})^2 = (2\pi)^2R^2, \ 
(n_z^{\rm eff})^2  = \frac{(v_{\varphi}^{\rm eff})^2(l_z^{\rm eff})^2}{(2\pi)^2R^2} = \frac{(v_{\varphi}^{\rm eff})^2\Delta^2}{(2\pi)^2R^2},
\end{eqnarray}
which form a self-consistent set. 
The expressions for the various components of $\mathcal{T}^{I}{}_{J}\sqrt{-g}$ may now be written entirely in terms of the new parameters since
\begin{eqnarray} \label{omegaAlt_3.2}
\omega^{-2} = \frac{a^2(l_z^{\rm eff})^2 + (2\pi)^2R^2(n_z^{\rm eff})^2}{a^2(l_z^{\rm eff})^2} = 1 +  \frac{(2\pi)^2R^2}{a^2(\lambda_z^{\rm eff})^2}.
\end{eqnarray}
In general, we have that
\begin{eqnarray} \label{lambda_z^eff_3.2}
a^2(\lambda_z^{\rm eff})^2 = (2\pi)^2 R^2 \left(\frac{\omega^2}{1-\omega^2}\right),
\end{eqnarray}
so that the tensionless condition $\omega^2=1/2$ is equivalent to $a^2(\lambda_z^{\rm eff})^2 = (2\pi)^2 R^2$ which, for a string which is `straight' with respect to the Minkowski directions, is clearly only possible when the windings are linear, $\lambda_z^{\rm eff} = \lambda_z$, thus yielding the condition obtained previously, (\ref{NoTens1_3.2}). 
It is straightforward to show that the consistency of the relations above also requires 
\begin{eqnarray} \label{ConsRel4_3.2}
(v_{\varphi}^{\rm eff})^2 = a^2 \left(\frac{1-\omega^2}{\omega^2}\right) \equiv \frac{v_{\perp}^2}{a^2}. 
\end{eqnarray}  
This can be interpreted physically since, for windings formed dynamically by the motion of the string in the compact space, rather than as an initial condition at the time of string formation, we expect 
\begin{eqnarray} \label{DynWind_3.2}
\frac{(v_{\varphi}^{\rm eff})^2}{a^2(v_{z}^{\rm eff})^2} = \frac{1-\omega^2}{\omega^2}.
\end{eqnarray} 
This is equivalent to the previous relation (\ref{ConsRel4_3.2}) since $(v_{z}^{\rm eff})^2=1$.\\
\indent
Next, we define the integrated (i.e. spatially-averaged) values of the parameters on the left-hand sides of Eqs. (\ref{l_z^eff_3.2})-(\ref{DispRel_3.2}) according to the averaging condition
\begin{eqnarray} \label{Q_3.2}
\langle Q^2 \rangle(t) = \frac{1}{2\pi}\int_{0}^{2\pi}(Q^{\rm eff})^2 {\rm d}\sigma.
\end{eqnarray}
The integrated values of any parameters appearing on the right-hand-sides are defined implicitly in terms of those on the left. 
Hence, we have 
\begin{eqnarray} \label{<l_z>_3.2}
\langle l_z^2 \rangle = \Delta^2, \ \ \ \langle n_z^2 \rangle = n_z^2/\alpha^2,
\end{eqnarray}
and
\begin{eqnarray} \label{<omega_z>_3.2}
\langle \omega_z^2 \rangle = \langle k_z^2 \rangle, \ \ \ \langle k_z^2 \rangle = \frac{(2\pi)^2}{\langle \lambda_z^2 \rangle} = \frac{(2\pi)^2}{\Delta^2}\langle (\partial_{\sigma}\varphi)^2 \rangle,
\end{eqnarray}
where 
\begin{eqnarray} \label{<lambda_z>_3.2}
\langle \lambda_z^2 \rangle = \alpha^2\lambda_z^2,
\end{eqnarray}
together with
\begin{eqnarray} \label{<omega_phi>_3.2}
\langle \omega_{z}^2\rangle = \langle \omega_{\varphi}^2 \rangle = \frac{\langle v_{\varphi}^2 \rangle }{R^2} = \langle \dot{\varphi}^2 \rangle.
\end{eqnarray}
Combining these expression as before gives
\begin{eqnarray} \label{<ConsRel>1_3.2}
\langle l_z^2 \rangle = \langle n_{z}^2\rangle \langle \lambda_{z}^2\rangle, \ \ \ 
\langle v_{\varphi}^2 \rangle \langle \lambda_z^2 \rangle = (2\pi)^2R^2, \ \ \ 
\langle n_z^2 \rangle = \frac{\langle v_{\varphi}^2 \rangle \langle l_{z}^2 \rangle}{(2\pi)^2R^2} = \frac{\langle v_{\varphi}^2 \rangle\Delta^2}{(2\pi)^2R^2},
\end{eqnarray}
and 
\begin{eqnarray} \label{<v_z>_3.2}
\langle v_{\varphi}^2\rangle = v_{\varphi}^2/\alpha^2.
\end{eqnarray}
In general, we have 
\begin{eqnarray} \label{<lambda_z>*_3.2}
a^2 \langle \lambda_z^2 \rangle = (2\pi)^2 R^2 \left(\frac{\Omega^2}{1-\Omega^2}\right).
\end{eqnarray}
The case of zero net tension occurs when the average value of the wavelength, along the entire string, is equal to the circumference of the compact space
\begin{eqnarray} \label{NoNetTens2_3.2}
a^2 \langle \lambda_z^2 \rangle = (2\pi)^2 R^2.
\end{eqnarray}
As before, the consistency of the above relations also requires
\begin{eqnarray} \label{<ConsRel>4_3.2}
\langle v_{\varphi}^2\rangle = a^2 \left(\frac{1-\Omega^2}{\Omega^2}\right) \equiv \frac{\langle v_{\perp}^2 \rangle}{a^2} = \frac{\Omega^2}{a^2}\langle v_{\varphi}^2\rangle.
\end{eqnarray} 
This follows from the expression for the ratio of the instantaneous velocities for dynamically formed windings. 
By integrating Eq. (\ref{ConsRel4_3.2}) over ${\rm d}\sigma$, we obtain
\begin{eqnarray} \label{<DynWind>_3.2}
\frac{\langle v_{\varphi}^2\rangle}{a^2\langle v_{z}^2\rangle} = \frac{1-\Omega^2}{\Omega^2},
\end{eqnarray} 
which is equivalent to Eq. (\ref{<ConsRel>4_3.2}), since $\langle v_{z}^2\rangle=1$. 
\\ \indent
The physical parameters $E$, $\mathcal{P}^{z}$, $\Pi$ and $\mathcal{T}^{z}$  can now be written in terms of the spatially-averaged variables as
\begin{eqnarray} \label{E5_3.2}
E = \mathcal{T} a \Delta\left(1 + \frac{(2\pi)^2R^2}{a^2 \langle \lambda_z^2 \rangle}\right),
\end{eqnarray}
\begin{eqnarray} \label{P^z5_3.2}
\mathcal{P}^{z}  = \pm \mathcal{T} a \Delta \times \frac{(2\pi)^2 R^2}{ a^2\langle \lambda_z^2 \rangle} = \pm \mathcal{T} \sqrt{\langle n_z^2 \rangle} \times \frac{(2\pi)^2 R^2}{a \sqrt{ \langle \lambda_z^2 \rangle}},
\end{eqnarray}
\begin{eqnarray} \label{Pi5_3.2}
\Pi = \mathcal{T} a \Delta\sqrt{1 + \frac{(2\pi)^2 R^2}{a^2\langle \lambda_z^2 \rangle}\left(\frac{2\langle n_z^2\rangle - n_z^2}{\langle n_z^2\rangle}\right)},
\end{eqnarray}
\begin{eqnarray} \label{T^z5_3.2}
\mathcal{T}^{z} = -\mathcal{T} a \Delta\left(1 - \frac{(2\pi)^2R^2}{a^2 \langle \lambda_z^2 \rangle}\right).
\end{eqnarray}
These expressions remain consistent with our earlier definitions, since
\begin{eqnarray} \label{OmegaAlt_3.2}
\Omega^{-2} = \frac{a^2\langle l_z^2 \rangle + (2\pi)^2R^2\langle n_z^2 \rangle}{a^2\langle l_z^2 \rangle} = 1 +  \frac{(2\pi)^2R^2}{a^2\langle \lambda_z^2 \rangle}.
\end{eqnarray}
In the linear case, we have 
\begin{eqnarray} \label{LinCasQ_3.2}
\langle Q^2 \rangle = (Q^{\rm eff})^2 = Q^2 = {\rm const.},
\end{eqnarray}
\begin{eqnarray} \label{LinCas_omega}
\omega^2 = \Omega^2 = {\rm const.}, \ \ \ \forall t,\sigma 
\end{eqnarray}
and $\alpha=1$, so that the expressions above reduce to
\begin{eqnarray} \label{LinCasRel1_3.2}
\Delta^2 = n_z^2\lambda_z^2, \ \ \ 
v_{\varphi}^2 \lambda_z^2 = (2\pi)^2R^2, \ \ \
n_z^2  = \frac{v_{\varphi}^2 \Delta^2}{(2\pi)^2R^2}.
\end{eqnarray}
\begin{eqnarray} \label{LinCasRel2_3.2}
\omega_z^2 = k_z^2 = \frac{(2\pi)^2}{\lambda_z^2} \equiv \omega_{\varphi}^2 = \frac{v_{\varphi}^2}{R^2}, 
\end{eqnarray}
and $v_z^2=1$. 
These last relations also hold true, for the parameters contained within the argument of the function $\varphi(t,z) \equiv \varphi(t,\sigma)$, even in the nonlinear case, for which $(Q^{\rm eff})^2 \neq \langle Q^2 \rangle \neq Q^2$.

\subsection{Circular $F$-string loops} \label{Sect2.3}
In this section, we consider circular loops, both with and without higher-dimensional windings. 
Loops with higher-dimensional windings are considered in detail in Sec. \ref{Sect2.3.2} and the general solution to the string EOM for arbitrary initial conditions is given. 
As it is instructive to first review the treatment of circular loops without windings, a brief analysis of this is presented in Sect. \ref{Sect2.3.1}.

\subsubsection{Circular $F$-string loops in $(3+1)$ dimensions} \label{Sect2.3.1}
To treat string loops we switch from Cartesian to cylindrical polar coordinates, so that the $(y,z)$-plane becomes the $(\rho,\sigma)$-plane
\begin{eqnarray} \label{Met_4.1}
{\rm d}s^2 = a^2({\rm d}t^2-{\rm d}x^2-{\rm d}y^2-{\rm d}z^2) - R^2{\rm d}\varphi^2 = a^2({\rm d}t^2-{\rm d}x^2-{\rm d}\rho^2-\rho^2{\rm d}\sigma^2) - R^2{\rm d}\varphi^2.
\end{eqnarray}
For circular loops with no higher-dimensional windings, the embedding is
\begin{eqnarray} \label{Emb_4.1}
X^I(\tau,\sigma)  \equiv X^I(t,\sigma) =\left(t(\tau)=\zeta \tau, x=0, \rho(t), \sigma, \varphi = 0\right),
\end{eqnarray}
and the action becomes
\begin{eqnarray} \label{Act_4.1}
S = -\mathcal{T}a^2 \int {\rm d}t {\rm d}\sigma \rho \sqrt{1-\dot{\rho}^2}.
\end{eqnarray}
This can be written in the standard form (\ref{ActWarp_2.4}) by setting 
\begin{eqnarray} \label{dl_4.1}
\frac{{\rm d}l}{{\rm d}\sigma} = a\rho, \ \ \ \frac{v_{\perp}^2}{a^2} = \dot{\rho}^2.
\end{eqnarray}
The EOM are
\begin{subequations} 
\begin{align}
\partial_{0}(T^{0}{}_{0}\sqrt{-g}) + \partial_{\sigma}(T^{\sigma}{}_{0}\sqrt{-g}) &= 0, \label{EOM_4.1a}\\
\partial_{0}(T^{0}{}_{\sigma}\sqrt{-g}) + \partial_{\sigma}(T^{\sigma}{}_{\sigma}\sqrt{-g}) &=  0, \label{EOM_4.1b}\\
\partial_{0}(T^{0}{}_{\rho}\sqrt{-g}) + \partial_{\sigma}(T^{\sigma}{}_{\rho}\sqrt{-g}) - \frac{1}{2}\partial_{\rho}g_{\sigma\sigma}T^{\sigma\sigma}\sqrt{-g}&=  0, \label{EOM_4.1c}
\end{align}
\end{subequations}
where the relevant components of $\mathcal{T}^{I}{}_{J}\sqrt{-g}$ are given by
\begin{subequations}
\begin{align}
\mathcal{T}^{0}{}_{0}\sqrt{-g} &= \mathcal{T} \frac{a^2\rho}{\sqrt{1-\dot{\rho}^2}}, \ \ \ \ \ \mathcal{T}^{\sigma}{}_{0}\sqrt{-g} = 0, \label{EMT_4.1a}\\
\mathcal{T}^{0}{}_{\sigma}\sqrt{-g} &= 0, \ \ \  \ \ \ \ \  \ \ \ \ \  \ \ \ \ \mathcal{T}^{\sigma}{}_{\sigma}\sqrt{-g} = \mathcal{T} a^2 \rho \sqrt{1-\dot{\rho}^2}, \label{EMT_4.1b}\\
\mathcal{T}^{0}{}_{\rho}\sqrt{-g} &= -\mathcal{T} \frac{a^2\rho\dot{\rho}}{\sqrt{1-\dot{\rho}^2}}, \ \ \mathcal{T}^{\sigma}{}_{\rho}\sqrt{-g} = 0, \ \ \mathcal{T}^{\sigma\sigma}\sqrt{-g} = -\mathcal{T} \frac{\sqrt{1-\dot{\rho}^2}}{\rho}. \label{EMT_4.1c}
\end{align}
\end{subequations}
It is straightforward to show that each of these is equivalent to 
\begin{eqnarray} \label{EOM2_4.1}
1 - \dot{\rho}^2 + \rho\ddot{\rho} = 0,
\end{eqnarray}
which represents the conservation of energy, since the only nonzero constant of motion is the Hamiltonian
\begin{eqnarray} \label{E_4.1}
E = 2\pi \mathcal{T}  \frac{a\rho}{\sqrt{1-\dot{\rho}^2}}.
\end{eqnarray} 
The integrated tensions are
\begin{eqnarray} \label{T^rho_4.1}
\mathcal{T}^{\rho} = -2\pi \mathcal{T}  \frac{a\rho\dot{\rho}}{\sqrt{1-\dot{\rho}^2}}, \ \ \ 
\mathcal{T}^{\sigma} = -2\pi \mathcal{T}  a\rho \sqrt{1-\dot{\rho}^2}.
\end{eqnarray}
and the only possible shear, between the $\sigma$- and $\rho$-directions, is zero
\begin{eqnarray} \label{S^sigrho_4.1}
\Sigma^{\rho\sigma} = 0,
\end{eqnarray}
since $\mathcal{T}^{\rho\sigma}\sqrt{-g} = 0$.
The expressions for the Hamiltonian and integrated tensions, and, in fact, for all the components of the energy momentum tensor $\mathcal{T}^{I}{}_{J}\sqrt{-g}$, become equivalent to those for the long straight string (with no higher-dimensional windings) for
\begin{eqnarray} \label{Delta1_4.1}
\Delta = 2\pi\rho(t_i), \ \ \ \dot{\rho}(t) = 0, \ \ \  \ddot{\rho}(t) = 0, \ \forall t,
\end{eqnarray}
where $t_i$ denotes the initial time, which is assumed to be the time of loop formation.
\\ \indent
Due to the circular symmetry of the system, the Hamiltonian density is independent of $\sigma$ and the EOM for $\rho(t)$ is directly integrable. 
We may solve it, subject to the generic boundary conditions
\begin{eqnarray} \label{BC1_4.1}
\rho(t)|_{t=t_i} = \rho(t_i), \ \  \ \dot{\rho}(t)|_{t=t_i} = \dot{\rho}(t_i),
\end{eqnarray}
to give
\begin{eqnarray} \label{GenSoln_4.1}
\rho(t)  = \frac{\rho(t_i)}{\sqrt{1- \dot{\rho}^2(t_i)}}\left|\cos\left(\frac{\sqrt{1- \dot{\rho}^2(t_i)}}{\rho(t_i)}(t-t_i)-\sin^{-1}(\dot{\rho}(t_i))\right)\right|.
\end{eqnarray}
Note that we must make use of the identity 
\begin{eqnarray} \label{Indent_4.1}
\cos(\sin^{-1}( x)) = \sqrt{1-x^2}
\end{eqnarray}
to recover the correct initial radius at $t=t_i$. 
The loop oscillates between two critical radii 
\begin{eqnarray} \label{CritRad_4.1}
\rho_{c1} =  \rho(t_i), \ \ \ \rho_{c2} = 0,
\end{eqnarray}
performing one full oscillation with a time period
\begin{eqnarray} \label{Period_4.1}
T_{c} = \frac{\rho(t_i)}{\sqrt{1- \dot{\rho}^2(t_i)}}\pi.  
\end{eqnarray}

\subsubsection{Circular $F$-string loops with higher-dimensional windings} \label{Sect2.3.2}
Using cylindrical polar coordinates, the general embedding for a circular loop with higher-dimensional windings is
\begin{eqnarray} \label{Emb_4.2}
X^I(\tau,\sigma)  \equiv X^I(t,\sigma) =\left(t(\tau)=\zeta \tau, x=0, \rho(t), \sigma, \varphi(\tau,\sigma) \equiv \varphi(t,\sigma) \right),
\end{eqnarray}
so that the action is
\begin{eqnarray} \label{Act_4.2}
S = -\mathcal{T} \int {\rm d}t {\rm d}\sigma  \sqrt{a^2(1-\dot{\rho}^2)(a^2\rho^2+R^2(\partial_{\sigma}\varphi)^2)-a^2\rho^2R^2\dot{\varphi}^2},
\end{eqnarray}
which can be written in the standard form (\ref{ActWarp_2.4}) with 
\begin{eqnarray} \label{dl_4.2}
\frac{{\rm d}l}{{\rm d}\sigma} = \sqrt{a^2\rho^2+R^2(\partial_{\sigma}\varphi)^2}, \ \ \ \frac{v_{\perp}^2}{a^2} = \dot{\rho}^2 + \frac{\rho^2R^2\dot{\varphi}^2}{a^2\rho^2+R^2(\partial_{\sigma}\varphi)^2}.
\end{eqnarray}
The EOM are
\begin{subequations} \label{++}
\begin{align}
\partial_{0}(T^{0}{}_{0}\sqrt{-g}) + \partial_{\sigma}(T^{\sigma}{}_{0}\sqrt{-g}) &= 0, \label{EOM_4.2a*}\\
\partial_{0}(T^{0}{}_{\sigma}\sqrt{-g}) + \partial_{\sigma}(T^{\sigma}{}_{\sigma}\sqrt{-g}) &=  0, \label{EOM_4.2b*}\\
\partial_{0}(T^{0}{}_{\rho}\sqrt{-g}) + \partial_{\sigma}(T^{\sigma}{}_{\rho}\sqrt{-g}) - \frac{1}{2}\partial_{\rho}g_{\sigma\sigma}T^{\sigma\sigma}\sqrt{-g}&=  0, \label{EOM_4.2c**}\\
\partial_{0}(T^{0}{}_{\varphi}\sqrt{-g}) + \partial_{\sigma}(T^{\sigma}{}_{\varphi}\sqrt{-g}) &= 0. \label{EOM_4.2d*}
\end{align}
\end{subequations}
Writing out the EOM in $\varphi(t,\sigma)$ explicitly in terms of $(-\gamma)$ and its derivatives, we have
\begin{eqnarray} \label{}
(-\gamma)[(1-\dot{\rho}^2)\partial_{\sigma}^2\varphi - \rho^2\ddot{\varphi} - 2\rho\dot{\rho}\dot{\varphi}] +  \frac{1}{2}\rho^2\dot{\varphi}\frac{\partial(-\gamma)}{\partial t} - \frac{1}{2}(1-\dot{\rho}^2)\partial_{\sigma}\varphi \frac{\partial(-\gamma)}{\partial \sigma}= 0,
\end{eqnarray}
which may be rearranged to give
\begin{eqnarray} \label{+}
& &  a^2R^2[(1-\dot{\rho}^2)\partial_{\sigma}^2\varphi - \rho^2\ddot{\varphi} -  2\rho\dot{\rho}\dot{\varphi}][(1-\dot{\rho}^2)(\partial_{\sigma}\varphi)^2 - \rho^2\dot{\varphi}^2]
\nonumber\\
&+& a^2R^2\rho^2\dot{\varphi}[(1-\dot{\rho}^2)\partial_{\sigma}\varphi\partial_{\sigma}\dot{\varphi} - \rho\ddot{\rho}(\partial_{\sigma}\varphi)^2 - \rho\dot{\rho}\dot{\varphi}^2 - \rho^2\dot{\varphi}\ddot{\varphi}]
\nonumber\\
&-& a^2R^2(1-\dot{\rho}^2)\partial_{\sigma}\varphi[(1-\dot{\rho}^2)\partial_{\sigma}\varphi\partial_{\sigma}^2\varphi - \rho^2\dot{\varphi}\partial_{\sigma}\dot{\varphi}]
\nonumber\\
&+& a^4\rho^2\left\{(1-\dot{\rho}^2)[(1-\dot{\rho}^2)\partial_{\sigma}^2\varphi - \rho^2\dot{\varphi}\partial_{\sigma}\dot{\varphi}] - \rho^2\dot{\rho}\ddot{\rho}\dot{\varphi}\right\}.
\end{eqnarray}
This is very complicated, but we may try to guess an equivalent compact form by noting that Eq. (\ref{+}) reduces to the EOM for $\varphi(t,z)$ for the long straight string, when the conditions in Eq. (\ref{Delta1_4.1}) are satisfied. 
Thus, let us assume that
\begin{eqnarray} \label{EOM1_4.2}
\dot{\varphi}^2 = \frac{(1-\dot{\rho}^2)}{\rho^2}(\partial_{\sigma}\varphi)^2.
\end{eqnarray}
Differentiating with respect to $t$ and $\sigma$, respectively, then gives
\begin{eqnarray} \label{EOM2_4.2}
(1-\dot{\rho}^2)\partial_{\sigma}\varphi\partial_{\sigma}\dot{\varphi} - \dot{\rho}\ddot{\rho}(\partial_{\sigma}\varphi)^2 - \rho\dot{\rho}\dot{\varphi}^2 - \rho^2\dot{\varphi}\ddot{\varphi} = 0,
\end{eqnarray}
\begin{eqnarray} \label{EOM3_4.2}
(1-\dot{\rho}^2)\partial_{\sigma}\varphi\partial_{\sigma}^2\varphi - \rho^2\dot{\varphi}\partial_{\sigma}\dot{\varphi} = 0,
\end{eqnarray}
and the complicated EOM (\ref{+}) simplifies to
\begin{eqnarray} \label{EOM4_4.2}
(1-\dot{\rho}^2)^2\partial_{\sigma}^2\varphi -  \rho^2\dot{\rho}\ddot{\rho}\dot{\varphi} - (1-\dot{\rho}^2)[\rho\dot{\rho}\dot{\varphi} + \rho^2\ddot{\varphi}] = 0.
\end{eqnarray}
Rearranging Eqs. (\ref{EOM2_4.2})-(\ref{EOM3_4.2}) to make $\partial_{\sigma}\dot{\varphi}$ the subject and equating the results, gives
\begin{eqnarray} \label{EOM5_4.2}
(\partial_{\sigma}\varphi)^2[(1-\dot{\rho}^2)^2\partial_{\sigma}^2\varphi -  \rho^2\dot{\rho}\ddot{\rho}\dot{\varphi}] - \rho^2\dot{\varphi}^2[\rho\dot{\rho}\dot{\varphi} + \rho^2\ddot{\varphi}] = 0.
\end{eqnarray}
Substituting again from the compact expression Eq. (\ref{EOM1_4.2}), we recover the simplified EOM Eq. (\ref{EOM4_4.2}). 
Hence, our guess (\ref{EOM1_4.2}) is genuinely equivalent to the original complicated EOM for $\varphi(t,\sigma)$, Eq. (\ref{+}). Eq. (\ref{EOM1_4.2}) is solved by any function of the form
\begin{eqnarray} \label{phiSoln1_4.2}
\varphi(t,\sigma) = \varphi\left(n_{\sigma}\sigma + \int \omega_{\sigma}(t){\rm d}t\right), 
\end{eqnarray}
where
\begin{eqnarray} \label{omega_sig_4.2} 
\omega_{\sigma}^2(t) = n_{\sigma}^2\left(\frac{1-\dot{\rho}^2}{\rho^2}\right), \ \ \ \omega_{\sigma}(t) = \pm n_{\sigma}\frac{\sqrt{1-\dot{\rho}^2}}{\rho},
\end{eqnarray}
which for convenience may again be split into linear and nonlinear parts
\begin{eqnarray} \label{phiSoln2_4.2}
\varphi(t,\sigma) = n_{\sigma}\sigma + \int \omega_{\sigma}(t){\rm d}t + \varphi_{\rm NL}\left(n_{\sigma}\sigma + \int \omega_{\sigma}(t){\rm d}t\right).
\end{eqnarray}
However, since we have imposed circular symmetry on the $(3+1)$-dimensional part of the ansatz, we must also impose this with respect to distribution of the the windings, from a $(3+1)$-dimensional perspective. Therefore, we may set $\varphi_{\rm NL}=0$, so that 
\begin{eqnarray} \label{phiSoln3_4.2}
\varphi(t,\sigma) = n_{\sigma}\sigma + \int \omega_{\sigma}(t){\rm d}t, 
\end{eqnarray}
which is equivalent to the condition
\begin{eqnarray} \label{Constraint_4.2}
\partial_{\sigma}\omega = 0.
\end{eqnarray}
\indent
Since we have imposed three constraints on a system of four PDEs, i.e. fixing the forms of $X^{0}=\zeta \tau$ and $X^{1}=\sigma$, together with the condition (\ref{Constraint_4.2}), we should now be left with a single remaining (independent) EOM in the Euler-Lagrange equations. 
In other words, if the requirement of circular symmetry for the windings is physically necessary, as symmetry suggests, it should, together with the constraints imposed so far, imply that any combination of Eqs. (\ref{EOM_4.2a*})-(\ref{EOM_4.2d*}) yields the same independent EOM.
\\ \indent
After substituting for $\dot{\varphi}^2$ from Eq. (\ref{EOM1_4.2}), the components of $\mathcal{T}^{I}{}_{J}\sqrt{-g}$ which occur in the EOM may be written as,
\begin{subequations}
\begin{align}
\mathcal{T}^{0}{}_{0}\sqrt{-g} &= \mathcal{T} \frac{a^2\rho}{\sqrt{1-\dot{\rho}^2}}\omega^{-2}, \ \ \ \ \ \ \ \ \ \mathcal{T}^{\sigma}{}_{0}\sqrt{-g} = \mp \mathcal{T} a^2\left(\frac{1-\omega^{2}}{\omega^{2}}\right), \label{EMT_4.2a*}\\
\mathcal{T}^{0}{}_{\sigma}\sqrt{-g} &=  \pm \mathcal{T} a^2\rho^2 \left(\frac{1-\omega^{2}}{\omega^{2}}\right) , \ \ \ \mathcal{T}^{\sigma}{}_{\sigma}\sqrt{-g} = \mathcal{T} a^2 \rho \sqrt{1-\dot{\rho}^2} \left(\frac{2\omega^{2}-1}{\omega^{2}}\right), \label{EMT_4.2b*}\\
\mathcal{T}^{0}{}_{\rho}\sqrt{-g} &= -\mathcal{T} \frac{a^2\rho\dot{\rho}}{\sqrt{1-\dot{\rho}^2}}\omega^{-2}, \ \ \ \ \ \ \ \mathcal{T}^{\sigma}{}_{\rho}\sqrt{-g} =  \pm  \mathcal{T} a^2\dot{\rho}\left(\frac{1-\omega^{2}}{\omega^{2}}\right) , \nonumber\\  \mathcal{T}^{\sigma\sigma}\sqrt{-g} &= -\mathcal{T} \frac{\sqrt{1-\dot{\rho}^2}}{\rho} \left(\frac{2\omega^{2}-1}{\omega^{2}}\right), \label{EMT_4.2c**}\\
\mathcal{T}^{0}{}_{\varphi}\sqrt{-g} &= \pm \mathcal{T} a \rho R \frac{\sqrt{1-\omega^{2}}}{\omega}, \ \ \ \ \ \ \mathcal{T}^{\sigma}{}_{\varphi}\sqrt{-g} = \mp  \mathcal{T} a \sqrt{1-\dot{\rho}^2} R \frac{\sqrt{1-\omega^{2}}}{\omega}, \label{EMT_4.2d*}
\end{align}
\end{subequations}
where
\begin{eqnarray}  \label{omega_4.2*}
\omega^{-2}(t) = \frac{a^2\rho^2 + R^2n_{\sigma}^2}{a^2\rho^2}.
\end{eqnarray}
We now define the \emph{normalised} components of the current density as
\begin{eqnarray} \label{J_compts*} 
\mathcal{J}^{0}\sqrt{-g}  = \frac{e}{2\pi}\frac{\sqrt{1-\dot{\rho}(t_i)}}{\rho(t_i)}\frac{\rho}{\sqrt{1-\dot{\rho}}}\dot{\varphi}, \ \ \ \mathcal{J}^{\sigma}\sqrt{-g}  = \frac{e}{2\pi}\frac{\sqrt{1-\dot{\rho}(t_i)}}{\rho(t_i)}\frac{\sqrt{1-\dot{\rho}}}{\rho}\partial_{\sigma}\varphi,
\end{eqnarray}
so that using the EOM in $\varphi$ gives
\begin{eqnarray} \label{J_reln*}
\mathcal{J}^{0}\sqrt{-g}  = \pm \frac{e}{2\pi}\frac{\sqrt{1-\dot{\rho}(t_i)}}{\rho(t_i)}\partial_{\sigma}\varphi = -\mathcal{J}^{\sigma}\sqrt{-g}  = \mp \frac{e}{2\pi}\frac{\sqrt{1-\dot{\rho}(t_i)}}{\rho(t_i)}\dot{\varphi}.
\end{eqnarray}
The components of the $4$-current may again be expressed in terms of $\mathcal{T}^{0}{}_{\varphi}\sqrt{-g} $ and $\mathcal{T}^{\sigma}{}_{\varphi}\sqrt{-g}$, now given by Eq. (\ref{EMT_4.2d*}), and it is straightforward to show that the current conservation equation is equivalent to Eq. (\ref{EOM_4.2d*}).
\\ \indent
By direct substitution of Eqs. (\ref{EMT_4.2a*})-(\ref{EMT_4.2c**}), together with Eq. (\ref{Constraint_4.2}), it may be verified that Eqs. (\ref{EOM_4.2b*}) and (\ref{EOM_4.2d*}) yield
\begin{eqnarray} \label{EOM6_4.2*}
 \frac{\rho}{\dot{\rho}}\frac{\dot{\omega}}{\omega} = 1-\omega^2,
\end{eqnarray}
and that Eqs. (\ref{EOM_4.2a*}) and (\ref{EOM_4.2c**}) give, respectively, 
\begin{eqnarray} \label{EOM7_4.2}
(1-\dot{\rho}^2)(2\omega^2-1) + \rho\ddot{\rho} + 2\dot{\rho}^2(1-\dot{\rho}^2)\left(1-\omega^2 - \frac{\rho}{\dot{\rho}}\frac{\dot{\omega}}{\omega}\right)= 0,
\end{eqnarray}
and
\begin{eqnarray} \label{EOM7*_4.2} 
(1-\dot{\rho}^2)\left(1-2\frac{\rho}{\dot{\rho}}\frac{\dot{\omega}}{\omega}\right) + \rho\ddot{\rho} = 0.
\end{eqnarray}
In fact, Eq. (\ref{EOM6_4.2*}) follows directly from the definition of $\omega^2$ under the assumption that $\omega^2 = \omega^2(t)$ (Eq. (\ref{omega_4.2*})) and is therefore equivalent to the condition 
$\partial_{\sigma}\omega = 0$ (\ref{Constraint_4.2}). 
Thus, the sole remaining independent EOM is
\begin{eqnarray}  \label{EOM8_4.2*}
(1-\dot{\rho}^2)(2\omega^2-1) + \rho\ddot{\rho} = 0.
\end{eqnarray}
In this system, the EOM in $\rho(t)$ is redundant and is satisfied identically according to the conservation of energy and angular momentum represented by the EOM in $X^0=t$, $\sigma$ and $\varphi$. 
It is also possible, though time consuming, to show that for $\partial_{\sigma}\omega \neq 0$ ($\varphi_{\rm NL} \neq 0$), Eqs. (\ref{EOM_4.2a*})-(\ref{EOM_4.2d*}) are \emph{not} self-consistent. 
Physically, this is because nonlinear fluctuations in the winding density would induce additional $\sigma$-dependence in the effective tension of the string, causing it to contract or expand at different rates at different points on the circumference of the loop. 
Therefore, even if the string began in an initially circular configuration in Minkowski space, this symmetry would immediately be broken if not reflected in the initial winding distribution. 
\\ \indent
Again utilising the EOM in $\varphi$, we may relate the parameter $\omega^2$ to the perpendicular velocity of the string
\begin{eqnarray} \label{v_perp_4.2}
\frac{{\rm d}l}{{\rm d}\sigma} = a\rho \omega^{-1}, \ \ \ \frac{v_{\perp}^2}{a^2} = \dot{\rho}^2 + (1-\dot{\rho}^2)(1-\omega^{2}). 
\end{eqnarray}
The $\sigma$-independence of the winding distribution also implies that 
\begin{eqnarray}  \label{omega_4.2}
\omega^{-2}(t) =  \Omega^{-2}(t) = \frac{a^2\rho^2 + R^2n_{\sigma}^2}{a^2\rho^2} = 1 +  \frac{(2\pi)^2R^2}{a^2 \lambda_{\sigma}^2},
\end{eqnarray}
where
\begin{eqnarray} \label{lambda_sig_4.2}
\lambda_{\sigma}(t) = \frac{2\pi \rho(t)}{n_{\sigma}},
\end{eqnarray}
which is analogous to the first relation in Eq. (\ref{LinCasRel1_3.2}) for straight strings, and we define the (time-dependent) wavenumber via
\begin{eqnarray} \label{k_sig_4.2}
k_{\sigma}(t) = \frac{2\pi}{\lambda_{\sigma}(t)}.
\end{eqnarray}
As for long straight strings with higher-dimensional windings, the EOM in $\varphi(t,\sigma)$ may be written as a dispersion relation
\begin{eqnarray} \label{DispRel_4.2}
\omega_{\sigma}^2 =(1 - \dot{\rho}^2)k_{\sigma}^2, \ \ \ \omega_{\sigma} = \pm \sqrt{1 - \dot{\rho}^2}k_{\sigma},
\end{eqnarray}
but now the instantaneous linear velocity of the string is given by 
\begin{eqnarray} \label{v_sig_4.2}
v_{\sigma}^2 = 1 - \dot{\rho}^2, \ \ \ v_{\sigma} = \pm \sqrt{1 - \dot{\rho}^2},
\end{eqnarray}
i.e., it depends on velocity of the string in the macroscopic dimensions and, also, implicitly on its curvature, since $\dot{\rho}$ is related to $\rho$.
\\ \indent
Further expressions, analogous to those in Eqs. (\ref{LinCasRel1_3.2})-(\ref{LinCasRel2_3.2}) which are valid for the straight string in the case of linear windings, also hold:
\begin{eqnarray} \label{omega_phi_4.2}
\omega_{\sigma}^2 = \omega_{\varphi}^2 = \frac{v_{\varphi}^2}{R^2}, \ \ \
v_{\varphi}^2 \lambda_{\sigma}^2 = (1 - \dot{\rho}^2)(2\pi)^2R^2, \ \ \
n_{\sigma}^2 = \frac{v_{\varphi}^2}{(1 - \dot{\rho}^2)} \frac{\rho^2}{R^2}.
\end{eqnarray}
In general, we have
\begin{eqnarray} \label{ConsRel3_4.2}
a^2\lambda_{\sigma}^2 = (2\pi)^2R^2\left(\frac{\Omega^2}{1-\Omega^2}\right),
\end{eqnarray}
so that  consistency requires
\begin{eqnarray} \label{ConsRel4_4.2}
v_{\varphi}^2 = a^2(1-\dot{\rho}^2)\left(\frac{1-\Omega^2}{\Omega^2}\right) \equiv \frac{v_{\perp}^2}{a^2}. 
\end{eqnarray} 
This is automatically satisfied for string loops with dynamically formed windings, for which the following relation holds
\begin{eqnarray} \label{DynWind_4.2}
\frac{v_{\varphi}^2}{a^2 v_{\sigma}^2} = \frac{1-\Omega^2}{\Omega^2},
\end{eqnarray} 
and which is equivalent to Eq. (\ref{ConsRel4_4.2}), since $v_{\sigma}^2 =1-\dot{\rho}^2$. 
Equations (\ref{ConsRel3_4.2})-(\ref{DynWind_4.2}) are the circular loop equivalents of Eqs. (\ref{lambda_z^eff_3.2})-(\ref{DynWind_3.2}), respectively. 
The string is everywhere effectively tensionless when
\begin{eqnarray} \label{NoTens1_4.2}
\frac{v_{\perp}^2}{a^2} = \omega^2 = \Omega^2 = 1/2, \ \forall t,
\end{eqnarray}
or equivalently when
\begin{eqnarray} \label{NoTens2_4.2}
a^2\lambda_{\sigma}^2 = (2\pi)^2R^2,
\end{eqnarray}
and, from Eq. (\ref{EOM8_4.2*}), it is immediately clear that under these conditions $\dot{\rho}=0$, $\ddot{\rho}=0 \ \forall t$, as expected. 
\\ \indent
In terms of the higher-dimensional string parameters, the constants of motion may be written as
\begin{eqnarray} \label{E1_4.2}
E  = \mathcal{T} \frac{2\pi a\rho}{\sqrt{1-\dot{\rho}^2}} \left(1 + \frac{(2\pi)^2R^2}{a^2 \lambda_{\sigma}^2}\right) =  \mathcal{T} \frac{2\pi a\rho}{\sqrt{1-\dot{\rho}^2}} \Omega^{-2},
\end{eqnarray}
\begin{eqnarray} \label{Lam^sig1_4.2}
\Lambda^{\sigma} = \pm 2\pi \mathcal{T}\frac{\rho}{\rho(t_i)}n_{\sigma} \times \frac{(2\pi)^2 R^2}{a \lambda_{\sigma}} = \pm 2\pi \mathcal{T}a \frac{\rho^2}{\rho(t_i)}\left(\frac{1-\Omega^{2}}{\Omega^{2}}\right), 
 \end{eqnarray}
\begin{eqnarray} \label{Lam^phi_4.2}
\Lambda^{\varphi} =  \pm 2\pi \mathcal{T} n_{\sigma} = \pm 2\pi\mathcal{T}\frac{a\rho}{R}\frac{\sqrt{1-\Omega^{2}}}{\Omega},
\end{eqnarray}
\begin{eqnarray} \label{Pi1_4.2}
\Pi = \mathcal{T} \frac{2\pi a\rho}{\sqrt{1-\dot{\rho}^2}}\Omega^{-1}\sqrt{\frac{1}{\Omega^2} - (1-\dot{\rho}^2)\left\{\frac{\rho^2}{\rho^2(t_i)}\frac{(1-\Omega^2)^2}{\Omega^2} + (1-\Omega^2)\right\}},
\end{eqnarray}
and the nonzero effective tensions are given by
\begin{eqnarray} \label{T^sig_4.2}
\mathcal{T}^{\sigma}  =  -2\pi \mathcal{T} a\rho \sqrt{1-\dot{\rho}^2} \left(1 - \frac{(2\pi)^2R^2}{a^2 \lambda_{\sigma}^2}\right)  =  -2\pi \mathcal{T} a\rho\sqrt{1-\dot{\rho}^2}\left(\frac{2\Omega^{2}-1}{\Omega^{2}}\right),
\end{eqnarray}
\begin{eqnarray} \label{T^rho_4.2}
\mathcal{T}^{\rho} = -\mathcal{T}\frac{a\rho \dot{\rho}^2}{\sqrt{1-\dot{\rho}^2}}\Omega^{-2} = -\dot{\rho}^2 E.
\end{eqnarray}
Again, $\mathcal{T}^{\varphi}=0$ since $\mathcal{T}^{\varphi\varphi}\sqrt{-g}=0$ by the EOM for $\varphi(t,\sigma)$. The nonzero shears between the $\sigma$- and $\rho$-,  $\sigma$- and $\varphi$-, and $\rho$- and $\phi$-directions are
\begin{eqnarray}  \label{S^sigrho_4.2}
\Sigma^{\rho\sigma} = \mp 2\pi \mathcal{T} a^2 \rho \dot{\rho} \left(\frac{1-\Omega^2}{\Omega^2}\right) = -a\rho(t_i)\frac{\dot{\rho}}{\rho}\Lambda^{\sigma}, 
\end{eqnarray}
\begin{eqnarray}  \label{S^sigphi_4.2}
\Sigma^{\sigma\varphi} = \mp 2\pi \mathcal{T} a^2 \rho \sqrt{1-\dot{\rho}^2}\frac{\sqrt{1-\Omega^2}}{\Omega^2} = -\frac{a}{R}\sqrt{1-\dot{\rho}^2}\Lambda^{\varphi}, 
\end{eqnarray}
\begin{eqnarray}  \label{rS^hophi_4.2}
\Sigma^{\rho\varphi} =  \mp 2\pi \mathcal{T} a^2 \rho \dot{\rho} \frac{\sqrt{1-\Omega^2}}{\Omega^2} = -a\rho(t_i)\frac{\dot{\rho}}{\rho}\Lambda^{\varphi}.
\end{eqnarray}
As expected, \emph{all} effective pressures and shears vanish locally when $\omega^2=\Omega^2=1/2$ and under these conditions we again have $\omega_{\sigma}^2 \equiv \omega_{\varphi}^2 = a^2/R^2$, as in the straight string case, (\ref{UniformRotn1}).
\\ \indent
We now note that, imposing the boundary conditions given in Eq. (\ref{Delta1_4.1}) and identifying $\lambda_{z} \leftrightarrow \lambda_{\sigma}(t_i)$, we may also identify
\begin{eqnarray}  \label{ChopOff1_4.2}
P^{z} \leftrightarrow \frac{l^{\sigma}}{2\pi \rho(t_i)},
\end{eqnarray}
which ensures that
\begin{eqnarray} \label{ChopOff2_4.2}
P^{z}P_{z} = l^{\sigma}l_{\sigma}(t_i),
\end{eqnarray}
and
\begin{eqnarray}  \label{ChopOff1_4.2*}
\mathcal{P}^{z} \leftrightarrow \Lambda^{\sigma}.
\end{eqnarray}
In this case, the total $4$-momentum, given by
\begin{eqnarray} \label{Pi2_4.2}
\Pi =  2\pi\mathcal{T} a\rho(t_i)\sqrt{1 + \frac{(2\pi)^2 R^2n_{\sigma}^2}{a^2 \lambda_{\sigma}^2(t_i)}} = 2\pi \mathcal{T} a\rho(t_i)\Omega^{-1}(t_i)
\end{eqnarray}
for $\dot{\rho}(t_i)=0$, is conserved when a long-string section chops off from the network to form a loop, despite the fact that linear momentum in the $z$-direction is `converted' into angular momentum in the $\sigma$-direction.\\
\indent
Finally, we consider the general solution of Eq. (\ref{EOM8_4.2*}). 
This was solved in \cite{LaYo12,LaWa10}, for the specific initial velocity $\dot{\rho}(t_i) =0$, by performing an Eulerian substitution of the second kind, giving
\begin{eqnarray} \label{EOMSoln5_4.2} 
\rho(t) = \rho(t_i)\sqrt{1 - \left(\frac{2\Omega^2(t_i)-1}{\Omega^4(t_i)}\right)\sin^2\left(\frac{\Omega^{2}(t_i)}{\rho(t_i)}(t-t_i)\right)}.
\end{eqnarray}
In this case the loop oscillates between two critical radii
\begin{eqnarray} \label{CritRad_4.1*} 
\rho_{c1} =  \rho(t_i), \ \ \ \rho_{c2} = \left(\frac{1-\Omega^{2}(t_i)}{\Omega^{2}(t_i)}\right)\rho(t_i),
\end{eqnarray}
performing one full oscillation with time period
\begin{eqnarray} \label{Period_4.1*}
T_{c} =  \rho(t_i)\Omega^{-2}(t_i) \pi. 
\end{eqnarray}
The same technique can be used to obtain the general solution for arbitrary initial velocities: $\rho(t)$ has the same functional form as in Eq. (\ref{EOMSoln5_4.2}),
\begin{eqnarray} \label{EOMSoln3_4.2}
\rho(t) = A\sqrt{1-B\sin(C(t-t_i) + D)},
\end{eqnarray}
but the constants $A-D$ are given by more complicated expressions involving $\dot{\rho}(t_i)$,
\begin{subequations}
\begin{align}
A &= \frac{1}{\sqrt{2}}\frac{\rho(t_i)}{\sqrt{1-\dot{\rho}^2(t_i)}}\Omega^{-2}(t_i) 
\nonumber\\
&\times
 \left[1 - 2(1-\dot{\rho}^2(t_i))\Omega^2(t_i)(1-\Omega^2(t_i)) + \sqrt{1 - 4(1-\dot{\rho}^2(t_i))\Omega^2(t_i)(1-\Omega^2(t_i))}\right]^{\frac{1}{2}}, \label{A}\\
B &= \frac{2\sqrt{1 - 4(1-\dot{\rho}^2(t_i))\Omega^2(t_i)(1-\Omega^2(t_i))}}{1 - 2(1-\dot{\rho}^2(t_i))\Omega^2(t_i)(1-\Omega^2(t_i)) + \sqrt{1 - 4(1-\dot{\rho}^2(t_i))\Omega^2(t_i)(1-\Omega^2(t_i))}}, \label{B}\\
C &= \frac{\sqrt{1-\dot{\rho}^2(t_i))}}{\rho(t_i)\Omega^{-2}(t_i)}, \label{C}\\
D &= -\sin^{-1}(B^{-1/2}\sqrt{1-\rho^2(t_i)/A^2}), \label{D}
\end{align}
\end{subequations}
In this case, the string oscillates with time period 
\begin{eqnarray} \label{Period_4.1*}
T_{c} =  \frac{\rho(t_i)}{\sqrt{1-\dot{\rho}^2(t_i)}}\Omega^{-2}(t_i) \pi. 
\end{eqnarray}
The second critical radius $\rho_{c2}$ may be deduced from Eq. (\ref{EOMSoln3_4.2}), but, since it is also a complicated function of the initial conditions we do not quote it here, explicitly, for the sake of brevity. 
In the limit $\Omega^2(t_i) \rightarrow 1$ ($\dot{\rho}^2(t_i) \geq 0$), we recover the previous solution for unwound circular loops, Eq. (\ref{GenSoln_4.1}) and for $\dot{\rho}^2(t_i) \rightarrow 0$ ($\Omega^2(t_i) < 1$), we recover Eq. (\ref{EOMSoln5_4.2}).

\subsection{Noncircular $F$-string loops} \label{Sect2.4}
In this section, we consider arbitrary planar loops, both with and without higher-dimensional windings. 
Since the EOM for this (most general) case are rather complicated in our chosen gauge, it is instructive to first review the analysis of noncircular loops without windings. 
This is done in Sect. \ref{Sect2.4.1}. 
Arbitrary planar loops with higher-dimensional windings are considered in detail in Sec. \ref{Sect2.4.2} and the general solution to the higher-dimensional part of the string EOM is given in terms of $(3+1)$-dimensional observables. 

\subsubsection{Noncircular $F$-string loops in $(3+1)$ dimensions} \label{Sect2.4.1}
The embedding for an arbitrary planar loop with no higher-dimensional windings, in cylindrical polar coordinates, is
\begin{eqnarray} \label{Emb_6.1}
X^I(\tau,\sigma)  \equiv X^I(t,\sigma) =\left(t(\tau)=\zeta \tau, x=0, \rho(t,\sigma), \sigma, \varphi=0 \right).
\end{eqnarray}
The action is 
\begin{eqnarray} \label{Act_6.1}
S = -\mathcal{T} a^2 \int {\rm d}t {\rm d}\sigma  \sqrt{(1-\dot{\rho}^2)\rho^2 + (\partial_{\sigma}\rho)^2},
\end{eqnarray}
which may be written in the standard form using
\begin{eqnarray} \label{dl_6.1}
\frac{{\rm d}l}{{\rm d}\sigma} = a \sqrt{\rho^2+(\partial_{\sigma}\rho)^2}, \ \ \ \frac{v_{\perp}^2}{a^2} = \dot{\rho}^2\left(1 - \frac{(\partial_{\sigma}\rho)^2}{\rho^2+(\partial_{\sigma}\rho)^2}\right).
\end{eqnarray}
Written terms of the energy-momentum tensor, the EOM are the same as for circular loops without windings, Eqs. (\ref{EOM_4.1a})-(\ref{EOM_4.1c}), but the components of $\mathcal{T}^{I}{}_{J}\sqrt{-g}$ are different:
\begin{subequations} 
\begin{align}
\mathcal{T}^{0}{}_{0}\sqrt{-g} = \mathcal{T}\frac{\rho^2+(\partial_{\sigma}\rho)^2}{\sqrt{(1-\dot{\rho}^2)\rho^2 + (\partial_{\sigma}\rho)^2)}}, \ \ \ 
\mathcal{T}^{\sigma}{}_{0}\sqrt{-g} =  -\mathcal{T}\frac{\dot{\rho}\partial_{\sigma}\rho}{\sqrt{(1-\dot{\rho}^2)\rho^2 + (\partial_{\sigma}\rho)^2)}}, \label{EMT_6.1a} \\
\mathcal{T}^{0}{}_{\sigma}\sqrt{-g} =  \mathcal{T}a^2\frac{\rho^2\dot{\rho}\partial_{\sigma}\rho}{\sqrt{(1-\dot{\rho}^2)\rho^2 + (\partial_{\sigma}\rho)^2)}}, \ \ \ 
\mathcal{T}^{\sigma}{}_{\sigma}\sqrt{-g} =  \mathcal{T}a^2\frac{(1-\dot{\rho}^2)\rho^2 }{\sqrt{(1-\dot{\rho}^2)\rho^2 + (\partial_{\sigma}\rho)^2)}}, \label{EMT_6.1b} \\
\mathcal{T}^{0}{}_{\rho}\sqrt{-g} =  -\mathcal{T}a^2\frac{\rho^2\dot{\rho}}{\sqrt{(1-\dot{\rho}^2)\rho^2 + (\partial_{\sigma}\rho)^2)}}, \ \ \ 
\mathcal{T}^{\sigma}{}_{\rho}\sqrt{-g} =  \mathcal{T}a^2\frac{\partial_{\sigma}\rho}{\sqrt{(1-\dot{\rho}^2)\rho^2 + (\partial_{\sigma}\rho)^2)}}, 
\nonumber\\
\mathcal{T}^{\sigma\sigma}\sqrt{-g} =  -\mathcal{T}a^2\frac{(1-\dot{\rho}^2)}{\sqrt{(1-\dot{\rho}^2)\rho^2 + (\partial_{\sigma}\rho)^2)}}. \label{EMT_6.1c}
\end{align}
\end{subequations}
By direct substitution, it is straightforward to show that each EOM is equivalent to
\begin{eqnarray} \label{noncircEOM}
(1-\dot{\rho}^2)\left(1-\frac{\partial^2_{\sigma}\rho}{\rho}\right) + \left(1+\frac{(\partial_{\sigma}\rho)^2}{\rho^2}\right)\rho\ddot{\rho} + 2\left(\frac{(\partial_{\sigma}\rho)^2}{\rho^2}-\frac{\dot{\rho}}{\rho}\partial_{\sigma}\rho\partial_{\sigma}\dot{\rho}\right) = 0,
\end{eqnarray}
which again expresses the conservation of energy, since the only nonzero constant of motion is the Hamiltonian. 
Formally, we may write down expressions for the constants of motion, integrated pressures and integrated shears, but these cannot be evaluated without adopting a more specific ansatz for $\rho(t,\sigma)$. 
It is sufficient for our purposes to note that in general $E \neq -\mathcal{T}^{\sigma}$, and that there is no way to obtain a static tensionless solution for an unwound string.

\subsubsection{Noncircular $F$-string loops with higher-dimensional windings} \label{Sect2.4.2}
In cylindrical polars, the general embedding for arbitrary planar loops with higher-dimensional windings is
\begin{eqnarray} \label{Emb_6.2}
X^I(\tau,\sigma)  \equiv X^I(t,\sigma) =\left(t(\tau)=\zeta \tau, x=0, \rho(t,\sigma), \sigma, \varphi(\tau,\sigma) \equiv \varphi(t,\sigma) \right).
\end{eqnarray}
The action is 
\begin{eqnarray} \label{Act_6.2}
S = -\mathcal{T} \int {\rm d}t {\rm d}\sigma \bigl[a^2(1-\dot{\rho}^2)(a^2\rho^2+R^2(\partial_{\sigma}\varphi)^2) + a^4(\partial_{\sigma}\rho)^2 
\nonumber\\
- (a^2\rho^2+a^2(\partial_{\sigma}\rho)^2)R^2\dot{\varphi}^2 + 2a^2\dot{\rho}\partial_{\sigma}\rho R^2\dot{\varphi}\partial_{\sigma}\varphi \bigr],
\end{eqnarray}
which may be written in the standard form by setting
\begin{eqnarray} \label{dl_6.2}
\frac{{\rm d}l}{{\rm d}\sigma} &=& \sqrt{a^2\rho^2+a^2(\partial_{\sigma}\rho)^2+R^2(\partial_{\sigma}\varphi)^2},
\nonumber\\
\frac{v_{\perp}^2}{a^2} &=& \dot{\rho}^2 + \frac{(\rho^2+(\partial_{\sigma}\rho)^2)R^2\dot{\varphi}^2-a^2\dot{\rho}^2(\partial_{\sigma}\rho)^2-2\dot{\rho}\partial_{\sigma}\rho R^2\dot{\varphi}\partial_{\sigma}\varphi}{a^2\rho^2+a^2(\partial_{\sigma}\rho)^2+R^2(\partial_{\sigma}\varphi)^2}.
\end{eqnarray}
The EOM are
\begin{subequations} 
\begin{align}
\partial_{0}(T^{0}{}_{0}\sqrt{-g}) + \partial_{\sigma}(T^{\sigma}{}_{0}\sqrt{-g}) &= 0, \label{EOM_6.2a}\\
\partial_{0}(T^{0}{}_{\sigma}\sqrt{-g}) + \partial_{\sigma}(T^{\sigma}{}_{\sigma}\sqrt{-g}) &=  0, \label{EOM_6.2b}\\
\partial_{0}(T^{0}{}_{\rho}\sqrt{-g}) + \partial_{\sigma}(T^{\sigma}{}_{\rho}\sqrt{-g}) - \frac{1}{2}\partial_{\rho}g_{\sigma\sigma}T^{\sigma\sigma}\sqrt{-g}&=  0, \label{EOM_6.2c}\\
\partial_{0}(T^{0}{}_{\varphi}\sqrt{-g}) + \partial_{\sigma}(T^{\sigma}{}_{\varphi}\sqrt{-g}) &= 0, \label{EOM_6.2d}
\end{align}
\end{subequations}
and Eq. (\ref{EOM_6.2d}) may be written in terms of $(-\gamma)$ and its derivatives as
\begin{eqnarray} \label{+x}
  & (-\gamma) \biggl[(1-\dot{\rho}^2)\partial_{\sigma}^2\varphi - (\rho^2+(\partial_{\sigma}\rho)^2)\ddot{\varphi} - 2(\rho\dot{\rho}+ \partial_{\sigma}\rho\partial_{\sigma}\dot{\rho})\dot{\varphi}   \nonumber\\
  &   (\ddot{\rho}\partial_{\sigma}\rho+\dot{\rho}\partial_{\sigma}\dot{\rho})\partial_{\sigma}\varphi + (\dot{\rho}\partial^2_{\sigma}\rho + \partial_{\sigma}\rho\partial_{\sigma}\dot{\rho})\dot{\varphi} - 2\dot{\rho}\partial_{\sigma}\dot{\rho}\partial_{\sigma}\varphi + 2\dot{\rho}\partial_{\sigma}\rho\partial_{\sigma}\dot{\varphi} \biggr]\nonumber\\
  & -  \frac{1}{2}\frac{\partial(-\gamma)}{\partial t}[\dot{\rho}\partial_{\sigma}\rho\partial_{\sigma}\varphi-(\rho^2+(\partial_{\sigma}\rho)^2)\dot{\varphi}] - \frac{1}{2}\frac{\partial(-\gamma)}{\partial \sigma}[(1-\dot{\rho}^2)\partial_{\sigma}\varphi + \dot{\rho}\partial_{\sigma}\rho\dot{\varphi}] = 0.
\end{eqnarray}
Writing out $(-\gamma)$ and its derivatives explicitly, this expression is even more complicated than Eq. (\ref{+}). 
However, bearing in mind our previous results, we may guess that Eq. (\ref{+x}) can be written in a compact form as 
\begin{eqnarray} \label{Const_6.2}
(1-\dot{\rho}^2)(\partial_{\sigma}\varphi)^2 -  (\rho^2+(\partial_{\sigma}\rho)^2)\dot{\varphi}^2 + 2\dot{\rho}\partial_{\sigma}\rho\dot{\varphi}\partial_{\sigma}\varphi = 0,
\end{eqnarray}
and that imposing this condition on the remaining EOM, Eqs. (\ref{EOM_6.2a})-(\ref{EOM_6.2c}), will lead to a self-consistent set of equations. 
In fact, it is far simpler to work directly with Eq. (\ref{EOM_6.2d}) as the EOM in $\phi(t,\sigma)$ and to impose the constraint in Eq. (\ref{Const_6.2}) \emph{before} multiplying out the derivatives. It is then straightforward to show that, subject to Eq. (\ref{Const_6.2}), 
\begin{eqnarray}
\mathcal{T}^{0}{}_{\varphi}\sqrt{-g} &=& \mp \mathcal{T}R^2 \partial_{\sigma}\varphi, 
\nonumber\\
\mathcal{T}^{\sigma}{}_{\varphi}\sqrt{-g} &=& \pm  \mathcal{T}R^2 \dot{\varphi},
\end{eqnarray}
so that Eq. (\ref{EOM_6.2d}) reduces to an identity. 
Thus, Eq. (\ref{+x}), which is equivalent to Eq. (\ref{EOM_6.2d}), is automatically satisfied by the imposition of the constraint, Eq. (\ref{Const_6.2}).
\\ \indent
As in the previous cases considered, for long straight strings and circular loops with higher-dimensional windings, imposing the constraint which satisfies the EOM in $\varphi(t,\sigma)$ implies that all higher-dimensional terms (i.e. terms dependent on $\varphi(t,\sigma)$ or its derivatives) cancel in $(-\gamma)$ at the level of the EOM. 
This condition also ensures $\mathcal{T}^{\varphi\varphi}\sqrt{-g}=0$, so that there is no effective pressure in either the $\varphi$-direction or the `$R$-direction' (i.e. ${\rm d}R=0$). 
Physically, this means that there is no effective pressure which can act to change the radius of the windings in the compact space. 
\\ \indent
For strings compactified on an genuine $M^4 \times S^1$ manifold, where the $S^1$ is of constant radius $R$, it is physically meaningless to talk about the string moving in the `$R$-direction'. 
However, the metric (\ref{Met_2.1}) is also valid as an effective metric for embeddings in more complex internal spaces, where the string wraps cycles of constant radius in the compact dimensions. 
For example, in \cite{LaYo12,LaWa10} and \cite{BlIg05}, strings wrapping great circles of the $S^3$ internal manifold which regularises the conifold at the tip of the Klebanov-Strassler geometry \cite{KlSt00} were considered. 
In such a geometry, the effective winding radius of the string may vary as a function of both $t$ and $\sigma$ and, since the windings are not topologically stabilised, they must be stabilised (if at all), dynamically.  
\\ \indent
As we have seen, for circular strings with higher-dimensional windings, nonzero effective pressure in the $\rho$-direction is associated with nonzero $\dot{\rho}$, but both these quantities become zero when there is no effective pressure in the $\sigma$-direction, $\mathcal{T}^{\sigma\sigma}\sqrt{-g}=0$. 
Though we did not explicitly include a term proportional to ${\rm d}R^2$ in the metric considered in this paper, or allow $R$ to be a function of the world-sheet coordinates in the embedding, we made such a restriction purely for simplicity and, in principle, the radius of the windings can evolve, in an arbitrary internal manifold, as a function of both space and time. 
Thus, nonzero $\dot{R}$ would be associated with nonzero $\mathcal{T}^{\varphi\varphi}\sqrt{-g}$, just as nonzero $\dot{\rho}$ is associated with nonzero $\mathcal{T}^{\sigma\sigma}\sqrt{-g}$. 
Viewed in this way, the condition (\ref{Const_6.2}) and the EOM for $\varphi(t,\sigma)$ and $\varphi(t,z)$ in Secs. \ref{Sect4.2} and \ref{Sect3.2}, respectively, result from, and are necessary conditions for, our initial assumption of a constant winding radius.  
\\ \indent
Interestingly, in each higher-dimensional case considered so far, the constraint equation which ensures $\mathcal{T}^{\varphi\varphi}\sqrt{-g}=0$ takes the form of a quadratic equation in $\dot{\varphi}/\partial_{\sigma}\varphi$ (or equivalently $\partial_{\sigma}\varphi/\dot{\varphi}$), whose descriminant is directly proportional to the value of $(-\gamma)$ \emph{after} the constraint has been applied, i.e. $(-\gamma) \propto (1-\dot{\rho}^2)\rho^2 +(\partial_{\sigma}\rho)^2$, or the equivalent thereof. 
Thus, from Eq. (\ref{Const_6.2}), we have that
\begin{eqnarray} \label{Const_6.2*}
\frac{\dot{\varphi}}{\partial_{\sigma}\varphi} = \frac{\dot{\rho}\partial_{\sigma}\rho \pm \sqrt{(1-\dot{\rho}^2)\rho^2 +(\partial_{\sigma}\rho)^2}}{\rho^2+(\partial_{\sigma}\rho)^2} = \frac{(1-\dot{\rho}^2)}{-\dot{\rho}\partial_{\sigma}\rho \pm \sqrt{(1-\dot{\rho}^2)\rho^2 +(\partial_{\sigma}\rho)^2}}
\end{eqnarray}
which reduces to the previous EOM in $\phi(t,\sigma)$ for circular loops in the limit $(\partial_{\sigma}\rho)^2 \rightarrow 0$ and to the EOM in $\varphi(t,z)$ for long straight strings when, in addition, $\dot{\rho} \rightarrow 0$ and $2\pi \rho = \Delta$. \\
\indent
We must now test the consistency of the Euler-Lagrange equations subject to this constraint. 
After applying Eqs. (\ref{Const_6.2})/(\ref{Const_6.2*}) to the components of the energy-momentum tensor, we have:
\begin{subequations}
\begin{align}
\mathcal{T}^{0}{}_{0}\sqrt{-g} &= \mathcal{T}a^2\frac{\rho^2+(\partial_{\sigma}\rho)^2}{\sqrt{(1-\dot{\rho}^2)\rho^2 + (\partial_{\sigma}\rho)^2}}\omega^{-2},  
\nonumber\\
\mathcal{T}^{\sigma}{}_{0}\sqrt{-g} &=  -\mathcal{T}a^2\frac{\dot{\rho}\partial_{\sigma}\rho}{\sqrt{(1-\dot{\rho}^2)\rho^2 + (\partial_{\sigma}\rho)^2}}\omega^{-2} \mp \mathcal{T}a^2\left(\frac{1-\omega^2}{\omega^2}\right), \label{EMT_6.2a*}\\
\mathcal{T}^{0}{}_{\sigma}\sqrt{-g} &=  \mathcal{T}a^2\frac{\rho^2\dot{\rho}\partial_{\sigma}\rho}{\sqrt{(1-\dot{\rho}^2)\rho^2 + (\partial_{\sigma}\rho)^2}}\omega^{-2} \pm  \mathcal{T}a^2\rho^2\left(\frac{1-\omega^2}{\omega^2}\right), 
\nonumber\\
\mathcal{T}^{\sigma}{}_{\sigma}\sqrt{-g} &=  \mathcal{T}a^2\frac{(1-\dot{\rho}^2)\rho^2 }{\sqrt{(1-\dot{\rho}^2)\rho^2 + (\partial_{\sigma}\rho)^2}}\left[1-\left(\frac{1-\omega^2}{\omega^2}\right)\frac{(\dot{\rho}\partial_{\sigma}\rho \pm \sqrt{(1-\dot{\rho}^2)\rho^2 +(\partial_{\sigma}\rho)^2})^2}{(1-\dot{\rho}^2)(\rho^2+(\partial_{\sigma}\rho)^2)}\right], \label{EMT_6.2b*}\\
\mathcal{T}^{0}{}_{\rho}\sqrt{-g} &=  -\mathcal{T}a^2\frac{\rho^2\dot{\rho}}{\sqrt{(1-\dot{\rho}^2)\rho^2 + (\partial_{\sigma}\rho)^2}}\omega^{-2} \mp \mathcal{T}a^2\partial_{\sigma}\rho\left(\frac{1-\omega^2}{\omega^2}\right), 
\nonumber\\
\mathcal{T}^{\sigma}{}_{\rho}\sqrt{-g} &=  \mathcal{T}a^2\frac{\partial_{\sigma}\rho}{\sqrt{(1-\dot{\rho}^2)\rho^2 + (\partial_{\sigma}\rho)^2}}\left[1-\left(\frac{1-\omega^2}{\omega^2}\right)\frac{(\dot{\rho}\partial_{\sigma}\rho \pm \sqrt{(1-\dot{\rho}^2)\rho^2 +(\partial_{\sigma}\rho)^2})^2}{(1-\dot{\rho}^2)(\rho^2+(\partial_{\sigma}\rho)^2)}\right]
\nonumber\\
&\pm  \mathcal{T}a^2\dot{\rho}\left(\frac{1-\omega^2}{\omega^2}\right)\frac{(\dot{\rho}\partial_{\sigma}\rho \pm \sqrt{(1-\dot{\rho}^2)\rho^2 +(\partial_{\sigma}\rho)^2}))^2}{(1-\dot{\rho}^2)(\rho^2+(\partial_{\sigma}\rho)^2)},
\nonumber\\
\mathcal{T}^{\sigma\sigma}\sqrt{-g} &=  -\mathcal{T}\frac{(1-\dot{\rho}^2)}{\sqrt{(1-\dot{\rho}^2)\rho^2 + (\partial_{\sigma}\rho)^2}}\left[1-\left(\frac{1-\omega^2}{\omega^2}\right)\frac{(\dot{\rho}\partial_{\sigma}\rho \pm \sqrt{(1-\dot{\rho}^2)\rho^2 +(\partial_{\sigma}\rho)^2})^2}{(1-\dot{\rho}^2)(\rho^2+(\partial_{\sigma}\rho)^2)}\right] \label{EMT_6.2c*}\\
\mathcal{T}^{0}{}_{\varphi}\sqrt{-g} &= \mp \mathcal{T} aR\sqrt{\rho^2 +(\partial_{\sigma}\rho)^2}\frac{\sqrt{1-\omega^2}}{\omega}, \nonumber\\ \mathcal{T}^{\sigma}{}_{\varphi}\sqrt{-g} &= \pm \mathcal{T} aR \left(\frac{\dot{\rho}\partial_{\sigma}\rho \pm \sqrt{(1-\dot{\rho}^2)\rho^2 +(\partial_{\sigma}\rho)^2}}{\sqrt{\rho^2+(\partial_{\sigma}\rho)^2}}\right) \frac{\sqrt{1-\omega^2}}{\omega}, \label{EMT_6.2d*}
\end{align}
\end{subequations}
where 
\begin{eqnarray} \label{omega_last}
\omega^{-2}(t,\sigma) = \frac{a^2(\rho^2+(\partial_{\sigma}\rho)^2) + R^2(\partial_{\sigma}\varphi)^2}{a^2(\rho^2+(\partial_{\sigma}\rho)^2) }.
\end{eqnarray}
In appropriate limits, these reduce to the equivalent expressions given in Secs. \ref{Sect2.2}-\ref{Sect2.3}. 
For noncircular loops, the normalized components of the current density are defined as
\begin{eqnarray} \label{J_compts**}
\mathcal{J}^{0}\sqrt{-g} &=& \frac{e}{2\pi} \frac{\dot{\rho}(t_i,\sigma)\partial_{\sigma}\rho(t_i,\sigma) \pm \sqrt{(1-\dot{\rho}^2(t_i,\sigma))\rho^2(t_i,\sigma) +(\partial_{\sigma}\rho(t_i,\sigma))^2}}{\rho^2(t_i,\sigma)+(\partial_{\sigma}\rho(t_i,\sigma))^2}
\nonumber\\
&\times& \frac{\rho^2+(\partial_{\sigma}\rho)^2}{\dot{\rho}\partial_{\sigma}\rho \pm \sqrt{(1-\dot{\rho}^2)\rho^2 +(\partial_{\sigma}\rho)^2}}\dot{\varphi},
\nonumber\\
\mathcal{J}^{\sigma}\sqrt{-g} &=& \frac{e}{2\pi} \frac{\dot{\rho}(t_i,\sigma)\partial_{\sigma}\rho(t_i,\sigma) \pm \sqrt{(1-\dot{\rho}^2(t_i,\sigma))\rho^2(t_i,\sigma) +(\partial_{\sigma}\rho(t_i,\sigma))^2}}{\rho^2(t_i,\sigma)+(\partial_{\sigma}\rho(t_i,\sigma))^2}
\nonumber\\
&\times& \frac{\dot{\rho}\partial_{\sigma}\rho \pm \sqrt{(1-\dot{\rho}^2)\rho^2 +(\partial_{\sigma}\rho)^2}}{\rho^2+(\partial_{\sigma}\rho)^2}\partial_{\sigma}\varphi,
\end{eqnarray}
so that using the EOM in $\varphi$ gives
\begin{eqnarray} \label{J_reln**}
&{}& \mathcal{J}^{0}\sqrt{-g} = \pm \frac{e}{2\pi}\frac{\dot{\rho}(t_i,\sigma)\partial_{\sigma}\rho(t_i,\sigma) \pm \sqrt{(1-\dot{\rho}^2(t_i,\sigma))\rho^2(t_i,\sigma) +(\partial_{\sigma}\rho(t_i,\sigma))^2}}{\rho^2(t_i,\sigma)+(\partial_{\sigma}\rho(t_i,\sigma))^2} \partial_{\sigma}\varphi 
\nonumber\\
= &{}&-\mathcal{J}^{\sigma}\sqrt{-g} = \mp \frac{e}{2\pi}\frac{\dot{\rho}(t_i,\sigma)\partial_{\sigma}\rho(t_i,\sigma) \pm \sqrt{(1-\dot{\rho}^2(t_i,\sigma))\rho^2(t_i,\sigma) +(\partial_{\sigma}\rho(t_i,\sigma))^2}}{\rho^2(t_i,\sigma)+(\partial_{\sigma}\rho(t_i,\sigma))^2} \dot{\varphi}.
\end{eqnarray}
The components of the $4$-current may again be expressed in terms of $\mathcal{T}^{0}{}_{\varphi}\sqrt{-g} $ and $\mathcal{T}^{\sigma}{}_{\varphi}\sqrt{-g}$, now given by Eq. (\ref{EMT_6.2d*}), and it is straightforward to show that the current conservation equation is equivalent to Eq. (\ref{EOM_6.2d}).
\\ \indent
It is tempting to think, based on the complexity of the expressions in Eqs. (\ref{EMT_6.2a*})-(\ref{EMT_6.2c*}), that Eqs. (\ref{EOM_6.2a})-(\ref{EOM_6.2c}) are intractably complex, at least analytically. 
However, as we will show in the remainder of this section, careful treatment of the remaining EOM will allow us to write the general solution of $\varphi(t,\sigma)$ in terms of $\rho(t,\sigma)$ and its first and second derivatives in $t$ and $\sigma$. 
As a first step towards demonstrating the self-consistency of Eqs. (\ref{EOM_6.2a})-(\ref{EOM_6.2c}), we note that each contains terms in $\omega$, $\dot{\omega}$ and $\partial_{\sigma}\omega$. 
In the case of circular loops, we had $\partial_{\sigma}\omega = 0$, so that the EOM contained terms in only $\omega$ and $\dot{\omega}$. 
However, we were able to eliminate the latter via appropriate manipulation, leaving us with a single self-consistent EOM in $\rho$ and $\omega$, in addition to the constraint on $\varphi$ (which also allowed $\varphi$ to be solved directly in terms of $\rho$). 
\\ \indent
Even in the more general case, we still have three EOM and two quantities (namely $\dot{\omega}$ and $\partial_{\sigma}\omega$) we wish to eliminate, so this poses no problem. 
One interesting aspect of the circular string case was that the EOM in $\rho$ was in fact redundant, being simply a `composite' of the expressions for the conservation of energy and momentum, which are equivalent to the EOM in $t$, $\sigma$ and $\varphi$ in our chosen gauge. 
As we shall see, the same is true even for noncircular loops, so that one of the (extremely complex) EOM above may be immediately ignored.
\\ \indent
We begin the elimination of $\dot{\omega}$ and $\partial_{\sigma}\omega$ by first writing each of the components of $\mathcal{T}^{I}{}_{J}\sqrt{-g}$ in Eqs. (\ref{EMT_6.2a*})-(\ref{EMT_6.2b*}) as the sum of two terms, one independent of the factor $(\omega^{-2}-1)$ and one directly proportional to $(\omega^{-2}-1)$. 
Thus, we set
\begin{subequations} 
\begin{align}
\mathcal{T}^{0}{}_{0}\sqrt{-g} = \beta + \beta(\omega^{-2}-1), \ \ \ \mathcal{T}^{\sigma}{}_{0}\sqrt{-g} = \gamma + \delta(\omega^{-2}-1), \label{GreekA*} \\
\mathcal{T}^{0}{}_{\sigma}\sqrt{-g} = \epsilon + \zeta(\omega^{-2}-1), \ \ \ \ \mathcal{T}^{\sigma}{}_{\sigma}\sqrt{-g} = \eta + \theta(\omega^{-2}-1), \label{GreekB*}
\end{align}
\end{subequations}
where the values of $\beta$, $\gamma$, $\delta$, $\epsilon$, $\zeta$, $\eta$ and $\theta$ are obtained by comparison of Eqs. (\ref{GreekA*})-(\ref{GreekB*}) with Eqs. (\ref{EMT_6.2a*})-(\ref{EMT_6.2b*}), i.e. 
\begin{eqnarray} \label{beta*}
\beta = \frac{\rho^2 + (\partial_{\sigma}\rho)^2}{\sqrt{(1-\dot{\rho}^2)\rho^2 + (\partial_{\sigma}\rho)^2}},
\end{eqnarray}
\begin{eqnarray} \label{delta*}
\gamma = \delta \mp 1, \ \ \ \delta = -\frac{\dot{\rho}\partial_{\sigma}\rho}{\sqrt{(1-\dot{\rho}^2)\rho^2 + (\partial_{\sigma}\rho)^2}} \pm 1.
\end{eqnarray}
\begin{eqnarray} \label{zeta*}
\epsilon = -\rho^2(\delta \mp 1), \ \ \  \zeta = -\rho^2\delta,
\end{eqnarray}
\begin{eqnarray} \label{theta*}
\eta = \frac{\rho^2(1-\dot{\rho})^2}{\sqrt{(1-\dot{\rho}^2)\rho^2 + (\partial_{\sigma}\rho)^2}}, \ \ \ \theta = -\frac{\rho^2\delta^2}{\beta}.
\end{eqnarray}
Substituting from Eqs. (\ref{GreekA*})-(\ref{GreekB*}) into Eqs. (\ref{EOM_6.2a})-(\ref{EOM_6.2b}), rearranging to make $\partial_{t}(\omega^{-2}-1)$ the subject and equating the results, gives
\begin{eqnarray} \label{GreekEOM1*}
(\zeta\delta-\beta\theta)\partial_{\sigma}(\omega^{-2}-1) + [\beta(\dot{\zeta}+\partial_{\sigma}\theta) - \zeta(\dot{\beta}+\partial_{\sigma}\delta)](\omega^{-2}-1) + \beta(\dot{\epsilon}+\partial_{\sigma}\eta) - \zeta(\dot{\beta}+\partial_{\sigma}\gamma) = 0. &{}&
\nonumber\\
&{}&
\end{eqnarray}
Then, substituting in for $\zeta$, $\delta$, $\beta$ and $\theta$ from Eqs. (\ref{beta*})-(\ref{theta*}), it may be shown that
\begin{eqnarray} \label{GreekEOM2*}
\zeta\delta-\beta\theta = 0,
\end{eqnarray}
and an EOM involving only $\omega$ can be obtained:
\begin{eqnarray} \label{GreekEOM3*}
[\beta(\dot{\zeta}+\partial_{\sigma}\theta) - \zeta(\dot{\beta}+\partial_{\sigma}\delta)](1-\omega^{2}) + [\beta(\dot{\epsilon}+\partial_{\sigma}\eta) - \zeta(\dot{\beta}+\partial_{\sigma}\gamma)]\omega^{2} = 0.
\end{eqnarray}
Next, we note that substituting for $(\dot{\epsilon}+\partial_{\sigma}\eta)$, $(\dot{\beta}+\partial_{\sigma}\gamma)$, $\beta$ and $\zeta$, the terms in the second set of square brackets in Eq. (\ref{GreekEOM3*}) 
are proportional to the expression on the left-hand-side of the EOM for noncircular strings \emph{without} higher-dimensional windings, Eq. (\ref{noncircEOM}). 
This is to be expected since, in the limit $\omega^2 \rightarrow 1$, consistency requires that we must recover the EOM for unwound strings. 
Specifically, we have that
\begin{eqnarray}
\beta(\dot{\epsilon}+\partial_{\sigma}\eta) - \zeta(\dot{\beta}+\partial_{\sigma}\gamma) = \left[\beta \partial_{\sigma}\rho - \rho^2\dot{\rho}\delta\right] \times \frac{\rho^3}{[(1-\dot{\rho}^2)\rho^2 + (\partial_{\sigma}\rho)^2]^{\frac{3}{2}}}\chi,
\end{eqnarray}
where
\begin{eqnarray} \label{chi*}
\chi = (1-\dot{\rho}^2)\left(1-\frac{\partial^2_{\sigma}\rho}{\rho}\right) + \left(1+\frac{(\partial_{\sigma}\rho)^2}{\rho^2}\right)\rho\ddot{\rho} + 2\left(\frac{(\partial_{\sigma}\rho)^2}{\rho^2}-\frac{\dot{\rho}}{\rho}\partial_{\sigma}\rho\partial_{\sigma}\dot{\rho}\right).
\end{eqnarray}
The first set of square brackets in Eq. (\ref{GreekEOM3*}) may also be written purely in terms of the quantities $\beta$ and $\delta$, i.e.  
\begin{eqnarray} \label{needlater}
\beta(\dot{\zeta}+\partial_{\sigma}\theta) - \zeta(\dot{\beta}+\partial_{\sigma}\delta) &=& \rho^2(\delta\dot{\beta}-\beta\dot{\delta}) - 2\delta(\rho\dot{\rho}\beta+\rho\partial_{\sigma}\rho\delta) + \frac{\rho^2\delta}{\beta}(\delta\partial_{\sigma}\beta-\beta\partial_{\sigma}\delta),
\end{eqnarray}
so that the final remaining independent EOM (involving only derivatives of $\rho$ and powers of $\omega$, but not $\dot{\omega}$ or $\partial_{\sigma}\omega$), may be written in a \emph{relatively} compact form as
\begin{eqnarray} \label{relcompact*}
\chi &+& \frac{\rho^3(\beta \partial_{\sigma}\rho - \rho^2\dot{\rho}\delta)}{[(1-\dot{\rho}^2)\rho^2 + (\partial_{\sigma}\rho)^2]^{\frac{3}{2}}}\left[\rho^2(\delta\dot{\beta}-\beta\dot{\delta}) - 2\delta(\rho\dot{\rho}\beta+\rho\partial_{\sigma}\rho\delta) + \frac{\rho^2\delta}{\beta}(\delta\partial_{\sigma}\beta-\beta\partial_{\sigma}\delta)\right]\left(\frac{1-\omega^2}{\omega^2}\right) 
\nonumber\\ &=& 0, 
\end{eqnarray}
where we make use of the definitions for $\chi$, $\beta$ and $\delta$, given in Eqs. (\ref{chi*}) and (\ref{beta*})-(\ref{delta*}). 
Hence, the term proportional to $(1-\omega^2)/\omega^2$ quantifies the contribution of the higher-dimensional motion to the macroscopic dynamics of the string. In the limit $\partial_{\sigma}\rho \rightarrow 0$, Eq. (\ref{relcompact*}) reduces to Eq. (\ref{EOM8_4.2*}), as required. 
It may be shown that $\omega^2=1/2$ corresponds to the tensionless case, for which $\dot{\rho} = 0$, $\partial_{\sigma}\dot{\rho}=0$ and $\ddot{\rho}=0$ for all $t$, by directly substituting each of these conditions into Eq. (\ref{relcompact*}). 
The EOM then reduces to $\omega^2 = 1-\omega^2$, which is satisfied if and only if $\omega^2=1/2$.
\\ \indent
Thus, Eqs. (\ref{relcompact*}) and (\ref{Const_6.2}) alone are sufficient to completely specify the dynamics of the string. 
The latter arises directly from the conservation of momentum in the compact space (the $\varphi$-direction) and the former combines the expressions for energy and momentum conservation in the whole space-time. 
As in the circular case, the EOM in $\rho$ is again redundant and was not used in the derivation of Eq. (\ref{relcompact*}).
\\ \indent
Finally, we may solve for $\varphi(t,\sigma)$ in terms of $\rho(t,\sigma)$ and its derivatives. 
For the sake of notational simplicity, we first rewrite Eq. (\ref{relcompact*}) as
\begin{eqnarray} \label{simples**}
-X\omega^2 + Y(1-\omega^2) = 0,
\end{eqnarray}
where the factors $X$ and $Y$ are independent of $\partial_{\sigma}\varphi$ or, equivalently
\begin{eqnarray} \label{simples+*}
\left(\frac{X}{Y}\right) =  \left(\frac{1-\omega^2}{\omega^2}\right),
\end{eqnarray}
where the factor $(1-\omega^2)/\omega^2$ is also independent of $\partial_{\sigma}\varphi$. 
Using the definition of $\omega^2(t,\sigma)$ (\ref{omega_last}), we may also rewrite $(\partial_{\sigma}\varphi)^2$ as 
\begin{eqnarray} \label{simples}
(\partial_{\sigma}\varphi)^2 = \frac{a^2}{R^2}(\rho^2+(\partial_{\sigma}\rho)^2)\left(\frac{1-\omega^2}{\omega^2}\right).
\end{eqnarray}
This expression has physical significance: as we we will see, it allows us to interpret the EOM in $\varphi(t,\sigma)$ as dispersion relation, as in Secs. \ref{Sect2.1} and \ref{Sect2.2}. 
From Eq. (\ref{Const_6.2*}) we then have 
\begin{eqnarray} \label{phi_dot}
\dot{\varphi}^2 = \frac{a^2}{R^2}\left(\dot{\rho}\partial_{\sigma}\rho \pm \sqrt{(1-\dot{\rho}^2)\rho^2 +(\partial_{\sigma}\rho)^2}\right)^2\left(\frac{1-\omega^2}{\omega^2}\right),
\end{eqnarray}
which allows us to reconstruct the required form of $\varphi(t,\sigma)$:
\begin{eqnarray} \label{phi_last*}
\varphi(t,\sigma) &=& \int \partial_{\sigma}\varphi {\rm d}\sigma + \int \dot{\varphi} {\rm d}t
\nonumber\\
&=& \frac{a}{R} \int \sqrt{\rho^2+(\partial_{\sigma}\rho)^2}\frac{\sqrt{1-\omega^2}}{\omega} {\rm d}\sigma \pm \frac{a}{R} \int \left(\dot{\rho}\partial_{\sigma}\rho \pm \sqrt{(1-\dot{\rho}^2)\rho^2 +(\partial_{\sigma}\rho)^2}\right)\frac{\sqrt{1-\omega^2}}{\omega} {\rm d}t.
\nonumber
\end{eqnarray}
\begin{eqnarray}
{}
\end{eqnarray}
For circular strings ($\partial_{\sigma}\rho=0$), Eqs. (\ref{simples+*}) and (\ref{simples**}) give
\begin{eqnarray} \label{simples*}
n_{\sigma}^2 = \frac{a^2}{R^2}\rho^2\left(\frac{1-\omega^2}{\omega^2}\right) = \frac{a^2}{R^2}\rho^2\left(\frac{1-\dot{\rho}^2 + \rho\ddot{\rho}}{1-\dot{\rho}^2 - \rho\ddot{\rho}}\right),
\end{eqnarray}
which is equivalent to Eq. (\ref{EOM8_4.2*}), using the expression for $\omega^2(t)$ in Eq. (\ref{omega_4.2}). 
This motivates the following definition for noncircular loops
\begin{eqnarray}\label{neff_last*}
(n_{\sigma}^{\rm eff})^2 = \frac{(v_{\varphi}^{\rm eff})^2(l_{\sigma}^{\rm eff})^2}{(v_{\sigma}^{\rm eff})^2(2\pi)^2R^2} = (\partial_{\sigma}\varphi)^2,
\end{eqnarray}
where 
\begin{eqnarray}\label{leff_last*}
(l_{\sigma}^{\rm eff})^2 = (2\pi)^2(\rho^2+(\partial_{\sigma}\rho)^2),
\end{eqnarray}
and
\begin{eqnarray}\label{enigma*}
\frac{(v_{\varphi}^{\rm eff})^2}{a^2(v_{\sigma}^{\rm eff})^2} = \left(\frac{1-\omega^2}{\omega^2}\right).
\end{eqnarray}
Together with Eqs. (\ref{neff_last*})-(\ref{enigma*}), the following definitions then form a self-consistent set:
\begin{eqnarray}\label{1*}
(l_{\sigma}^{\rm eff})^2 = (n_{\sigma}^{\rm eff})^2(\lambda_{\sigma}^{\rm eff})^2, \ \ \ 
(v_{\varphi}^{\rm eff})^2(\lambda_{\sigma}^{\rm eff})^2 = (v_{\sigma}^{\rm eff})^2(2\pi)^2R^2,
\end{eqnarray}
\begin{eqnarray}\label{3*}
(k_{\sigma}^{\rm eff})^2 = \frac{(2\pi)^2}{(\lambda_{\sigma}^{\rm eff})^2},
\end{eqnarray}
\begin{eqnarray}\label{4*}
(\omega_{\sigma}^{\rm eff})^2 = (v_{\sigma}^{\rm eff})^2(k_{\sigma}^{\rm eff})^2 \equiv (\omega_{\varphi}^{\rm eff})^2 =  \frac{(v_{\varphi}^{eff})^2}{R^2} = \dot{\varphi}^2.
\end{eqnarray}
Combining Eqs. (\ref{enigma*}) and (\ref{1*}), we then have
\begin{eqnarray}\label{5*}
a^2(\lambda_{\sigma}^{\rm eff})^2 = (2\pi)^2R^2\left(\frac{1-\omega^2}{\omega^2}\right).
\end{eqnarray}
We would like to be able to verify that the expressions for $(l_{\sigma}^{\rm eff})^2$, $(n_{\sigma}^{\rm eff})^2$, $(v_{\sigma}^{\rm eff})^2$, $(v_{\varphi}^{\rm eff})^2$, $(\lambda_{\sigma}^{\rm eff})^2$ and 
$(\omega_{\sigma}^{\rm eff})^2 \equiv (\omega_{\varphi}^{\rm eff})^2$ above, reduce to those for $l_{\sigma}^2$, $n_{\sigma}^2$, $v_{\sigma}^2$, $v_{\varphi}^2$ and $\omega_{\sigma}^2 \equiv \omega_{\varphi}^2$ for circular strings, and to those for $(l_{z}^{\rm eff})^2$, $(n_{z}^{\rm eff})^2$, $(v_{z}^{\rm eff})^2$, $(v_{\varphi}^{\rm eff})^2$, $(\lambda_{z}^{\rm eff})^2$ and $(\omega_{z}^{\rm eff})^2 \equiv (\omega_{\varphi}^{\rm eff})^2$ for straight strings, in appropriate limits. 
However, from Eqs. (\ref{neff_last*})-(\ref{5*}) alone, we do not have enough information to write down explicit expressions for $(\omega_{\sigma}^{\rm eff})^2$, $(v_{\sigma}^{\rm eff})^2$, or $(v_{\varphi}^{\rm eff})^2$. 
The missing piece comes from identifying $(\omega_{\sigma}^{\rm eff})^2$ with the integrand of the integral over ${\rm d}t$ in Eq. (\ref{phi_last*}). 
Hence, we have that
\begin{eqnarray}\label{6}
\varphi(t,\sigma) = \int n_{\sigma}^{\rm eff} {\rm d}\sigma + \int \omega_{\sigma}^{\rm eff} {\rm d}t,
\end{eqnarray}
where
\begin{eqnarray}\label{7}
(n_{\sigma}^{\rm eff})^2 = \frac{a^2}{R^2}(\rho^2+(\partial_{\sigma}\rho)^2)\left(\frac{1-\omega^2}{\omega^2}\right),
\end{eqnarray}
\begin{eqnarray}\label{8}
(\omega_{\sigma}^{\rm eff})^2 = \frac{a^2}{R^2}\left(\frac{(\dot{\rho}\partial_{\sigma}\rho \pm \sqrt{(1-\dot{\rho}^2)\rho^2 +(\partial_{\sigma}\rho)^2})^2}{\rho^2 +(\partial_{\sigma}\rho)^2}\right)\left(\frac{1-\omega^2}{\omega^2}\right),
\end{eqnarray}
and the sign of $\omega_{\sigma}^{\rm eff}$ relative to $n_{\sigma}^{\rm eff}$ depends on the sign of $\dot{\varphi}$ relative to $\partial_{\sigma}\varphi$.
\\ \indent
It may then be shown explicitly that the expressions for $(l_{\sigma}^{\rm eff})^2$, $(n_{\sigma}^{\rm eff})^2$, $(v_{\sigma}^{\rm eff})^2$, $(v_{\varphi}^{\rm eff})^2$, $(\lambda_{\sigma}^{\rm eff})^2$ and 
$(\omega_{\sigma}^{\rm \rm eff})^2 \equiv (\omega_{\varphi}^{\rm eff})^2$, defined in Eqs. (\ref{neff_last*})-(\ref{leff_last*}) and (\ref{1*})-(\ref{4*}), do in fact reduce to their counterparts for circular strings and long straight strings in the limits $\partial_{\sigma}\rho \rightarrow 0$ and $(2\pi)^2\rho \rightarrow \Delta$, respectively. 
Yet again, we see that the tensionless condition, $\omega^2=1/2$, corresponds to uniform rotation of the windings, with angular frequency $\omega_{\sigma}^2=a^2/R^2$, and the existence of a constant local wavelength, $a^2(\lambda_{\sigma}^{\rm eff})^2=a^2\lambda_{\sigma}=(2\pi)^2R^2$. 
Using Eqs. (\ref{Const_6.2}) and (\ref{simples**}), we may also express the perpendicular velocity of the string as
\begin{eqnarray} \label{vperp_last}
\frac{v_{\perp}^2}{a^2} = (1-\dot{\rho}^2)(1-\omega^2) + \dot{\rho}^2\left(1 - \frac{(\partial_{\sigma}\rho)^2\omega^2}{\rho^2+(\partial_{\sigma}\rho)^2}\right),
\end{eqnarray}
which may be seen as a straightforward generalisation of the cases considered previously. 
As in Sec. \ref{Sect2.4.1}, formally we may write down exact expressions for the constants of motion and integrated pressures and shears, but we are unable to evaluate them explicitly without further specifying the ansatz for $\rho(t,\sigma)$. 
It is therefore sufficient for our purposes to write down the remaining nonzero components of the energy-momentum tensor, from which, in conjunction with Eqs. (\ref{EMT_6.2a*})-(\ref{EMT_6.2d*}), all physical properties of the string can be calculated:
\begin{align*} 
\mathcal{T}^{\rho\rho}\sqrt{-g} &= \frac{-\mathcal{T}[(1-\dot{\rho}^2)(\rho^2 + (\partial_{\sigma}\rho)^2)-\rho^2]}{\sqrt{(1-\dot{\rho}^2)\rho^2 + (\partial_{\sigma}\rho)^2}} \\
&+ \frac{\mathcal{T}}{\sqrt{(1-\dot{\rho}^2)\rho^2 + (\partial_{\sigma}\rho)^2}}\biggl[\dot{\rho}^2(\rho^2 + (\partial_{\sigma}\rho)^2)-2\dot{\rho}\partial_{\sigma}\rho(\dot{\rho}\partial_{\sigma} \pm \sqrt{(1-\dot{\rho}^2)\rho^2 + (\partial_{\sigma}\rho)^2}) \\
&+  \frac{(\partial_{\sigma}\rho)^2(\dot{\rho}\partial_{\sigma} \pm \sqrt{(1-\dot{\rho}^2)\rho^2 + (\partial_{\sigma}\rho)^2})^2}{\rho^2 + (\partial_{\sigma}\rho)^2}\biggr]\left(\frac{1-\omega^2}{\omega^2}\right),
\end{align*}
\begin{eqnarray}\label{T1}
\mathcal{T}^{\rho\varphi}\sqrt{-g} = \pm \mathcal{T} \frac{a}{R} \frac{\rho^2\dot{\rho}}{\sqrt{(1-\dot{\rho}^2)\rho^2 + (\partial_{\sigma}\rho)^2}}\left(\frac{\dot{\rho}\partial_{\sigma}\rho \pm \sqrt{(1-\dot{\rho}^2)\rho^2 + (\partial_{\sigma}\rho)^2}}{\sqrt{\rho^2 + (\partial_{\sigma}\rho)^2}}\right)\frac{\sqrt{1-\omega^2}}{\omega}.
\end{eqnarray}
From Eqs. (\ref{EMT_6.2a*})-(\ref{EMT_6.2d*}) and (\ref{T1}), we see explicitly that \emph{all} effective pressures and shears vanish locally for $\omega^2=1/2$, as expected.
\\ \indent
Finally we note that, though it is beyond the scope of the current work, it would be useful to define a parameter $\alpha(t)$, the generalization of the constant $\alpha$ defined for long straight strings, which quantifies the degree of nonlinearity in the windings, at a given moment in time, for noncircular loops. 
By analogy with our previous results, we could then use this to determine the spatially-averaged values of the parameters defined in Eqs. (\ref{neff_last*})-(\ref{leff_last*}) and (\ref{1*})-(\ref{4*}), namely $\langle l_{\sigma}^2\rangle(t)$, $\langle n_{\sigma}^2\rangle(t)$, $\langle v_{\sigma}^2\rangle(t)$, $\langle v_{\varphi}^2\rangle(t)$, $\langle \lambda_{\sigma}^2\rangle(t)$ and $\langle \omega_{\sigma}^2\rangle(t) \equiv \langle \omega_{\varphi}^2\rangle(t)$, as well as a generalized expression for $\Omega^2(t)$, which is valid for arbitrary loop configurations. 
Based on our previous findings, we expect the condition of zero average tension to correspond to $\Omega^2= 1/2$, for which $a^2\langle \lambda_{\sigma}^2\rangle  = (2\pi)^2R^2$ and $\langle \omega_{\sigma}^2\rangle = a^2/R^2$ for all $t$. 
Such a definition would allow us to write the constants of motion and \emph{bulk} properties of noncircular strings in terms of spatially-averaged parameters. 
This could be useful for describing the bulk properties of string networks and help yield further insights into the relation between their microscopic and macroscopic dynamics. 
However, for the sake of brevity, we do not quote the explicit expressions here. 

\subsection{Summary of the wound-string model} \label{Sect2.5}
We have shown that the EOM corresponding to the higher-dimensional embedding of strings, with windings of constant radius in an internal manifold, take the form of canonical dispersion relations. 
The local propagation velocity of these extra-dimensional `waves', as seen from a $(3+1)$-dimensional perspective, is a function of the perpendicular velocity and curvature of the string in the large dimensions. 
Via the formal correspondence between higher-dimensional dynamics and an effective $(3+1)$-dimensional world-sheet current \cite{Ni79}, these dispersion relations may be regarded as governing the local current flow, at a given point on the string, at a given time $t$. 
For arbitrary planar loops, the EOM for the embedding coordinates in the large dimensions and the compact dimensions are separable, in an intuitive gauge in which the world-sheet parameters are directly identified with the cylindrically polar space-time coordinates. 
The macroscopic string dynamics can therefore be determined without reference to the motion of the string in the compact space, and the latter can be determined (in principle) from $(3+1)$-dimensional observables.
\\ \indent
The constants of motion of the string and all observable $(3+1)$-dimensional quantities, such as pressures, shears, etc, may be written in a compact form in terms of `microscopic' variables related to the higher-dimensional embedding, together with the fundamental model parameters such as the intrinsic tension $\mathcal{T}$ and compactification radius $R$. 
Particularly useful is the parameter $\omega^2(t,\sigma)$, which represents the local fraction of the string length lying in the large dimensions. 
The results obtained for the wound-string model may be compared with those obtained for current carrying defect strings in the Abelian-Higgs model, presented in Sec. \ref{Sect5}. 
Such a comparison suggests that $R$ plays the role of an effective four-dimensional thickness and that the higher-dimensional embedding parameter $\omega^2(t,\sigma)$ corresponds to the embedding of the lines of constant phase, or, equivalently, magnetic field lines, within the defect-string core. 
\\ \indent
In principle, a similar analysis to that conducted in Sec. \ref{Sect5} for Abelian-Higgs strings may be conducted for any defect string species, including chiral strings \cite{Martins&Shellard(1998),Carter&Davis(2000)}, electroweak strings \cite{Vachaspati:1992uz,James:1992zp} and GUT scale strings \cite{Jeannerot:2003qv,Rocher:2004rc,Sakellariadou:2004rf}. 
The analysis presented here may also be extended to include more complex warped geometries or internal manifolds and varieties of $F$- and $D$-string bound states \cite{Polchinski_Intro,CMP_FD1}. Thus, by writing the EOM for higher-dimensional dynamics in the form of canonical dispersion relations for waves `in' strings, and comparing these with genuine dispersion relations governing the flow of current in topological defects, we are able to predict the conditions under which a given $F/D$-string species, embedded in a higher-dimensional space, can be successfully mimicked by a field-theoretic string living in $(3+1)$-dimensions. 
In Sec. \ref{Sect5} we will see that, mathematically, the {\it exact} EOM governing the evolution of the of the microscopic `internal' variables for current-carrying defect stings are most naturally expressed in terms of effective Finsler and generalised Finsler metrics, for the large space-time dimensions. 

\section{Topological defect strings with internal currents}\label{Sect3}
In this section we review the simplest field theory theory that gives rise to vortex-strings, i.e., the Abelian-Higgs model. 
The action, EOM and constants of motion for the Abelian-Higgs fields, coupled to an external source of charge, are reviewed in Sec. \ref{Sect3.1.1} and the uncharged vortex solution, known as the Nielsen-Olesen string \cite{NO}, is reviewed in Sec. \ref{Sect3.1.2}. 
In Sec. \ref{Sect3.2} we briefly review the standard derivation of the effective actions for both non-superconducting and superconducting strings, in the zero-width limit. 

\subsection{The Abelian-Higgs model and the Nielsen-Olesen string}\label{Sect3.1}
In this Section, we give a brief overview of the Abelian-Higgs action, with an additional `external' current coupled to the electromagnetic sector. 
The EOM and energy-momentum tensor in the general case, $j^{\mu} \neq 0$, are obtained, and a summary of the Nielsen-Olesen vortex solution, for $j^{\mu} = 0$, is presented. 

\subsubsection{The Abelian-Higgs model with externally coupled charge ($j^{\mu} \neq 0$)} \label{Sect3.1.1}
In natural units, using the metric convention $(+---)$, the Abelian-Higgs action with an additional current $j^{\mu}$ coupled to the electromagnetic sector is
\begin{eqnarray} \label{AH_Act}
S = \int {\rm d}^4x \sqrt{-g} \left\{D_{\mu}\phi \overline{D}^{\mu}\overline{\phi} - \frac{1}{4}F_{\mu\nu}F^{\mu\nu} - V(|\phi|) - j^{\mu}A_{\mu}\right\},
\end{eqnarray}
where $\mu, \nu \in \left\{0,1,2,3\right\}$ and $V(|\phi|)$ is the potential term, which is determined by the symmetry breaking energy scale, $\eta$, and the scalar coupling, $\lambda$:
\begin{eqnarray} \label{AH_Pot}
V(|\phi|) = \frac{\lambda}{4}(|\phi|^2-\eta^2)^2.
\end{eqnarray}
The gauge covariant derivative $D_{\mu}$ and the electromagnetic field tensor $F_{\mu\nu}$ are defined according to
\begin{eqnarray} \label{Cov_Deriv}
D_{\mu} = \partial_{\mu} - ieA_{\mu}, \ \ \ F_{\mu\nu} = \partial_{\mu}A_{\nu} - \partial_{\nu}A_{\mu},
\end{eqnarray}
where $e$ is the vector coupling, and the covariant EOM are
\begin{eqnarray} \label{Cov_EOMs}
\frac{1}{\sqrt{-g}}D_{\mu}\left(\sqrt{-g}D^{\mu}\phi\right) + \lambda\phi\left(|\phi|^2-\eta^2\right) = 0,
\end{eqnarray}
\begin{eqnarray} \label{Cov_EOMv}
\frac{1}{\sqrt{-g}}\partial_{\nu}\left(\sqrt{-g}F^{\mu\nu}\right) + ie\left(\overline{\phi}D^{\mu}\phi-\phi\overline{D}^{\mu}\overline{\phi}\right) - j^{\nu} = 0.
\end{eqnarray}
The Abelian-Higgs part of the action (\ref{AH_Act}) is invariant under local $U(1)$ transformations of the form
\begin{eqnarray} \label{U(1)_invar}
\phi \rightarrow \phi' = \phi e^{i\Lambda(x)}, \ \ \ A_{\mu} \rightarrow A_{\mu}' = A_{\mu} + \frac{1}{e}\partial_{\mu}\Lambda(x),
\end{eqnarray}
where $\Lambda(x)$ is any single-valued real function, giving rise to the conserved current $\mathcal{J}^{\mu} =  -ie\left(\overline{\phi}D^{\mu}\phi-\phi\overline{D}^{\mu}\overline{\phi}\right)$.
The total conserved current, including the externally coupled charge is then given by
\begin{eqnarray} \label{Cons_current}
J^{\mu} = \mathcal{J}^{\mu} + j^{\mu} =  -ie\left(\overline{\phi}D^{\mu}\phi-\phi\overline{D}^{\mu}\overline{\phi}\right) + j^{\mu}.
\end{eqnarray}
The conservation of $J^{\mu}$ is expressed directly via the vector EOM, Eq. (\ref{Cov_EOMv}), and the corresponding conserved charge is
\begin{eqnarray} \label{Q}
Q = \int J^{0} \sqrt{-g} {\rm d}^3x.
\end{eqnarray}
The energy-momentum tensor is defined implicitly by varying the action with respect to the metric,
\begin{eqnarray} \label{T_mu_nu*}
\delta S = \int T^{\mu\nu} \delta g_{\mu\nu} \sqrt{-g}{\rm d}^4x,
\end{eqnarray}
so that 
\begin{eqnarray} \label{T_mu_nu}
T^{\mu\nu} = \frac{-2}{\sqrt{-g}}\frac{\partial (\mathcal{L}\sqrt{-g})}{\partial g_{\mu\nu}} = D^{\mu}\phi\overline{D}^{\nu}\overline{\phi} + \overline{D}^{\mu}\overline{\phi}D^{\nu}\phi - F^{\mu}{}_{\alpha}F^{\nu\alpha} 
- \frac{1}{2}\left(j^{\mu}A^{\nu} + j^{\nu}A^{\mu}\right) - g^{\mu\nu}\mathcal{L},
\end{eqnarray}
where  $\mathcal{L}$ is the Lagrangian density, that is, the sum of terms inside the curly brackets in Eq. (\ref{AH_Act}).

\subsubsection{The Nielsen-Olesen solution ($j^{\mu} = 0$)}\label{Sect3.1.2}
In cylindrical polar coordinates $(t,r,\theta,z)$ and assuming a Minkowski background 
\begin{eqnarray} \label{Metric1}
{\rm d}s^2 = \eta_{\mu\nu}dx^{\mu}{\rm d}x^{\nu} = {\rm d}t^2 - {\rm d}r^2 - r^2{\rm d}\theta^2 - {\rm d}z^2,
\end{eqnarray}
the ansatz for the Nielsen-Olesen string is \cite{NO}
\begin{eqnarray} \label{NO_ansatz}
\phi(r,\theta) = \eta f(r) e^{in\theta}, \ \ \ A_{\theta} = \frac{n}{e}\alpha_{\theta}(r), \ \ \ A_r=A_z=A_0 = 0,
\end{eqnarray}
with $j^{\mu}=0$, where $f(r)$ and $a_{\theta}(r)$ are dimensionless functions which obey the boundary conditions
\begin{equation} \label{NO_f_bc}
f(r) = \left \lbrace
\begin{array}{rl}
0,& \ r=0 \\
1,& \ r \rightarrow \infty,
\end{array}\right.
\end{equation}
\begin{equation} \label{NO_a_{theta}_bc}
a_{\theta}(r) = \left \lbrace
\begin{array}{rl}
0,& \ r=0 \\
1,& \ r \rightarrow \infty.
\end{array}\right.
\end{equation}
The EOM may be written as \cite{ViSh00}
\begin{eqnarray} \label{NO_specific_scalar_EOM} 
\frac{{\rm d}^2 f}{{\rm d} R_{s|n|}^2} + \frac{1}{R_{s|n|}} \frac{{\rm d} f}{{\rm d} r} - \frac{n^2 f}{R_{s|n|}^2}(1-\alpha_{\theta})^2 - \frac{1}{2}\left(\frac{r_{s|n|}}{r_s}\right)^2f(f^2-1) = 0,
\end{eqnarray}
\begin{eqnarray} \label{NO_specific_vector_EOM_theta}
\frac{{\rm d}^2 a_{\theta}}{{\rm d} R_{v|n|}^2} - \frac{1}{R_{v|n|}} \frac{{\rm d} a_{\theta}}{{\rm d} R_{v|n|}} + 2\left(\frac{r_{v|n|}}{r_v}\right)^2f^2(1 - a_{\theta}) - j^{\nu} = 0,
\end{eqnarray}
where we have defined the dimensionless coordinates
\begin{eqnarray} \label{R}
R_{i|n|} = \frac{r}{r_{i|n|}}, \ i \in \left\{s,v\right\},
\end{eqnarray}
and where $r_{s|n|}$ and $r_{v|n|}$ denote the values of the scalar and vector core radii for an $|n|$-vortex string, respectively.
We also use the simplified notation, $r_s$ and $r_v$, to refer to the scalar and vector core radii of $|n|=1$ strings and define the parameters
\begin{eqnarray}
\beta_{|n|} = \frac{r_{v|n|}^2}{r_{s|n|}^2}, \ \ \ \beta = \frac{r_{v}^2}{r_{s}^2},
\end{eqnarray}
for later convenience. 
Equations (\ref{NO_specific_scalar_EOM})-(\ref{NO_specific_vector_EOM_theta}) are solved, to leading order in the uncoupled regime, by \cite{Thesis}
\begin{equation} \label{NO_f}
f(r) \simeq \left \lbrace
\begin{array}{rl}
(r/r_{s|n|})^{|n|},& \ r \lesssim r_{s|n|} \\
1,& \ r \gtrsim r_{s|n|},
\end{array}\right.
\end{equation}
\begin{equation} \label{NO_alpha}
a_{\theta}(r) \simeq \left \lbrace
\begin{array}{rl}
(r/r_{v|n|})^{2},& \ r \lesssim r_{v|n|} \\
1,& \ r \gtrsim r_{v|n|}.
\end{array}\right.
\end{equation}
Strictly, the power law solutions for $f(r)$ and $a_{\theta}(r)$ given in Eqs. (\ref{NO_f})-(\ref{NO_alpha}) are valid only in the ranges $r \ll r_{i|n|}$, $i \in \left\{s,v\right\}$, while the asymptotic forms defined in Eqs. (\ref{NO_f_bc})-(\ref{NO_a_{theta}_bc}) hold only for $r \gg r_{i|n|}$. 
However, by assuming that each holds approximately, up to the limiting value $r = r_{i|n|}$, and ignoring the discontinuity of the derivatives, we can easily obtain order of magnitude estimates for the physical parameters of the string without the need for detailed numerical calculations.
\\ \indent
The topological winding number $n \in \mathbb{Z}$ is given by
\begin{eqnarray} \label{core_radii}
n = \frac{1}{2\pi}\int_{0}^{2\pi} \frac{\partial \vartheta}{\partial\theta}\Bigg|_{z = {\rm const.}} {\rm d}\theta,
\end{eqnarray}
where $\vartheta$ is the phase of the scalar field $\phi$. In general, the core radii are of the order of the Compton wavelengths of the associated bosons, $m_s^{-1}$ and $m_v^{-1}$, but may also depend on $|n|$. The $|n|$-dependence is expected to take a simple form \cite{Thesis,Bo:76A,Bo:76B} so that here we assume
\begin{eqnarray} \label{core_radii}
r_{s|n|} &\simeq& |n|^{\xi} m_s^{-1} = |n|^{\xi} (\sqrt{\lambda}\eta)^{-1},
\nonumber\\
r_{v|n|} &\simeq& |n|^{\epsilon} m_v^{-1} = |n|^{\epsilon} (e \eta)^{-1},
\end{eqnarray}
where $\xi \geq 0$, $\epsilon \geq 0$ are constants.\\ 
\indent
Substituting the ansatz (\ref{NO_ansatz}) together with the approximate solutions in Eqs. (\ref{NO_f})-(\ref{NO_alpha}) into Eq. (\ref{T_mu_nu}), the only nonzero component of the energy-momentum tensor is $T^{00}$ and the only constant of motion is the Hamiltonian $E_{|n|}$, which depends on the absolute value of the topological winding number. 
For $r_{s|n|}$ and $r_{v|n|}$ given by Eq. (\ref{core_radii}), it is straight forward to show that, assuming $r_v \geq r_s$ ($\beta \geq 1$), which corresponds to a Type II superconducting regime \cite{ViSh00}, the (constant) mass-energy per unit length of the string, $\mu_{|n|} = \int T^{00}r {\rm d}r {\rm d}\theta$, is:
\begin{eqnarray} \label{NO_mu2}
\mu_{|n|} = 2\pi \eta^2 \left[|n| + |n|^{2-2\epsilon} + |n|^{2\xi} + |n|^2\ln\left(\sqrt{\beta_{|n|}}\right)\right].
\end{eqnarray}
However, Bogomol'nyi showed that for a Nielsen-Olesen string at critical coupling $e = \sqrt{\lambda}$, for which
\begin{eqnarray} \label{critical_coupling}
r_{s|n|} = r_{s|n|} \equiv r_{c|n|}, \quad (\beta_{|n|} = 1), 
\end{eqnarray}
must satisfy the condition \cite{Bo:76A,Bo:76B}
\begin{eqnarray} \label{NO_mu2}
\mu_{|n|} \geq 2\pi \eta^2 |n|,
\end{eqnarray}
also known as the Bogomol'nyi-Prasad-Sommerfield (BPS) bound \cite{PrSo:75}. Since stability implies saturation of the bound \cite{ViSh00,deVe:76}, we require
\begin{eqnarray} \label{ParamBounds1}
0 \leq \xi \leq 1/2, \ \ \  \epsilon \geq 1/2,
\end{eqnarray}
for $|n| \sim \mathcal{O}(1)$. Alternatively, we may set
\begin{eqnarray} \label{ParamBounds2}
\xi = \epsilon = 1/2,
\end{eqnarray}
for arbitrary $|n|$. It is common in the literature to take $\xi=0$ and $\epsilon = 1/2$, following Bogomol'nyi's original assumptions \cite{Bo:76A,Bo:76B}, but in the present work we instead adopt the conditions given in Eq.  (\ref{ParamBounds2}) to allow for large $|n|$. 
For either choice of parameters, the order of magnitude estimate for the mass-energy per unit length of a Nielsen-Olesen string with winding number $n$, at critical coupling, is
\begin{eqnarray} \label{NO_mu3}
\mu_{|n|} = 2\pi \eta^2 |n|.
\end{eqnarray}
Though, in general, the effective tensions/pressures in the string $\mathcal{T}^{r}{}_{r} = \int T^{r}{}_{r}r {\rm d}r {\rm d}\theta$, $\mathcal{T}^{\theta}{}_{\theta} = \int T^{\theta}{}_{\theta}r {\rm d}r {\rm d}\theta$ and 
$\mathcal{T}^{z}{}_{z} = \int T^{z}{}_{z}r {\rm d}r {\rm d}\theta$, are not necessarily conserved, is straightforward to show that, for the Nielsen-Olesen string
\begin{eqnarray}
\mu_{|n|} = -\mathcal{T}^{z}{}_{z},
\end{eqnarray}
for any value of $\beta_{|n|}$, and
\begin{eqnarray} \label{T^{rr}**}
\mathcal{T}^{r}{}_{r} = \mathcal{T}^{\theta}{}_{\theta} = -2\pi \eta^2 \left[|n|^{2\xi} - |n|^{2-2\epsilon} - |n|^2\ln\left(\sqrt{\beta_{|n|}}\right)\right].
\end{eqnarray}
Hence, for $\xi = \epsilon = 1/2$, as in Eq. (\ref{ParamBounds2}), and at critical coupling, the only nonzero component of the tension is $\mathcal{T}^{z}{}_{z}$, which is equal to minus the energy per unit length of the string.
Substituting the ansatz (\ref{NO_ansatz}) with $j^{\mu} = 0$ into Eq. (\ref{Cons_current}), it is straightforward to show that $J^{\mu} = 0$ for all $\mu \in \left\{0,1,2,3\right\}$ so that the string is uncharged.

\subsection{String effective actions, with and without currents}\label{Sect3.2}
An effective action for the Nielsen-Olesen string can be constructed by switching to a set of dimensionless world-sheet coordinates $\zeta^{a}$, $a \in \left\{0,1\right\}$, where $\zeta^0 = \tau$ is time-like and $\zeta^1 = \sigma$ is space-like, as for $F$-strings. 
These parameterize the two-dimensional sheet swept out by the line $\langle \phi \rangle = 0$, which represents the central axis of the string core. 
Erecting two normals to this sheet, denoted $n_{\alpha}{}^{\mu}$, we may describe the small volume of space-time swept out by the finite-width string using the coordinates $Y^{\mu} = X^{\mu}(\tau,\sigma) + n_{\alpha}{}^{\mu}\rho^{\alpha}$, where $X^{\mu}$ are the embedding coordinates of the core central axis and $\rho^{\alpha}$ probe the core region, e.g. $\rho^{\alpha} \in \left\{r,\theta \right\}$ where $r \in [0,r_{c|n|}]$ and $\theta \in [0,2\pi)$ \cite{CoHiTu87,Copeland:1987yv}. 
\\ \indent
If the curvature radius of the string is much greater than its thickness, we may approximate the integral over this volume as ${\rm d}^4x\sqrt{-g} = {\rm d}^2\zeta {\rm d}^2\rho\sqrt{-\tilde{g}}$, where $\tilde{g}_{\mu\nu}(\tilde{\zeta})$ is the induced metric with respect to the coordinates $\tilde{\zeta}^{\mu} = \left\{\zeta^{a},\rho^{\alpha}\right\}$. 
To zeroth order in $\rho^{\alpha}$, this metric is given by a block diagonal matrix, such that $\tilde{g}_{\mu\nu} =$ diag$(\gamma_{ab},\delta_{\alpha\beta}) + \mathcal{O}(\rho)$, where $\delta_{\alpha\beta}$ is the usual Kronecker delta symbol and $\gamma_{ab}$ is the induced metric on the world-sheet defined by $\langle \phi \rangle = 0$, $\gamma_{ab}(\zeta) = g_{\mu\nu}(X)(\partial X^{\mu}/\partial \zeta^{a})(\partial X^{\mu}/\partial \zeta^{b})$. 
Substituting from (\ref{NO_ansatz}) and (\ref{NO_f})-(\ref{NO_alpha}) and performing the transverse integration over ${\rm d}^2\rho$, then gives \cite{CoHiTu87,Copeland:1987yv}
\begin{eqnarray} \label{NG_act}
S = -\mu \int {\rm d}^2\zeta \sqrt{-\gamma} + \ . \ . \  .
\end{eqnarray}
where $\mu \simeq \mu_{|n|}$ is given by Eq. (\ref{NO_mu3}) and we have again assumed critical coupling, $r_{s|n|} = r_{v|n|} = r_{c|n|}$ and $|n| \sim \mathcal{O}(1)$, for the sake of simplicity.
Hence, to leading order, the effective action for a Nielsen-Olesen vacuum string, in the Abelian-Higgs model, is simply the Nambu-Goto action for an $F$-string with no electric or magnetic world-sheet fluxes \cite{Go71,And02,Na77}.
\\ \indent
The presence of additional world-sheet fluxes then generates an additional contribution to the action (\ref{NG_act}), given by \cite{LaHa17}
\begin{eqnarray} \label{chiral_DeltaS-2}
\Delta S = \frac{1}{e^2\mu}\int {\rm d}^2\zeta \sqrt{-\gamma}\gamma_{ab} J^{a}J^{b},
\end{eqnarray}
where $J^{a}$ denotes the world-sheet components of the physical current. 
The total effective action may then be written as
\begin{eqnarray} \label{chiral_S}
S \simeq -\mu  \int {\rm d}^2\zeta \sqrt{-\gamma}(1 - \gamma_{ab} j^{a}j^{b}),
\end{eqnarray}
where $j^{a} = e^{-1}J^{a}$ is the dimensionless current. 
The world-sheet energy momentum tensor for the superconducting string is defined as $\theta^{ab} = 2j^{a}j^{b} - \gamma^{ab}(\gamma_{cd}j^{c}j^{d})$, so that the physical energy-momentum tensor of the effective model is given by 
\begin{eqnarray} \label{chiral_EMT}
T^{\mu\nu} = \frac{-2}{\sqrt{-g}}\int {\rm d}^2\zeta \sqrt{-\gamma}(\gamma^{ab} + \theta^{ab})\partial_{a}X^{\mu}\partial_{b}X^{\nu}\delta^{4}(x-X),
\nonumber
\end{eqnarray}
where $X^{\mu}$ denotes a string embedding coordinate, as before, and $x^{\mu}$ denotes a space-time background coordinate \cite{CoHiTu87,Copeland:1987yv,LaHa17}.
\\ \indent
In \cite{Witten:1985}, Witten outlined, from a microphysical perspective, how strings can carry fermionic currents, with fermionic charge carriers trapped as zero modes along the string. 
However, due to Bose-Fermi equivalence in $(1+1)$ dimensions, the effective action (\ref{chiral_S}) developed in \cite{CoHiTu87,Copeland:1987yv} is valid for strings with both bosonic and fermionic currents. 
It is therefore valid in general, as a first order approximation, for any species of superconducting string. 
\\ \indent
Before concluding this section, we note that, for superconducting defect-string loops, the persistent current is linked to the existence of a second topological invariant $N \neq n$, given by
\begin{eqnarray} \label{chiral_N}
N = \frac{1}{2\pi}\int_{0}^{2\pi}\frac{\partial \vartheta}{\partial \sigma}\Bigg|_{\theta = {\rm const.}} {\rm d}\sigma,
\end{eqnarray}
where $\sigma \in [0,2\pi)$ again denotes the space-like parameter that parameterizes the string length. 
Thus, $|N|$ gives the number of twists in the phase $\vartheta$ of $\phi$ within the loop and its conservation follows directly from the imposition of periodic boundary conditions to ensure continuity. 
For long strings, $N$ is no longer a topological invariant, but the imposition of periodic boundary conditions over a finite section of string still ensures the existence of a nonzero integer winding number. 
\\ \indent
In the zero-width approximation (\ref{chiral_S}), the role of $N$ is obscured, but, in Sec. \ref{Sect5}, we will explore the relationship between the dynamics of these `internal' field lines and the macroscopic dynamics of the string in detail. 
Using Finsler and generalised Finsler constructions we derive the same macroscopic EOM as the effective action approach, using a microphysical ansatz, and without the need for dimensional reduction or (approximate) effective actions.

\section{Finsler and generalised Finsler geometries} \label{Sect4}
In this section we give a brief review of Finsler geometry, which is used to model circular superconducting string loops in Sec. \ref{Sect5.2}, and the generalised Finsler geometry used to model noncircular loops in Sec. \ref{Sect5.3}. 
For both brevity and clarity, we introduce a minimum of technical detail, concentrating only on the specific properties of Finsler and generalised Finsler spaces that will be utilised in our description of current-carrying defect strings, presented in Sec. \ref{Sect6}. 
The interested reader is referred to the many excellent technical summaries of Finsler geometry, such as \cite{Chern-2,Chern-3,Antonelli}, for further details.  

\subsection{Finsler geometry} \label{Sect4.1}
Finsler geometry is a natural generalisation of Riemannian geometry, which, roughly speaking, is equivalent to ``Riemannian geometry without the quadratic restriction'' \cite{Chern-1}. 
Hence, while the action for a non-relativistic point-particle in Riemannian space is
\footnote{Here, the multiplicative constant with units of mass has been absorbed into the definition of the metric tensor (\ref{ds_Riemann}).}
\begin{eqnarray} \label{S_Riemann}
S = \int {\rm d}s,
\end{eqnarray}
where the line-element ${\rm d}s$ is quadratic in form,
\begin{eqnarray} \label{ds_Riemann}
{\rm d}s^2 = g_{ij}(x) {\rm d}x^i {\rm d}x^j, \quad i,j \in \left\{1,2,3, \dots \right\},
\end{eqnarray}
the corresponding action in Finsler space is given by the general functional,
\begin{eqnarray} \label{S_Finsler}
S = \int F(x,\partial_{\tau}x).
\end{eqnarray}
The metric tensor $g_{ij}(x)$ in (\ref{ds_Riemann}) is a function of the spatial coordinates {\it only} whereas the Finlser space action (\ref{S_Finsler}) depends on both $x$ and $\partial_{\tau}x$, where $\tau$ is an `internal' dynamical variable, and the derivative terms belong to the tangent space at $x$, $\partial_{\tau}x \in T_{x}M$. 
Thus, in Finsler geometry, the more familiar Riemannian space is `non-localised' by allowing the line-element to depend a single vector, which is defined on the tangent bundle \cite{Carvalho:2022sdz,Pfeifer:2019tyy}.  
\\ \indent
However, strictly speaking, standard Finsler geometries are defined as generalised Riemannian spaces and the question of how to extend Finslerian structures to manifolds with pseudo-Riemannian signatures is nontrivial and by no means settled (see \cite{Huang:2007en,Vacaru:2007ng,Chang:2008,Vacaru:2008qs,Tavakol:2009zz,Pfeifer:2013gha,Pfeifer:2019wus} for further discussion). 
Nonetheless, it seems reasonable to generalise the standard space-time line-element, 
\begin{eqnarray} \label{dS_pseuo-Riemann}
{\rm d}s^2 = g_{\mu\nu}(x){\rm d}x^{\mu}{\rm d}x^{\nu}, \quad \mu,\nu \in \left\{0,1,2,3, \dots \right\},
\end{eqnarray}
to 
\begin{eqnarray} \label{dS_Finsler}
{\rm d}\tilde{s}^2 = \tilde{g}_{\mu\nu}(x,\partial_{\tau}x){\rm d}x^{\mu}{\rm d}x^{\nu}.
\end{eqnarray}
This corresponds to a minimally modified Riemannian structure, in which the space-time interval is `almost' quadratic in the coordinate intervals, but is not exactly so due to the presence of an effective {\it Finsler metric}, $\tilde{g}_{\mu\nu}(x,\partial_{\tau}x)$.
In this case, the internal time-like variable $\tau$ may, or may not, be identified with the external space-time coordinate, depending on the choice of embedding for the particle, string, or other extended object (e.g., $D$-brane). 
\\ \indent
As we will show, explicitly, in Sec. \ref{Sect5.2}, it is precisely this freedom that allows us to describe the embedded motion of circular superconducting string loops, in a {\it physical} pseudo-Riemannian space-time, in terms of an effective Finsler metric (\ref{dS_Finsler}). 
However, this is only possible because, in the case of circular symmetry, the space-like world-sheet coordinate, $\sigma$, does not appear explicitly in the embedding ansatz. 
Therefore, in order to treat non-circular loops in a similar way, we must consider a form of `generalised Finsler geometry', with an effective metric that depends on both $\partial_{\tau}x$ and $\partial_{\sigma}x$, as well as $x$. 

\subsection{Generalised Finsler geometry} \label{Sect4.2}
Based on our preliminary discussions in Sec. \ref{Sect4.1}, we introduce the generalised Finsler geometry in which geodesic motion is characterised by the action
\footnote{Here, again, the multiplicative constant with units of mass has been absorbed into the definition of the effective metric, $\tilde{g}_{\mu\nu}$.}
\begin{eqnarray} \label{S_gFinsler}
S = \int {\rm d}\tilde{s} = \int \tilde{g}_{\mu\nu}(x,\partial_{\tau}x,\partial_{\sigma}x){\rm d}x^{\mu}{\rm d}x^{\nu}.
\end{eqnarray}
Here, the time- and space-like internal variables, $\tau$ and $\sigma$, may or may not be identified with space-time background coordinates, according to the chosen embedding. 
In Sec. \ref{Sect5.2}, we will identify them with the string world-sheet coordinates. 
This allows us to describe non-circular superconducting string loops, in a {\it physical} pseudo-Riemannian space-time, in terms of an effective generalised Finsler metric, $\tilde{g}_{\mu\nu}(x,\partial_{\tau}x,\partial_{\sigma}x)$. 
\\ \indent
However, before continuing, we address a possible source of confusion that arises through the inconsistent use of terminology in the mathematical literature. 
What we have called ‘generalised Finsler geometry’, for want of a better term, is sometimes referred to simply as Finsler geometry (see for example \cite{Stavrinos-Ikeda-2006}), though, strictly speaking, it is not. 
We stress that, in Finsler geometry proper, the line-element functional depends on a single vector, which is defined on the tangent bundle. 
By contrast, in further extensions of the canonical Riemannanian structure, this can be generalised to include multiple vectors, which may or may not be defined on the tangent space, as well as spinors or other mathematical objects. 
For the rigorous mathematical treatment of these Finsler-inspired spaces the interested reader is referred to \cite{Stavrinos-Ikeda-2006}, and references therein, for further details.

\section{Topological defect strings with internal currents, revisited} \label{Sect5}
In this Section, we consider the effect of additional charged matter, represented by the current $j^{\mu} \neq 0$, which becomes localized within the Abelian-Higgs string core. 
This is not intended as a realistic model of a superconducting string but, rather, an illustrative one, which captures certain generic features of the string phenomenology. In particular, results obtained by Nielsen \cite{Ni79,NiOl87} suggest that, generically, the $(3+1)$-dimensional dynamics of higher-dimensional wound-strings should be equivalent to the dynamics of superconducting defect strings under dimensional reduction. 
\\ \indent
With this in mind, we aim to go one step further than the dimensional reduction program: we search for ansatzes for the Abelian-Higgs fields which yield EOM for the macroscopic string dynamics that are \emph{identical} to those obtained previously for both long (i.e., formally infinite) wound strings, and wound-string loops \cite{LaYo12,YaLa15}. 
We show that an appropriate ansatz for the scalar and vector fields (and the `external' current $j^{\mu}$), describing long superconducting defect strings, exists in the usual sense. 
That is, ignoring gravitational effects, such strings live in a globally Minkowski space-time and the field ansatz is independent of the background metric. 
However, in order to obtain EOM for circular loops of superconducting defect string, which are the same as those obtained previously for wound-string loops, we must consider a generalization of the geometry to an effective Finsler metric. 
To treat noncircular superconducting loops, an appropriately generalised Finsler geometry is required. 
\\ \indent
Remarkably, by selecting an appropriate ansatz/metric combination, and combining the Euler-Lagrange equations for the Abelian-Higgs fields with the conservation of the energy-momentum tensor, we obtain EOM describing the macroscopic (external) string dynamics which are identical to those for wound-strings obtained in Sec. \ref{Sect3}, {\it without} the need for dimensional reduction. 
Furthermore, we also obtain EOM for the microscopic (internal) field evolution within the string core, which are identical, under the exchange of field theory and embedding coordinate variables, to those which determine the evolution of the $F$-string in the compact extra dimensions of the wound-string model. 
\\ \indent
We argue that the selection of our ansatz / Finsler or generalised Finsler metric combination is valid, in that it reflects an alternative geometric formulation of the Abelian-Higgs theory, in terms of an extended (Finsler or generalised Finsler) space, which is physically \emph{equivalent} to standard formulation in pseudo-Riemannian space. 
Although this is an unconventional approach, we note that it does not require the formulation of an effective action and that, therefore, if the method itself is valid, the resulting string EOM are \emph{exact}.

\subsection{Long defect strings: Riemann geometry} \label{Sect5.1}
Let us first consider a metric with line element 
\begin{eqnarray} \label{Metric2}
{\rm d}s^2 = P^2(\tau){\rm d}\tau^2 - {\rm d}r^2 - r^2{\rm d}\theta^2 - Q^2(\tau){\rm d}\sigma^2 = P^2(t){\rm d}t^2 - {\rm d}r^2 - r^2{\rm d}\theta^2 - Q^2(t){\rm d}\sigma^2,
\end{eqnarray}
where we define the dimensionless world-sheet variables $\tau \in [0,\infty)$ and $\sigma \in [0,2\pi)$ and the second equality follows from the identifications ${\rm d}t = \zeta{\rm d}\tau$, $P(t) = \zeta^{-1} P(\tau)$ and $Q(t) = Q(\tau)$. 
These, in turn, follow from the adoption of the static gauge for the string, $X^{0} = t = \zeta \tau$, and the final spatial coordinate is directly proportional to the space-like world-sheet coordinate $\sigma$ (\ref{Emb/2.3}). 
This metric is more general than the one we need to treat long strings, either with or without currents, but it is useful to derive the more general EOM first and then to impose more restrictive conditions later. 
The corresponding Jacobian is $\sqrt{-g} = rP(t)Q(t)$. 
\\ \indent
Now consider an ansatz of the form
\begin{eqnarray} \label{New_ansatz1}
\phi(r,\theta,z,t) = \eta f(r) e^{in\theta + in\Gamma(z,t)}, \ \ \ A_{\theta} = \frac{n}{e}a_{\theta}(r), \ \ \ A_r = 0,
\nonumber\\
A_z(r,z,t) = \frac{n}{e}a_{\theta}(r)\Gamma'(z,t), \ \ \ A_0(r,z,t) = \frac{n}{e}a(r)\dot{\Gamma}(z,t),
\end{eqnarray}
where we have defined the coordinate $z(\sigma,t) = Q(t)\sigma$, for convenience, and a dash and a dot represent differentiation with respect to $z$ and $t$, respectively. 
(Clearly, this denotes the true Cartesian $z$-axis only when $Q = {\rm const.}$)
The functions $f(r)$ and $a_{\theta}(r)$ obey the boundary conditions in Eqs. (\ref{NO_f_bc})-(\ref{NO_a_{theta}_bc}), and the new radial function $a(r)$ obeys analogous conditions, i.e. 
\begin{equation} \label{a_bc}
a(r) = \left \lbrace
\begin{array}{rl}
0,& \ r=0 \\
1,& \ r \rightarrow \infty.
\end{array}\right.
\end{equation}
Equivalently, we may write the ansatz (\ref{New_ansatz1}) as
\begin{eqnarray} \label{New_ansatz2}
\phi(r,\theta,\sigma,\tau) = \eta f(r) e^{in\theta + in\Gamma(\sigma,\tau)}, \ \ \ A_{\theta} = \frac{n}{e}a_{\theta}(r), \ \ \ A_r = 0,
\nonumber\\
A_{\sigma}(r,\sigma,\tau) = \frac{n}{e}a_{\theta}(r)Q^{-1}(\tau)\partial_{\sigma}\Gamma(\sigma,\tau), \ \ \ A_{\tau}(r,\sigma,\tau) = \frac{n}{e}a(r)P^{-1}(\tau)\partial_{\tau}\Gamma(\sigma,\tau).
\end{eqnarray}
The two notations are completely interchangeable and, in the analysis that follows, we use whichever one is most convenient for our purpose.
\\ \indent
Defining the world-sheet gauge covariant derivatives and electromagnetic field tensor via
\begin{eqnarray} \label{WS_D}
D_{a} = \partial_{a} - ieA_{a}, \ \ \ F_{a\nu} = \partial_{a}A_{\nu} - \partial_{\nu}A_{a},
\end{eqnarray}
where $a \in \left\{\tau,\sigma\right\}$, the scalar EOM (\ref{Cov_EOMs}) may be written as
\begin{eqnarray} \label{Scalar_1}
\frac{1}{r}\partial_{r}(r\partial_{r}\phi) + \frac{1}{r^2}D_{\theta}(D_{\theta}\phi) + \frac{1}{PQ}D_{\sigma}\left(\frac{P}{Q}D_{\sigma}\phi\right) - \frac{1}{PQ}D_{\tau}\left(\frac{Q}{P}D_{\tau}\phi\right) - \frac{\lambda}{2}\phi(|\phi|^2 - \eta^2) = 0,
\end{eqnarray}
and the vector EOM (\ref{Cov_EOMv}) becomes
\begin{eqnarray} \label{Vector_1}
\frac{1}{r}\partial_{r}(rF_{r}{}^{\nu}) + \frac{1}{r^2}\partial_{\theta}(F_{\theta}{}^{\nu}) + \frac{1}{PQ}\partial_{\sigma}\left(\frac{P}{Q}F_{\sigma}{}^{\nu}\right) - \frac{1}{PQ}\partial_{\tau}\left(\frac{Q}{P}F_{\tau}{}^{\nu}\right) + ie(\overline{\phi}D^{\nu}\phi - \phi \overline{D}^{\nu}\phi) - j^{\nu} = 0,
\end{eqnarray}
where $j^{\mu}$ again denotes the current sourced by the externally coupled charge, which is confined to the string core. 

\subsubsection{Long defect strings without internal currents} \label{Sect5.1.1}
We may consider a finite section $\Delta = {\rm const.}$ of a (formally) infinitely long string by setting
\begin{eqnarray} \label{consts1}
P(t) = 1, \ \ \ Q(t) = (2\pi)^{-1}\Delta,
\end{eqnarray}
so that 
\begin{eqnarray} \label{dtdz_1}
{\rm d}t = \zeta {\rm d}\tau, \ \ \ {\rm d}z = (2\pi)^{-1}\Delta {\rm d}\sigma.
\end{eqnarray}
In this case, the scalar EOM separates into real and imaginary parts
\begin{eqnarray} \label{Scalar_2A*}
\frac{{\rm d}^2f}{{\rm d}r^2} + \frac{1}{r}\frac{{\rm d}f}{{\rm d}r} - \frac{n^2f}{r^2}(1-a_{\theta})^2 - n^2f(1-a)^2[\Gamma'^2 - \dot{\Gamma}^2] - \frac{1}{2r_s^2}f(f^2-1) = 0,
\end{eqnarray}
\begin{eqnarray} \label{Scalar_2B*}
inf(1-a)[\Gamma'' - \ddot{\Gamma}] = 0.
\end{eqnarray}
The $\theta-$, $r-$, $z-$ and $t-$components of the vector EOM are:
\begin{eqnarray} \label{Vector_2_theta}
\frac{{\rm d}^2a_{\theta}}{{\rm d}r^2} - \frac{1}{r}\frac{{\rm d}a_{\theta}}{{\rm d}r} + \frac{2f^2}{r_v^2}(1-a_{\theta})+ j_{\theta} = 0
\end{eqnarray}
\begin{eqnarray} \label{Vector_2_r}
\frac{n}{e}\frac{{\rm d}a}{{\rm d}r}[\Gamma'' - \ddot{\Gamma}] + j_r = 0,
\end{eqnarray}
\begin{eqnarray} \label{Vector_2_z}
\frac{{\rm d}^2a}{{\rm d}r^2} + \frac{1}{r}\frac{{\rm d}a}{{\rm d}r} + \frac{2f^2}{r_v^2}(1-a) + \frac{e}{n}\Gamma'^{-1}j_z = 0,
\end{eqnarray}
\begin{eqnarray} \label{Vector_2_t}
\frac{{\rm d}^2a}{{\rm d}r^2} + \frac{1}{r}\frac{{\rm d}a}{{\rm d}r} + \frac{2f^2}{r_v^2}(1-a) + \frac{e}{n}\dot{\Gamma}^{-1}j_0 = 0,
\end{eqnarray}
respectively. The consistency of the ansatz then requires $\Gamma'^2 - \dot{\Gamma}^2 = {\rm const}$.  
However, as we will show explicitly at the end of the present section, the conservation of energy and momentum requires us to set
\begin{eqnarray} \label{Scalar_2C}
\dot{\Gamma}^2 = \Gamma'^2, \ \ \  \dot{\Gamma} = \pm \Gamma',
\end{eqnarray}
This automatically implies
\begin{eqnarray} \label{Scalar_2D}
\Gamma'' - \ddot{\Gamma} = 0,
\end{eqnarray}
but Eq. (\ref{Scalar_2C}) represents the stronger condition and Eqs. (\ref{Scalar_2C})-(\ref{Scalar_2D}) may be compared with the analogous conditions (\ref{phi(t,z)EOM_3.2})-(\ref{phi(t,z)EOM*_3.2}) for the `internal' angular coordinate in the wound-string model. 
\\ \indent
The simplest way to impose (\ref{Scalar_2C}) is to set $\Gamma(z,t) = 0$, which corresponds to the zero current scenario, $j^{\mu} = 0$ \cite{LaYo12}. 
The EOM (\ref{Scalar_2A*})-(\ref{Scalar_2B*}) and (\ref{Vector_2_theta})-(\ref{Vector_2_t}) then reduce to the standard EOM for the Nielsen-Olesen string the energy and tension of the finite string section are,
\begin{eqnarray} \label{}
E = \mu_{|n|} \Delta, \quad \mathcal{T}^{z} = -\mu_{|n|} \Delta,
\end{eqnarray}
and all other effective pressures and shears are zero, as expected. 
Clearly, this is exactly analogous to the corresponding $F$-string case, with the effective warped-geometry string tension $a\mathcal{T}$ replaced by $\mu_{|n|}$. 

\subsubsection{Long defect strings with internal currents} \label{Sect5.1.2}
To construct the long-string solution with internal current, we now note that Eq. (\ref{Scalar_2C}) is satisfied by any function of the form
\begin{eqnarray} \label{Scalar_2D*}
\Gamma(z,t) = \Gamma(k_z z + \omega_z t) \equiv \Gamma(\sigma,t) = \Gamma(m\sigma + \omega_z t), \ \ \ m \in \mathbb{Z}
\end{eqnarray}
where 
\begin{eqnarray} \label{DispRel1}
\omega_z^2 = k_z^2, \ \ \ \omega_z = \pm k_z,
\end{eqnarray}
but that it does {\it not} admit superpositions of left and right movers, $\Gamma(z,t) = \Gamma_L(k_z z + \omega_z t) + \Gamma_R(k_z z - \omega_z t)$, which are solutions to the relativistic wave equation (\ref{Scalar_2D}). 
The constant $k_z$ is naturally interpreted as a wave number and the associated wavelength is defined via
\begin{eqnarray} \label{k_z}
k_z = 2\pi/\lambda_z
\end{eqnarray}
The imposition of periodic boundary conditions over the length $\Delta$ then implies 
\begin{eqnarray} \label{Delta}
\Delta = m\lambda_z.
\end{eqnarray}
The expression for $\Gamma(z,t)$ in Eq. (\ref{Scalar_2D*}) may be compared with (\ref{phi(t,z)_3.2}) and (\ref{DispRel1})-(\ref{Delta}) are formally identical to (\ref{DispRel_3.2*}), (\ref{WavNo_3.2}) and (\ref{WavLen_3.2}), since we have used the same notation to refer to analogous, but physically different, variables. 
The physical interpretation of the parameters $\omega_z$, $k_z$ and $\lambda_z$ in the defect-string model is considered later in this section.
\\ \indent
By analogy with Sec. \ref{Sect3}, we note that different plane wave modes in the Fourier expansion of $\Gamma(z,t)$ correspond to different frequencies $\omega_j$ and wavelengths $\lambda_j$, such that $\omega_j = \pm k_j = \pm 2\pi/\lambda_j = \pm 2\pi m_j/ \Delta$, for some $m_j \in \mathbb{Z}$. 
By writing $\Gamma(z,t)$ in the form (\ref{Scalar_2D}), which implies $\partial \Gamma/\partial t = \omega_z {\rm d}\Gamma/{\rm d}u$,  $\partial \Gamma/\partial z = k_z {\rm d}\Gamma/{\rm d}u$,  $\partial \Gamma/\partial \sigma = m {\rm d}\Gamma/{\rm d}u$, where $u = k_z z + \omega_z t = m\sigma + \omega_z t$, we select a single mode as being characteristic of the wave form. For functions with only nonlinear terms in $u$, the natural choice is the mode with the highest amplitude, which then gives the approximate wavelength of any fluctuation in current density. 
If $\Gamma$ contains linear terms in $z$ and $t$, it is most natural to use the associated $k_z$ and $\omega_z$ to give the characteristic wavelength and frequency. Linear terms in $\Gamma(z,t)$ correspond to a uniform current and any additional nonlinear terms describe local fluctuations in current density around the mean value. 
\\ \indent
The relationship between the model parameters, in particular $\lambda_z$ and $m$, and the internal structure of the string (that is, the configuration of the scalar and vector fields) will be discussed in detail later in this section. 
We argue that the function $\Gamma(z,t)$ determines the space-time embedding of lines of constant phase in the scalar field $\phi$ and that these are analogous to the space-time embedding of the angular coordinate for wound strings in higher-dimensional theories \cite{LaYo12,LaHa17}. 
We aim to show that equivalent phase-line/angular coordinate embeddings give rise to identical macroscopic EOM for different string species. 
{\it This enables us to specify precisely the conditions under which topological defect strings, originating in field theory, and fundamental Nambu-Goto strings, originating in string theory, are phenomenologically equivalent from an observational point of view.} 
An advantage of this method is that it does not require us to make any approximations. 
In particular, we need not resort to an effective action to determine the EOM for the superconducting defect string.
\\ \indent
If we now set
\begin{eqnarray} \label{j_1}
j_{\theta} = 0, \ \ \  j_{r} = 0,  
\end{eqnarray}
\begin{eqnarray} \label{j_2}
j_{z}(r,z,t) = \frac{n}{e}\Gamma'(t,z) j(r), \ \ \ j_{0}(r,z,t) = \frac{n}{e}\dot{\Gamma}(t,z) j(r),
\end{eqnarray}
the EOM for the current-carrying string reduce to Eqs.  (\ref{NO_specific_scalar_EOM})-(\ref{NO_specific_vector_EOM_theta}), as in the Nielsen-Olesen case, plus a new EOM in $a(r)$,
\begin{eqnarray} \label{EOM_in_a_1}
\frac{{\rm d}^2a}{{\rm d}r^2} + \frac{1}{r}\frac{{\rm d}a}{{\rm d}r} + \frac{2f^2}{r_v^2}(1-a) + j = 0.
\end{eqnarray}
\indent
It may be shown that a small distribution of `external' charge, that is, charge not generated specifically by the configuration of the Abelian-Higgs fields, located near the centre of the string,
\begin{equation} \label{j(r)}
j(r) = \left \lbrace
\begin{array}{rl}
k,& \ 0 \leq r \leq \delta \\
0,& \ r > \delta.
\end{array}\right.
\end{equation}
where $k  < 0$ is a constant and $\delta \ll r_{s} \leq r_{v}$, is sufficient to prevent divergence of the electric and magnetic field densities (i.e. the divergence of $a(r)$) as $r \rightarrow 0$ \cite{LaYo12}. 
Alternatively, a phenomenologically equivalent model may be obtained by taking the pure Abelian-Higgs action, together with the ansatz Eq. (\ref{New_ansatz1}), and imposing a cut-off $r_{\rm min} = \delta \ll r_{s} \leq r_{v}$, assumed to be due to quantum gravity effects at small distances \cite{Svetovoy:1997dk}. 
This allows the pure Abelian-Higgs string to support a zero-mode, so that no additional `external' charge is required. 
\\ \indent
In the additional current model, $f(r)$ and $a_{\theta}(r)$ are given by Eqs. (\ref{NO_f})-(\ref{NO_alpha}) and the approximate solution for $a(r)$ is \cite{LaYo12}
\begin{equation} \label{a_soln*}
a(r) \simeq \left \lbrace
\begin{array}{rl}
-kr^{2},& \ r \lesssim \delta \\
1 - CK_0((r/r_{c|n|})^{1+|n|}),& \ \delta \lesssim r \lesssim r_{c|n|}, \\
1,& \  r \gtrsim r_{c|n|},
\end{array}\right.
\end{equation}
where $C$ is a constant of integration, which must be fixed via the matching condition at $r = \delta$, and $K_0$ denotes the zeroth order modified Bessel function of the second kind. 
(Here, again, we assume critical coupling $r_{s|n|} = r_{v|n|} = r_{c|n|}$, for simplicity.)
Typically, $j(r)$ gives contributions to the relevant components of the energy-momentum tensor of order $\pm r_{c|n|}^2aj\Gamma'^2$. 
Integrating over $r{\rm d}r$ then yields $\pm r_{c|n|}^2k^2\delta^4\Gamma'^2$. For $|k| \lesssim \delta^{-2}$, which is automatically implied by Eq. (\ref{a_soln*}), these are subdominant to all other contributions. 
In this case, $J^{0} \simeq \mathcal{J}^{0} \equiv J^{z} \simeq  \mathcal{J}^{z}$ since $j^{0}$, $j^{z}$ give subdominant contributions to constants of motion and effective pressures and shears inside the string, as well as to the total current $J^{\mu} + \mathcal{J}^{\mu} + j^{\mu}$. 
\\ \indent
We now define the parameters $\omega^2(z,t)$ and $\Omega^2(t)$ via
\begin{eqnarray} \label{omega^2_1}
\omega^{-2}(z,t) = \frac{\Delta^2 + (2\pi)^2r_{c|n|}^2(\partial_{\sigma}\Gamma)^2}{\Delta^2}, \ \ \  \Omega^{-2}(t) = \frac{1}{2\pi}\int_{0}^{2\pi} \omega^{-2}(z,t) {\rm d}\sigma,
\end{eqnarray}
by analogy with (\ref{omega_3.2}). 
For the non-superconducting Nielsen-Olesen string, both the lines of constant phase in $\phi$ and the magnetic field lines are parallel to the central axis of the string whereas, in the superconducting case, both become twisted. 
In the constant current case, the magnetic field lines form static helices and the phase lines are helices that rotate around the central axis at a constant angular velocity. 
Thus, nonlinear terms in $\Gamma(z,t)$ induce local fluctuations in angular velocity and, hence, local compressions and rarefractions in the both the phase and magnetic field lines. 
The integer $m$ which characterises the linear term in $\sigma$ in the function $\Gamma$, if present, may then be interpreted as the number of windings or `twists' in the field lines over the length $\Delta$. Physically, the parameter $\omega^2(z,t)$ represents the local fraction of the total length of a line of constant phase that lies parallel to the string axis. 
It is therefore analogous to the parameter, defined for wound strings in higher dimensions \cite{LaYo12,LaWa10,Thesis,Lake:2009nq,Avgoustidis:2005vm}, which gives the local fraction of the total string length in the non-compact directions. 
The parameter $\Omega^2(t)$ represents the spatially-averaged (and, in principle, still time-dependent) value of $\omega^2(z,t)$.
\\ \indent
In order to make the analogy with the wound string case more explicit, it is useful to define the physical observables $U^{\mu}{}_{\nu}$ as integrals (over the space-like world-sheet parameter ${\rm d}\sigma$) of components of an effective energy-momentum tensor for the string, $\mathcal{T}^{\mu}{}_{\nu}\sqrt{-g}$. 
This must therefore be defined in terms of the fundamental energy-momentum tensor for the fields $T^{\mu}{}_{\nu}\sqrt{-g}$ and the relevant Killing vectors, so that
\begin{eqnarray} \label{U-T-mathcal{T}_1}
U^{\mu}{}_{\nu} = \int \mathcal{T}^{\mu}{}_{\nu}\sqrt{-g}{\rm d}\sigma = \int k_{(\nu)\alpha} T^{\mu\alpha}\sqrt{-g} {\rm d}r {\rm d}\theta {\rm d}z,
\end{eqnarray}
where $k_{(\nu)\alpha}$ denotes a set of vectors labelled by the index $\nu$, not a two-index tensor. 
We then have, for example, $E_{|n|} \equiv U^{0}{}_{0} = \int \mathcal{T}^{0}{}_{0}{\rm d}\sigma= \int k_{(0)\alpha}T^{0\alpha}\sqrt{-g} {\rm d}r {\rm d}\theta {\rm d}z$, etc. 
In this model, the relevant Killing vectors are
\begin{eqnarray} \label{Killing_1}
&k_{(0)\alpha} = [1,0,0,0] = [\eta_{00},0,0,0] = \eta_{(0)\alpha},
\nonumber\\
&k_{(\theta)\alpha} = [0,0,-r^2,0] = [0,0,\eta_{\theta\theta},0] = \eta_{(\theta)\alpha},
\nonumber\\
&k_{(z)\alpha} = [0,0,0,-1] = [0,0,0,\eta_{zz}] = \eta_{(z)\alpha},
\end{eqnarray}
which may be written in a compact form as
\begin{eqnarray} \label{Killing_2}
k_{(\nu)\alpha} = \eta_{(\nu)\alpha},
\end{eqnarray}
where $\eta_{\mu\nu}$ is the Minkowski metric, here expressed in the coordinates $\left\{t,r,\theta,z\right\}$.

In terms of $\omega^2(z,t)$, the nonzero components of the string energy-momentum tensor may be written in a compact form as: 
\begin{subequations} \label{+++}
\begin{align}
\mathcal{T}^{0}{}_{0}\sqrt{-g} &= \eta^2|n| \Delta \omega^{-2}, \ \ \ \ \ \ \ \ \ \ \ \ \ \mathcal{T}^{z}{}_{0}\sqrt{-g} = \mp \eta^2|n| \Delta \left(\frac{1-\omega^{2}}{\omega^{2}}\right), \label{EMT_3.2a}\\
\mathcal{T}^{0}{}_{z}\sqrt{-g} &= \pm \eta^2|n| \Delta \left(\frac{1-\omega^{2}}{\omega^{2}}\right), \ \ \mathcal{T}^{z}{}_{z}\sqrt{-g} = \eta^2|n| \Delta \left(\frac{2\omega^{2}-1}{\omega^{2}}\right), \label{EMT_3.2b}\\
\mathcal{T}^{0}{}_{\theta}\sqrt{-g} &= \pm \eta^2|n| \Delta r_{c|n|} \frac{\sqrt{1-\omega^{2}}}{\omega}, \ \ \ \mathcal{T}^{z}{}_{\theta}\sqrt{-g} = \mp  \eta^2|n| \Delta r_{c|n|} \frac{\sqrt{1-\omega^{2}}}{\omega}. \label{EMT_3.2c}
\end{align}
\end{subequations} 
Clearly, these expressions should be compared with Eqs. (\ref{EMT_3.2a*})-(\ref{EMT_3.2c*}) in the wound-string case. 
Strictly, they represent the order of magnitude values of the components of the energy-momentum tensor but the approximate equalities simply reflect the fact that we have used the approximate solutions, Eqs. (\ref{NO_f})-(\ref{NO_alpha}) and (\ref{a_soln*}), when performing the integrals over ${\rm d}r$, together with the condition $|k| \lesssim \delta^{-2}$. 
The expressions may be made {\it exact} by determining the appropriate numerical solutions for scalar and vector EOM functions, $f(r)$, $a_{\theta}(r)$ and $a(r)$, but this does not affect the macroscopic dynamics of the string since each component $\mathcal{T}^{I}{}_{J}\sqrt{-g}$ is scaled by the same multiplicative factor of order unity.
\\ \indent
Interestingly, the EOM (\ref{Scalar_2C}) is precisely the condition required to ensure that the Lagrangian $\mathcal{L}$ is \emph{independent} of $\dot{\Gamma}$ and $\Gamma'$. 
In fact, imposing (\ref{Scalar_2C}) ensures that $\mathcal{L}$ reduces \emph{exactly} to the Lagrangian of the Nielsen-Olesen string, at the level of the energy-momentum tensor. 
This is analogous to the wound string case considered in Sec. \ref{Sect2}, where the EOM governing the motion of the windings was found to be precisely the condition required to ensure that the string Lagrangian is independent of the higher-dimensional embedding coordinate.
\\ \indent
The conservation equation for the fundamental fields
\begin{eqnarray} \label{ConsLaw1}
\nabla_{\mu}T^{\mu}{}_{\nu}\sqrt{-g} = \frac{\partial}{\partial x^{\mu}}\left(T^{\mu}{}_{\nu}\sqrt{-g}\right)  - \frac{1}{2} \frac{\partial g_{\alpha\beta}(x)}{\partial x^{\nu}}T^{\alpha\beta}\sqrt{-g} = 0,
\end{eqnarray}
implies the conservation law for the string
\begin{eqnarray} \label{ConsLaw2}
\nabla_{\mu}\mathcal{T}^{\mu}{}_{\nu}\sqrt{-g} = \frac{\partial}{\partial X^{\mu}}\left(\mathcal{T}^{\mu}{}_{\nu}\sqrt{-g}\right)  - \frac{1}{2} \frac{\partial g_{\alpha\beta}(X)}{\partial X^{\nu}}\mathcal{T}^{\alpha\beta}\sqrt{-g} = 0,
\end{eqnarray}
where $T^{\mu}{}_{\nu}\sqrt{-g}$ and $\mathcal{T}^{\mu}{}_{\nu}\sqrt{-g}$ are related via Eq. (\ref{U-T-mathcal{T}_1}) and the $x^{\nu}$ denote the space-time background coordinates whereas the $X^{\nu}$ denote embedding coordinates. 
For long strings, this gives 
\begin{subequations} 
\begin{align}
\partial_{0}(\mathcal{T}^{0}{}_{0}\sqrt{-g}) + \partial_{z}(\mathcal{T}^{z}{}_{0}\sqrt{-g}) &= 0, \label{EOM_3.2a}\\
\partial_{0}(\mathcal{T}^{0}{}_{z}\sqrt{-g}) + \partial_{z}(\mathcal{T}^{z}{}_{z}\sqrt{-g}) &=  0, \label{EOM_3.2b}\\
\partial_{0}(\mathcal{T}^{0}{}_{\theta}\sqrt{-g}) + \partial_{z}(\mathcal{T}^{z}{}_{\theta}\sqrt{-g}) &= 0.\label{EOM_3.2c}
\end{align}
\end{subequations}
each of which, it may be verified directly, is equivalent to Eq. (\ref{Scalar_2C}). 
(See Eqs. (\ref{EOM_3.2a})-(\ref{EOM_3.2c}), for comparison, in the wound-string case.) 
Equation (\ref{EOM_3.2c}) explicitly ensures the conservation of electric charge, since
\begin{eqnarray} \label{J_omega}
J^{0}\sqrt{-g} = \frac{e}{(2\pi)^2 \Delta r_{c|n|}^2  \eta^2}\mathcal{T}^{0}{}_{\theta}\sqrt{-g} = -J^{z}\sqrt{-g} = \frac{e}{(2\pi)^2 \Delta r_{c|n|}^2  \eta^2}\mathcal{T}^{z}{}_{\theta}\sqrt{-g}.
\end{eqnarray}
Finally, for convenience, we split $\Gamma(\sigma,t)$ into linear and nonlinear parts
\begin{eqnarray} \label{}
\Gamma(\sigma,t) = m\sigma + \omega_z t + \Gamma_{NL}(m\sigma + \omega_z t )
\end{eqnarray}
and introduce the parameter $\alpha^2$, which quantifies the nonlinearity of the windings in the field lines, via
\begin{eqnarray} \label{alpha_3.2}
\alpha^{-2} = \frac{1}{2\pi m} \int_{0}^{2\pi}(\partial_{\sigma}\Gamma)^2 {\rm d}\sigma = \frac{1}{2\pi m} \int_{0}^{2\pi}\left(1 + \frac{\partial_{\sigma}\Gamma_{NL}}{m}\right)^2 {\rm d}\sigma,
\end{eqnarray}
by analogy with (\ref{alpha_3.2*}). 
The number of complete windings over the length $\Delta$ is given by
\begin{eqnarray} \label{}
n_z = \frac{1}{2\pi}\int_{0}^{2\pi}\partial_{\sigma}\Gamma {\rm d}\sigma,
\end{eqnarray}
so that the nonlinear terms in $\Gamma$ make no contribution and we see that 
\begin{eqnarray} \label{}
n_z = m,
\end{eqnarray}
as in the wound-string case. However, here, $n_z$ refers to the number of complete twists in the lines on constant phase, over the interval $z \in [0,\Delta]$ at $\theta = {\rm const.}$, as opposed to the number of windings in the compact internal space.
The effective local and spatially averaged values of the model parameters, $(Q^{\rm eff})^2(\sigma,t)$ and $\langle Q^2 \rangle(t)$, are related as in the wound-string case, by Eq. (\ref{Q_3.2}),
and it is possible to show that 
\begin{eqnarray} \label{<l_z>_3.2*}
\langle \lambda_z^2 \rangle = \alpha^2\lambda_z^2, \ \ \ \langle n_z^2 \rangle = n_z^2/\alpha^2,
\end{eqnarray}
\begin{eqnarray} \label{<omega_z>_3.2*}
\langle \omega_z^2 \rangle = \langle k_z^2 \rangle, \ \ \ \langle k_z^2 \rangle = \frac{(2\pi)^2}{\langle \lambda_z^2 \rangle}. 
\end{eqnarray}
The constants of motion and integrated pressures and shears inside the string may then be written in a compact form as:
\begin{eqnarray} \label{E4_3.2*}
E_{|n|} = 2\pi\eta^2|n| \Delta\left(1 + \frac{(2\pi)^2R^2}{\langle \lambda_z^2 \rangle}\right) = 2\pi\eta^2|n| \Delta \Omega^{-2},
\end{eqnarray}
\begin{eqnarray} \label{P^z4_3.2*}
P^{z}  = \pm 2\pi\eta^2|n| \langle n_z \rangle \times \frac{(2\pi)^2 r_{c|n|}^2}{\langle \lambda_z \rangle} = \pm 2\pi\eta^2|n| \Delta \left(\frac{1-\Omega^{2}}{\Omega^{2}}\right),
\end{eqnarray}
\begin{eqnarray} \label{Lam^phi4_3.2*}
\Lambda^{\theta} = \pm (2\pi)^2 \eta^2|n| r_{c|n|} n_z = \pm 2\pi\eta^2|n| \Delta \frac{ \sqrt{1-\Omega^{2}}}{\Omega}\alpha
\end{eqnarray}
\begin{eqnarray} \label{Pi4_3.2*}
\Pi_{|n|} = 2\pi\eta^2|n| \Delta\sqrt{1 + \frac{(2\pi)^2 r_{c|n|}^2}{\lambda_z^2}(2\alpha^{-2} - 1)} =  2\pi\eta^2|n| \Delta \Omega^{-1}\sqrt{2-\alpha^2 - (1-\alpha^2)\Omega^2},
\end{eqnarray}
\begin{eqnarray} \label{T^z_3.2*}
\mathcal{T}^{z} = -2\pi\eta^2|n|\Delta\left(1 - \frac{(2\pi)^2r_{c|n|}^2}{\langle \lambda_z^2 \rangle}\right) = -2\pi\eta^2|n|\Delta \left(\frac{2\Omega^{2}-1}{\Omega^{2}}\right),
\end{eqnarray}
which should be compared with (\ref{E4_3.2})-(\ref{T^z_3.2}), using (\ref{alpha_3.2*}) where necessary. 
\\ \indent
Here, $E_{|n|}$ denotes the total energy, $P^{z}$ the linear momentum of the current, $\Lambda^{\theta}$ the total angular momentum in the $\theta-$direction, and $\Pi_{|n|}$ the total $4-$momentum of the string in the length $\Delta$, while $\mathcal{T}^{z} = U^{z}{}_{z}$ is the integrated pressure in the $z-$direction. 
The integrated pressure in the $\theta-$direction is $\mathcal{T}^{\theta} = U^{\theta}{}_{\theta} = 0$, as expected for a string with a stable fixed radius. Note that, technically, $\Lambda^{\theta}$, as given in (\ref{Lam^phi4_3.2}), is defined as the geometric mean  $\Lambda^{\theta} = \pm\sqrt{-U^{0}{}_{\theta}U^{0\theta}}$, for convenience. 
However,  all three quantities $\Lambda^{\theta}$, $U^{0}{}_{\theta}$ and $U^{0\theta}$ are conserved, differing only by factors of $r_{c|n|}$.
\\ \indent
Analogous expressions also hold for the energy and momentum densities and local pressures and shears. 
These may be obtained by replacing $\Omega^2(t)$ with $\omega^2(z,t)$ or by replacing the quantities $\langle Q^2 \rangle$ with $(Q^{\rm eff})^2$ and setting $\alpha^2=1$ in Eqs. (\ref{E4_3.2})-(\ref{T^z_3.2}). 
In the simple linear case, corresponding to constant current, we have $\langle Q^2 \rangle = (Q^{\rm eff})^2 = Q^2$. 
In particular, we note the existence of a critical tensionless solution for 
\begin{eqnarray} \label{Tensionless1}
\omega^2(t,z) = \Omega^2(t) = 1/2, \ \forall z,t
\end{eqnarray}
or, equivalently
\begin{eqnarray} \label{Tensionless2}
\lambda_{z} = \sqrt{\langle \lambda_{z}^2\rangle} = 2\pi r_{c|n|},
\end{eqnarray}
whereas zero \emph{net} tension requires only the weaker condition
\begin{eqnarray} \label{Tensionless2}
\sqrt{\langle \lambda_{z}^2\rangle} = 2\pi r_{c|n|} \leftrightarrow  \Omega^2(t) = 1/2, \forall t.
\end{eqnarray} 
The parallels with the wound-string case are obvious. 
For $\omega^2(t,z) > 1/2$ the local `tension' of the string becomes a repulsive pressure. 
However, though such a solution exists in the classical model, quantum mechanical considerations suggest it can never be realised physically, since superconducting strings with currents exceeding the threshold value
\begin{eqnarray} \label{J_threshold}
J_{\rm max} \simeq \frac{e}{2\pi r_{c|n|},},
\end{eqnarray}
where $e$ is the coupling (i.e. charge) and $r_{c|n|}$ is the string radius, become unstable and decay \cite{ViSh00}. 
In our formulation, this threshold is equivalent to a minimum wavelength for the phase- and magnetic field-line `twists', 
\begin{eqnarray} \label{lambda_threshold}
\lambda_{z}^{\rm min} = 2\pi r_{c|n|},
\end{eqnarray}
which is precisely the wavelength required to satisfy the tensionless condition (\ref{Tensionless1}).
\\ \indent
The fundamental equations of motion and the expressions for the physical properties of the (long) current-carrying Abelian-Higgs string are completely analogous to those obtained for wound $F$-strings in 
Sec. \ref{Sect2}, under the identifications 
\begin{eqnarray} \label{phi-Gamma}
\Gamma(z,t) \leftrightarrow \varphi(z,t), \ \ \ r_{c|n|} \leftrightarrow R,
\end{eqnarray}
where $\varphi_(z,t)$ represents the angular embedding coordinate in the higher-dimensional space and $R$ is compactification radius (and, where necessary, the identifications $\Delta \leftrightarrow a\Delta$, $\lambda_z \leftrightarrow a\lambda_z$ etc, where $a$ denotes the `warp factor' of the wound-string background geometry). 
In some sense, this is unsurprising, since the motion of higher-dimensional windings is formally equivalent to an effective world-sheet current in $(3+1)$ dimensions under dimensional reduction \cite{Ni79,YaLa15,NiOl87}, as stated previously. 
Nonetheless, it is common in the literature to construct effective actions for either current-carrying defect strings \cite{CoHiTu87,Spergel:1986uu,Hindmarsh:1987vh,Copeland:1987yv} or dimensionally reduced strings \cite{NiOl87} in order to determine (and solve) their approximate equations of motion. 
The approach outlined here demonstrates an {\it exact} equivalence between the two (albeit in a specific case, that of long straight strings) without the need for an effective action. 
In the following sections, we extend this result for circular, then arbitrary planar loop configurations, though a complete generalization to arbitrary configurations in $(3+1)$ dimensions is left to future work. 
A similar approach may, in principle, be extended to any other species of defect string in which the flow of current is associated with `twists' in an order parameter, for example chiral strings or vorton models \cite{LaYo12,LaWa10,Martins&Shellard(1998),Carter&Davis(2000),BlOlVi01,Test1,Test2,Carter(1990),VaVa:90,Carter&Martin(1993),Larsen(1993),Vortons(2013)}.  

\subsection{Circular defect-string loops: Finsler geometry} \label{Sect5.2}
To treat circular superconducting loops, we again consider the line-element (\ref{Metric2}) but, instead of imposing the condition (\ref{consts1}), we instead allow $P(t)$ and $Q(t)$ to be genuine functions of time. 

\subsubsection{Circular defect-string loops without currents} \label{Sect5.2.1}
As in the case of long strings, it is in many ways simpler to obtain the zero-current solution as a particular limit of the more general case, in which a nonzero superconducting current exists. 
This avoids the needless repetition of similar calculations. 
We therefore refer the reader to the following subsection, in which the EOM for the non-superconducting circular string loop is recovered by substituting $\Gamma(\sigma,t) = 0$ ($j^{\mu} = 0$) into the more general formulae that are valid for  $\Gamma(\sigma,t) \neq 0$ ($j^{\mu} \neq 0$).

\subsubsection{Circular defect-string loops with currents} \label{Sect5.2.2}

We may consider a circular loop by interpreting $\sigma \in [0,2\pi)$ as an angular coordinate and $Q(t)$ as a (time-dependent) radial factor in (\ref{Metric2}),
so that the background metric contains a genuinely time-dependent $g_{00}$ component. 
The real and imaginary parts of the scalar EOM then become
\begin{eqnarray} \label{Scalar_2A}
\frac{{\rm d}^2f}{{\rm d}r^2} + \frac{1}{r}\frac{{\rm d}f}{{\rm d}r} + \frac{n^2f}{r^2}(1-a_{\theta})^2 + n^2f(1-a)^2\left[\frac{1}{Q^2}(\partial_{\sigma}\Gamma)^2 - \frac{1}{P^2}\dot{\Gamma}^2\right] + \frac{1}{2r_s^2}f(f^2-1) = 0,
\end{eqnarray}
\begin{eqnarray} \label{Scalar_2B}
inf(1-a)\left[\frac{1}{Q^2}\partial_{\sigma}^2\Gamma + \frac{1}{P^2}\ddot{\Gamma} + \frac{1}{PQ}\frac{{\rm d}}{{\rm d}t}\left(\frac{Q}{P}\right)\dot{\Gamma}\right] = 0.
\end{eqnarray}
Adopting the conditions in Eq. (\ref{j_1}), the $r-$ and $\sigma-$components of the vector EOM are:
\begin{eqnarray} \label{Vector_3_r}
\frac{n}{e}\frac{{\rm d}a}{{\rm d}r}\left[\frac{1}{Q^2}\partial_{\sigma}^2\Gamma + \frac{1}{P^2}\ddot{\Gamma} + \frac{1}{PQ}\frac{{\rm d}}{{\rm d}t}\left(\frac{Q}{P}\right)\dot{\Gamma}\right] = 0,
\end{eqnarray}
\begin{eqnarray} \label{Vector_3_z}
\frac{{\rm d}^2a}{{\rm d}r^2} + \frac{1}{r}\frac{{\rm d}a}{{\rm d}r} + \frac{2f^2}{r_v^2}(1-a) + \frac{e}{n}(\partial_{\sigma}\Gamma)^{-1}j_{\sigma} = 0.
\end{eqnarray}
The $\theta-$component is again as in the Nielsen-Olesen case and the $t-$component is identical to Eq. (\ref{Vector_2_t}). 
We then define 
\begin{eqnarray} \label{j_3}
j_{\sigma}(r,\sigma,t) = \frac{n}{e}\partial_{\sigma}\Gamma(t,\sigma) j(r), \ \ \ j_{0}(r,\sigma,t) = \frac{n}{e}\dot{\Gamma}(t,\sigma) j(r),
\end{eqnarray}
and impose the condition
\begin{eqnarray}  \label{Scalar_3A}
\dot{\Gamma}^2 = \frac{P^2}{Q^2}(\partial_{\sigma}\Gamma)^2, \ \ \ \dot{\Gamma} = \pm \frac{P}{Q}\partial_{\sigma}\Gamma,
\end{eqnarray}
which automatically implies
\begin{eqnarray}  \label{Scalar_3B}
\partial_{\sigma}\Gamma^2 + \frac{Q^2}{P^2}\ddot{\Gamma} + \frac{Q}{P}\frac{{\rm d}}{{\rm d}t}\left(\frac{Q}{P}\right)\dot{\Gamma} = 0.
\end{eqnarray}
This may be shown explicitly by differentiating Eq. (\ref{Scalar_3A}) with respect to both $\sigma$ and $t$, then eliminating $\partial_{\sigma}\dot{\Gamma}$. 
Equations (\ref{Scalar_3A})-(\ref{Scalar_3B}) represent generalisations of Eqs. (\ref{Scalar_2C})-(\ref{Scalar_2D}) and, as we will soon show, can be cast in ana analogous form to (\ref{EOM1_4.2}) for an appropriate choice of $P(t)$ and $Q(t)$.
\\ \indent
We again note that (\ref{Scalar_3A}) is precisely the condition required to ensure that the Lagrangian $\mathcal{L}$ reduces to that of the Nielsen-Olesen string at the level of the energy-momentum tensor, being independent of $\dot{\Gamma}$, $\partial_{\sigma}\Gamma$ and $P^2$, $Q^2$. 
Thus, subject to Eq. (\ref{Scalar_3A}), the remaining EOM in the radial functions $f(r)$, $a_{\theta}(r)$ and $a(r)$ reduce to those obtained in Sec. \ref{Sect3.1}. 
Adopting the same ansatz for $j(r)$, Eq. (\ref{j(r)}), the solutions for all three functions, plus $j(r)$ itself, are those obtained previously.
\\ \indent
The parameters $\omega^2(\sigma,t)$ and $\Omega^2(t)$ are now defined as
\begin{eqnarray} \label{omega^2_2}
\omega^{-2}(\sigma,t) = \frac{Q^2 + r_{c|n|}^2(\partial_{\sigma}\Gamma)^2}{Q^2}, \ \ \ \Omega^{-2}(t) = \frac{1}{2\pi}\int_{0}^{2\pi}\omega^{-2}(\sigma,t){\rm d}\sigma,
\end{eqnarray}
and the physical properties and effective energy-momentum tensor for the string may be defined using only a slight modification of Eq. (\ref{U-T-mathcal{T}_1}):
\begin{eqnarray} \label{U-T-mathcal{T}_2}
U^{\mu}{}_{\nu} = \int \mathcal{T}^{\mu}{}_{\nu}\sqrt{-g}{\rm d}\sigma = \int \kappa_{(\nu)\alpha}T^{\mu\alpha}\sqrt{-g} {\rm d}r {\rm d}\theta {\rm d}\sigma.
\end{eqnarray}
Here we have replaced the integral over ${\rm d}z$ with an integral over ${\rm d}\sigma$ and the set of vectors $\kappa_{(\nu)\alpha}$ are defined by analogy with Eqs. (\ref{Killing_1})-(\ref{Killing_2}), but with an important difference, related to the $g_{\sigma\sigma}$ component of the new metric. 
Thus, we set
\begin{eqnarray} \label{kappa_1}
&\kappa_{(0)\alpha} = [1,0,0,0] = [\eta_{00},0,0,0] = \eta_{(0)\alpha},
\nonumber\\
&\kappa_{(\theta)\alpha} = [0,0,-r^2,0] = [0,0,\eta_{\theta\theta},0] = \eta_{(\theta)\alpha},
\nonumber\\
&\kappa_{(\sigma)\alpha} = [0,0,0,-Q^2] = [0,0,0,g_{\sigma\sigma}] = g_{(\sigma)\alpha},
\end{eqnarray}
or, equivalently
\begin{eqnarray} \label{kappa_2}
\kappa_{(\nu)\alpha} = \eta_{(\nu)\alpha} +\delta^{(\sigma)}{}_{(\nu)}(g_{(\sigma)\alpha} - \eta_{(\sigma)\alpha}).
\end{eqnarray}
We note that, for $Q^2(t)=1$, Eqs. (\ref{kappa_1}) and (\ref{kappa_2}) reduce to Eqs. (\ref{Killing_1}) and (\ref{Killing_2}), respectively. 
Though the definition of $\kappa_{(\sigma)\alpha}$, especially, in Eqs. (\ref{kappa_1})-(\ref{kappa_2}) may appear arbitrary, we will show that it is in fact \emph{necessary}, in order to yield a self-consistent set of conservation equations.
\\ \indent
However, as we shall see, the consistency of the Euler-Lagrange equations and the conservation of the energy-momentum tensor also require us to impose a condition relating the metric component $g_{00} = P^2(t)$ to the `$Q$-coordinate' velocity, namely $P^2 = 1- \dot{Q}^2$. 
Imposing this condition, \emph{and} interpreting $Q(t)$ as a genuine space-time coordinate, the metric becomes Finsler and, strictly speaking, we cannot call the set $\kappa_{(\nu)\alpha}$ `Killing vectors', since the \emph{original} concept of a Killing vector, defined by Killing's equation (see, for example \cite{Nakahara:1990th}), applies only to Riemannian and pseudo-Riemannian geometries. 
However, the concept of a Killing-like vector field, which characterizes an isometry, has been extended to Finsler spaces in \cite{NumberWang!,MisraMisra}. 
\\ \indent
Using the definitions in Eqs. (\ref{kappa_1})-(\ref{kappa_2}), the components of $\mathcal{T}^{\mu}{}_{\nu}\sqrt{-g}$ that are analogous to those in Eqs. (\ref{EMT_3.2a})-(\ref{EMT_3.2c}) are
\begin{subequations}
\begin{align}
\mathcal{T}^{0}{}_{0}\sqrt{-g} &= 2\pi\eta^2|n| \frac{Q}{P}\omega^{-2}, \ \ \ \ \ \ \ \ \ \ \ \ \ \ \  \mathcal{T}^{\sigma}{}_{0}\sqrt{-g} = \pm 2\pi\eta^2|n| \left(\frac{1-\omega^{2}}{\omega^{2}}\right), \label{EMT_4.2a}\\
\mathcal{T}^{0}{}_{\sigma}\sqrt{-g} &= \pm 2\pi\eta^2|n| Q^2 \left(\frac{1-\omega^{2}}{\omega^{2}}\right) , \ \ \ \mathcal{T}^{\sigma}{}_{\sigma}\sqrt{-g} = 2\pi\eta^2|n| PQ \left(\frac{2\omega^{2}-1}{\omega^{2}}\right), \label{EMT_4.2b}\\
\mathcal{T}^{0}{}_{\theta}\sqrt{-g} &= \pm 2\pi\eta^2|n| Q r_{c|n|} \frac{\sqrt{1-\omega^{2}}}{\omega}, \ \ \mathcal{T}^{\sigma}{}_{\theta}\sqrt{-g} = \mp  2\pi\eta^2|n| P r_{c|n|} \frac{\sqrt{1-\omega^{2}}}{\omega}, \label{EMT_4.2d}
\end{align}
\end{subequations}
and the relevant conservation equations are
\begin{subequations} \label{++}
\begin{align}
\partial_{0}(\mathcal{T}^{0}{}_{0}\sqrt{-g}) + \partial_{\sigma}(\mathcal{T}^{\sigma}{}_{0}\sqrt{-g}) &= 0, \label{EOM_4.2a}\\
\partial_{0}(\mathcal{T}^{0}{}_{\sigma}\sqrt{-g}) + \partial_{\sigma}(\mathcal{T}^{\sigma}{}_{\sigma}\sqrt{-g}) &=  0, \label{EOM_4.2b}\\
\partial_{0}(\mathcal{T}^{0}{}_{\theta}\sqrt{-g}) + \partial_{\sigma}(\mathcal{T}^{\sigma}{}_{\theta}\sqrt{-g}) &= 0. \label{EOM_4.2d}
\end{align}
\end{subequations}
%
%
Substituting from Eqs. (\ref{EMT_4.2a})-(\ref{EMT_4.2d}) into Eqs. (\ref{EOM_4.2a})-(\ref{EOM_4.2d}), it may be verified that Eqs. (\ref{EOM_4.2b}) and (\ref{EOM_4.2d}) are each equivalent to
\begin{eqnarray} \label{EOM6_4.2} 
 \frac{Q}{\dot{Q}}\frac{\dot{\omega}}{\omega} = 1-\omega^2,
\end{eqnarray}
and that Eq. (\ref{EOM_4.2a}) gives 
\begin{eqnarray} \label{EOM7_4.2}
P^2\left(1-\omega^2 - 2\frac{Q}{\dot{Q}}\frac{\dot{\omega}}{\omega}\right) + P\dot{P}\frac{Q}{\dot{Q}} = 0.
\end{eqnarray}
In fact, Eq. (\ref{EOM6_4.2}) follows directly from the definition of $\omega^2$ under the assumption that $\omega^2 = \omega^2(t)$ or, equivalently, $\partial_{\sigma}\omega = 0$. 
For circular loops, this is required by symmetry and $\Gamma(\sigma,t)$ contains no nonlinear terms, so that 
\begin{eqnarray}  \label{}
\Gamma(\sigma,t) = n_{\sigma}\sigma + \omega_{\sigma}t, \ \ \ (n_{\sigma} \in \mathbb{Z})
\end{eqnarray}
Eqs. (\ref{EOM6_4.2})-(\ref{EOM7_4.2}) together imply
\begin{eqnarray}  \label{EOM8_4.2}
P(2\omega^2-1) + P\dot{P}\frac{Q}{\dot{Q}} = 0.
\end{eqnarray}
\indent
We now note that this is equivalent to the EOM for a circular wound string loops obtained in Sec. \ref{Sect2} (see also \cite{LaYo12,Lake:2009nq}), \emph{if} we impose the conditions
\begin{eqnarray} \label{P-Q}
P^2(t) = 1 - \dot{Q}^2(t), \ \ \ P(t) = \pm \sqrt{1 - \dot{Q}^2(t)},
\end{eqnarray}
which, together with Eq.(\ref{EOM6_4.2}), implies 
\begin{eqnarray} \label{}
\frac{Q}{\dot{Q}}\frac{\dot{P}}{P} = \frac{Q\ddot{Q}}{P^2} = 2\omega^2-1.
\end{eqnarray}
The two unique EOM that satisfy both the fundamental Abelian-Higgs EOM and the conservation equations (\ref{EOM_4.2a})-(\ref{EOM_4.2c}), and which are sufficient to determine the two free functions $Q(t)$ and $\Gamma(\sigma,t)$, may then be written as
\begin{subequations} \label{FINAL}
\begin{align}
(1-\dot{Q}^2)(2\omega^2-1) + Q\ddot{Q} = 0, \label{FINALa}\\
\dot{\Gamma}^2 = \frac{(1-\dot{Q}^2)}{Q^2}(\partial_{\sigma}\Gamma)^2. \label{FINALb}
\end{align}
\end{subequations}
Equations (\ref{FINALa})-(\ref{FINALb}) are, clearly, formally analogous the two EOM that determine the dynamics of the circular wound-string loop, i.e., Eqs. (\ref{EOM1_4.2}) under the identification $Q(t) \leftrightarrow \rho(t)$. 
Together with the {\it Finsler condition} (\ref{P-Q}) this implies $P(t) \leftrightarrow \sqrt{1-\dot{\rho}^2}$ and Eqs. (\ref{EOM6_4.2}), (\ref{EOM7_4.2}) and (\ref{EOM8_4.2}) are formally equivalent to 
Eqs. (\ref {EOM6_4.2*}), (\ref{EOM7*_4.2}) and (\ref{EOM8_4.2*}), respectively. 
%
%
\\ \indent
The \emph{normalized} components of the current density are 
\begin{eqnarray} 
J^{0}\sqrt{-g} = \frac{e}{2\pi}\frac{\sqrt{1-\dot{Q}(t_i)}}{Q(t_i)}\frac{Q}{\sqrt{1-\dot{Q}}}\dot{\Gamma}, \ \ \ 
J^{\sigma}\sqrt{-g} = \frac{e}{2\pi}\frac{\sqrt{1-\dot{Q}(t_i)}}{Q(t_i)}\frac{\sqrt{1-\dot{Q}}}{Q}\partial_{\sigma}\Gamma,
\end{eqnarray}
by analogy with (\ref{J_compts*}), so that using the EOM in $\Gamma$ gives
\begin{eqnarray} 
J^{0}\sqrt{-g} = \pm \frac{e}{2\pi}\frac{\sqrt{1-\dot{Q}(t_i)}}{Q(t_i)}\partial_{\sigma}\Gamma = -J^{\sigma}\sqrt{-g} = \mp \frac{e}{2\pi}\frac{\sqrt{1-\dot{Q}(t_i)}}{Q(t_i)}\dot{\Gamma},
\end{eqnarray}
by analogy with (\ref{J_reln*}).
The components of the $4$-current may be expressed in terms of $\mathcal{T}^{0}{}_{\theta}\sqrt{-g} $ and $\mathcal{T}^{\sigma}{}_{\theta}\sqrt{-g}$, given by Eq. (\ref{EMT_4.2d}), and it is straightforward to show that the current conservation equation is equivalent to Eq. (\ref{EOM_4.2d}). 
As for long strings, the EOM in $\Gamma(\sigma,t)$ may be written as a dispersion relation
\begin{eqnarray} \label{DispRel_4.2}
\omega_{\sigma}^2 = P^2k_{\sigma}^2, \ \ \ \omega_{\sigma} = \pm P k_{\sigma},
\end{eqnarray}
where we have defined the (time-dependent) wavenumber and wavelength as
\begin{eqnarray} \label{lambda_sig_4.2}
k_{\sigma}(t) = \frac{2\pi}{\lambda_{\sigma}(t)}, \ \ \ \lambda_{\sigma}(t) = \frac{2\pi Q(t)}{n_{\sigma}}.
\end{eqnarray}
\indent
However, since we have decided to interpret $\sigma$ as an angular coordinate, and hence to interpret $Q(t)$ as a radial coordinate, we are free to redefine the metric such that
\begin{eqnarray} \label{Metric3}
{\rm d}s^2 = P^2(t){\rm d}t^2 - {\rm d}r^2 - r^2{\rm d}\theta^2 - Q^2(t){\rm d}\sigma^2 \rightarrow P^2(t){\rm d}t^2 - {\rm d}x^2 - {\rm d}Q^2(t) - Q^2(t){\rm d}\sigma^2,
\end{eqnarray}
where $x$ is a standard (static) Cartesian coordinate. 
Note that, here, $Q$ is really a function of the `internal' world-sheet coordinate $\tau$, but that this is identified directly with the physical time coordinate $t$ via the choice of the static gauge for the string embedding, $t = \zeta\tau$. 
Hence, both $\left\{t,r,\theta,\sigma\right\}$ and $\left\{t,x,Q,\sigma\right\}$ are legitimate coordinate systems for the string. 
The former contains the coordinates that are most useful to describe the string core, $r$ and $\theta$, but it is more convenient to adopt the latter to describe the macroscopic string dynamics.
Using the conditions (\ref{P-Q}), the effective Finsler-type metric may then be written as:
\begin{eqnarray} \label{Finsler}
{\rm d}s^2 = (1-\dot{Q}^2(t)){\rm d}t - dx^2 - {\rm d}Q^2(t) - Q^2(t){\rm d}\sigma^2.
\end{eqnarray}
\indent
In the new coordinate system the components of the effective energy-momentum tensor involving only $t$ and $\sigma$ are the same as before, but there also exist additional components whose conservation equation,
\begin{eqnarray}  \label{EOM_4.2c}
\partial_{0}(T^{0}{}_{Q}\sqrt{-g}) + \partial_{\sigma}(T^{\sigma}{}_{Q}\sqrt{-g}) - \frac{1}{2}\partial_{Q}g_{\sigma\sigma}T^{\sigma\sigma}\sqrt{-g} =  0,
\end{eqnarray}
must be satisfied by the macroscopic dynamics of the string. 
Although $\mathcal{T}^{\sigma\sigma}\sqrt{-g}$ is well defined (albeit implicitly) by Eqs. (\ref{T_mu_nu}) and (\ref{U-T-mathcal{T}_2}), the additional components $T^{0}{}_{Q}\sqrt{-g}$ and $T^{\sigma}{}_{Q}\sqrt{-g}$ cannot be specified using these. 
In order to apply the same method as that used to determine $\mathcal{T}^{\sigma\sigma}\sqrt{-g}$, etc, we would first need to write down a new ansatz for the field variables $\phi$, $A_{\mu}$ (and $j^{\mu}$) in the new coordinate system, after the transformation $\left\{t,r,\theta,\sigma\right\} \rightarrow \left\{t,x,Q,\sigma\right\}$. 
\\ \indent
Nonetheless, we \emph{do} know that, whatever the correct definitions of these quantities, the resulting conservation equation must be compatible with (\ref{EOM8_4.2}). 
Based on the results derived in Sec. \ref{Sect2}, we propose the set
\begin{eqnarray} \label{EMT_4.2c*}
\mathcal{T}^{0}{}_{Q}\sqrt{-g} = -2\pi\eta^2|n| \frac{Q\dot{Q}}{P}\omega^{-2}, \ \ \ 
\mathcal{T}^{\sigma}{}_{Q}\sqrt{-g} = \pm 2\pi\eta^2|n|\dot{Q}\left(\frac{1-\omega^{2}}{\omega^{2}}\right), 
\nonumber\\ 
\mathcal{T}^{\sigma\sigma}\sqrt{-g} = -2\pi\eta^2|n| \frac{P}{Q} \left(\frac{2\omega^{2}-1}{\omega^{2}}\right). 
\end{eqnarray}
by analogy with Eq. (\ref{EMT_4.2c**}). 
The conservation equation (\ref{EOM_4.2c}) then gives
\begin{eqnarray}  \label{EOM_4.2c*}
P^2(2\omega^2-1) + Q\ddot{Q} + \dot{Q}^2\left(1 - \frac{Q}{\dot{Q}}\frac{\dot{P}}{P} - 2\frac{Q}{\dot{Q}}\frac{\dot{\omega}}{\omega}\right) = 0.
\end{eqnarray}
Applying the `Finsler conditions' (\ref{P-Q}) / (\ref{Finsler}), Eq. (\ref{EOM_4.2c*}) reduces to (\ref{FINALa}). Our definitions of $T^{0}{}_{Q}\sqrt{-g}$ and $T^{\sigma}{}_{Q}\sqrt{-g}$ are therefore consistent with both the Euler-Lagrange equations of the Abelian-Higgs model \emph{and} the conservation equations for the energy-momentum tensor corresponding to the previous set of coordinates $\left\{t,r,\theta,\sigma\right\}$, in which the microscopic field ansatz was defined.
\\ \indent
It is straightforward to verify that the general solution to Eq. (\ref{FINALb}) is of the form 
\begin{eqnarray} \label{Gamma_soln}
\Gamma(\sigma,t) = \Gamma\left(n_{\sigma}\sigma \pm n_{\sigma}\int \frac{\sqrt{1-\dot{Q}^2}}{Q}{\rm d}t\right),
\end{eqnarray}
whereas, at least for $\dot{Q}(t_i)=0$, the solution of Eq. (\ref{FINALa}) takes the simple form
\begin{eqnarray} \label{Q_soln}
Q(t) = Q(t_i)\sqrt{1 - \left(\frac{1-2\omega^2(t_i)}{\omega^4(t_i)}\right)\sin^2\left(\frac{\omega^2(t_i)}{Q(t_i)}(t-t_i)\right)}.
\end{eqnarray}
These solutions have been studied in detail in \cite{LaYo12,Thesis,Lake:2009nq} and Eqs. (\ref{Gamma_soln}) and (\ref{Q_soln}) should be compared with Eqs. (\ref{phiSoln1_4.2})-(\ref{omega_sig_4.2}) and (\ref{EOMSoln5_4.2}), respectively. 
\\ \indent
After applying the Finsler conditions, both the EOM (\ref{FINALa})-(\ref{FINALb}) and the expressions for the components of the energy momentum tensor (\ref{EOM_4.2a})-(\ref{EOM_4.2c}) are equivalent to those obtained in Sec. \ref{Sect2}, for circular wound string loops with time-dependent radius $\rho(t)$, under the identifications
\begin{eqnarray} \label{phi-Gamma}
\Gamma(\sigma,t) \leftrightarrow \varphi(\sigma,t), \ \ \ Q(t)  \leftrightarrow \rho(t), \ \ \ r_{c|n|}  \leftrightarrow R,
\end{eqnarray}
or, strictly, $Q(t) = a\rho(t)$ and $P(t) = a\sqrt{1-\dot{\rho}^2(t)}$ if $a < 1$ in the wound string model. 
\\ \indent
This, in turn, justifies our `guess' for the $T^{0}{}_{Q}\sqrt{-g}$ and $T^{\sigma}{}_{Q}\sqrt{-g}$ components of the effective string energy-momentum tensor. 
The definitions proposed in Eq. (\ref{EMT_4.2c*}), together with the conditions (\ref{P-Q}), ensure we obtain a self-consistent set of Euler-Lagrange and conservation equations for the superconducting defect string, which turn out to be formally equivalent to the EOM obtained in the wound-string case, despite the fact that they are derived from radically different actions, i.e., the fundamental action for the Abelian-Higgs fields (\ref{AH_Act}) and the Nambu-Goto action for the $F$-string (\ref {Act_2.1}).
\\ \indent
We note that this result was by no means \emph{certain}. 
The formal correspondence in \cite{Ni79} demonstrates that the motion of windings in the compact extra dimensions can be interpreted as a charge density on the effective $(3+1)$-dimensional world sheet of the string. 
It does \emph{not} show that the resulting EOM will be identical to those obtained for any particular defect string model. In principle, any differences in the macroscopic evolution equations could give rise to specific higher-dimensional signatures in the wound-string case. 
This could have occurred, for example, if it were not possible to reconcile Eq. (\ref{EOM_4.2c}) with Eqs. (\ref{EOM_4.2a})-(\ref{EOM_4.2d}) to form a self-consistent set.
\\ \indent
Using the procedure above, we obtain precisely the same EOM as in the wound-string case, but derived from the fundamental Euler-Lagrange and conservation equations for the Abelian-Higgs fields. 
In particular, our results suggest that the EOM coupling the evolution of the windings to the $(3+1)$-dimensional dynamics of the $F$-string are \emph{precisely} equivalent to the EOM coupling the internal field evolution to the macroscopic dynamics of the superconducting $U(1)$ string. 
This makes sense if we consider the configuration of the $F$-string in the internal space as determining its `internal' structure from an effective $(3+1)$-dimensional perspective.
\\ \indent
Thus, together with $f(r)$, $a_{\theta}(r)$, $a(r)$ and $j(r)$, the functions $\Gamma(\sigma,t)$ and $Q(t) = \rho(t)$ determine both the microscopic string structure and its macroscopic evolution. 
It follows that the expressions for the constants of motion also take the same form as those given in Sec. \ref{Sect2}. 
We may study the behavior of $f(r)$, $a_{\theta}(r)$, $a(r)$, $j(r)$ and $\Gamma(\sigma,t)$ in the coordinate system covering the internal structure of the string core, $\left\{t,r,\theta,\sigma\right\}$, whereas $Q(t)$ determines the macroscopic motion in the $\left\{t,x,Q,\sigma\right\}$ coordinate system. Eqs. (\ref{FINALa})-(\ref{FINALb}) therefore describe the \emph{interplay} between the large scale dynamics of the string, as a composite structure, and the microscopic configuration of its constituent fields. 
In practice, we can reconstruct the former from the latter. Since the EOM in $Q(t)$ depends only on the constants $\partial_{\sigma}\Gamma = n_{\sigma}$ and $r_{c|n|}$, it may be solved independently. 
We then substitute the solution into the second EOM to determine $\Gamma(\sigma,t)$ and, together with the solutions (either approximate or numerical) for $f(r)$, $a_{\theta}(r)$, $a(r)$ and $j(r)$, the microscopic field evolution is precisely determined.
\\ \indent
It is intriguing, and suggestive of a more fundamental link, that the same set of equations describing current-carrying string loops can be obtained from such different fundamental theories (without the need for approximations or effective actions) when one theory, first defined in Minkowski space, then has a `Finsler condition' imposed upon it at the level of the EOM. 
As we shall see in the next Section, the EOM governing arbitrary planar loops, in the wound string case, may also be obtained from the fundamental Abelian-Higgs equations. In this model, one first defines these in Minkowski space-time, before imposing a `generalised Finsler condition'.

\subsection{Noncircular defect-string loops: generalised Finsler geometry} \label{Sect5.3}
To treat noncircular superconducting loops, we must consider a generalisation of the line-element (\ref{Metric2}), in which the free functions depend also on an `internal' space-like variable, i.e., such that $P(t) \rightarrow P(t,\sigma)$ and $Q(t) \rightarrow Q(t,\sigma)$. 

\subsubsection{Noncircular defect-string loops without currents} \label{Sect5.3.1}
As for circular string loops, it is again simpler to obtain the zero-current solution as a particular limit of a more general scenario, which in this case corresponds to an arbitrary planar loop with nonzero current. 
We derive the EOM for this configuration in the following subsection, and the corresponding non-superconducting solution is again recovered by substituting $\Gamma(\sigma,t) = 0$ ($j^{\mu} = 0$). 

\subsubsection{Noncircular defect-string loops with currents} \label{Sect5.3.2}
Let us now consider a metric with line element of the form
\begin{eqnarray} \label{ds^2-3}
{\rm d}s^2 = P^2(\sigma,t){\rm d}t^2 - {\rm d}r^2 - r^2{\rm d}\theta^2 - 2U(t,\sigma){\rm d}t{\rm d}\sigma - Q^2(\sigma,t){\rm d}\sigma^2,
\end{eqnarray}
giving $\sqrt{-g} = r\sqrt{P^2Q^2+U^2}$. 
Using the same field ansatz, Eq. (\ref{New_ansatz1}), but identifying $\Gamma(z,t) \equiv \Gamma(\sigma,t)$ and substituting $A_{\sigma}(r,\sigma,t) = (n/e)a(r)\partial_{\sigma}\Gamma$ for $A_z(r,z,t) = (n/e)a(r)\Gamma'$, the scalar EOM again separates into real and imaginary parts
\begin{eqnarray} 
\frac{{\rm d}^2f}{{\rm d}r^2} + \frac{1}{r}\frac{{\rm d}f}{{\rm d}r} - \frac{n^2f}{r^2}(1-a_{\theta})^2 - \frac{n^2f(1-a)^2}{(-X)}[P^2(\partial_{\sigma}\Gamma)^2 - Q^2\dot{\Gamma}^2 + 2U\dot{\Gamma}\partial_{\sigma}\Gamma] - \frac{1}{2r_s^2}f(f^2-1) = 0, \nonumber
\end{eqnarray}
\begin{eqnarray} \label{Kawa-scalar-real}
{}
\end{eqnarray}
and
\begin{eqnarray} \label{Kawa-scalar-imag} 
(-X)[Q^2\ddot{\Gamma} - P^2\partial_{\sigma}^2\Gamma - 2U\partial_{\sigma}\dot{\Gamma} - \partial_{\sigma}U\dot{\Gamma} - \dot{U}\partial_{\sigma}\Gamma + 2Q\dot{Q}\dot{\Gamma} - 2P\partial_{\sigma}P\partial_{\sigma}\Gamma] 
\nonumber\\
- \frac{1}{2}\frac{\partial(-X)}{\partial t}[Q^2\dot{\Gamma} - U\partial_{\sigma}\Gamma] - \frac{1}{2}\frac{\partial(-X)}{\partial \sigma}[-P^2\partial_{\sigma}\Gamma - U\dot{\Gamma}] = 0,
\end{eqnarray}
respectively, where we have denoted $(-X) = P^2Q^2+U^2$. 
\\ \indent
Next, we again adopt the conditions (\ref{j_1}) and define in the remaining nonzero components of $j^{\mu}$ as in Eq. (\ref{j_3}). 
Under these conditions, it may be also shown that, \emph{if} Eq. (\ref{Kawa-scalar-imag}) is satisfied, the  $r-$ component of the vector EOM vanishes identically, while the $\sigma-$, and $t-$components both reduce to (\ref{EOM_in_a_1}) and the $\theta-$component again reduces to that obtained for the Nielsen-Olesen string.
\\ \indent
For the new metric (\ref{ds^2-3}), we modify the definition of the effective energy-momentum tensor for the string in Eq. (\ref{U-T-mathcal{T}_2}), so that 
\begin{eqnarray} \label{U-T-mathcal{T}_3}
U^{\mu}{}_{\nu} = \int \mathcal{T}^{\mu}{}_{\nu}\sqrt{-g}{\rm d}\sigma = \int \tilde{\kappa}_{(\nu)\alpha}T^{\mu\alpha}{}_{\nu}\sqrt{-g} {\rm d}r {\rm d}\theta {\rm d}\sigma.
\end{eqnarray}
The new set of vectors $\tilde{\kappa}_{(\nu)\alpha}$ are defined by analogy with Eqs. (\ref{kappa_1})-(\ref{kappa_2}), but with an important difference, related to the $g_{\sigma\sigma}$ and $g_{0\sigma}$ components of the new metric. Thus, we set
\begin{eqnarray} \label{kappa_1*}
&\tilde{\kappa}_{(0)\alpha} = [1,0,0,-U(\sigma,t)] = [\eta_{00},0,0,0] = \eta_{(0)\alpha},
\nonumber\\
&\tilde{\kappa}_{(\theta)\alpha} = [0,0,-r^2,0] = [0,0,\eta_{\theta\theta},0] = \eta_{(\theta)\alpha},
\nonumber\\
&\tilde{\kappa}_{(\sigma)\alpha} = [0,0,0,[-Q^2+(P^2Q^2+U^2)]/(1-P^2)]  = [0,0,0,(g_{\sigma\sigma} +\eta_{\theta\theta}(-g))/(1-g_{00})],
\end{eqnarray}
or, equivalently
\begin{eqnarray} \label{kappa_2*}
\tilde{\kappa}_{(\nu)\alpha} = \eta_{(\nu)\alpha} +\delta^{(\sigma)}{}_{(\nu)}\delta^{\sigma}{}_{\alpha}\left[(g_{(\sigma)\sigma} + \eta_{\theta\theta}(-g))/(1-g_{00}) - \eta_{(\sigma)\sigma}\right],
\end{eqnarray}
where $g_{\mu\nu}$ refers to the metric giving the line element (\ref{ds^2-3}). 
Note that, when $U(\sigma,t)=0$ and $Q(\sigma,t)=Q(t)$, Eqs. (\ref{kappa_1*}) and (\ref{kappa_2*}) reduce to Eqs. (\ref{kappa_1}) and (\ref{kappa_2}), respectively. 
Again, the definition of $\tilde{\kappa}_{(\sigma)\alpha}$ may appear arbitrary at first, but we will show that it is necessary in order to obtain a consistent set of conservation equations. 
\\ \indent
If we now {\it assume} that
\begin{eqnarray} \label{Assume}
P^2(\partial_{\sigma}\Gamma)^2 - Q^2\dot{\Gamma}^2 + 2U\dot{\Gamma}\partial_{\sigma}\Gamma = 0,
\end{eqnarray}
the scalar EOM (\ref{Kawa-scalar-real}) reduces to the usual one for the Nielsen-Olesen string, as before. 
Equation (\ref{Assume}) may be solved in the variable ($\dot{\Gamma}/\partial_{\sigma}\Gamma$), so that
\begin{eqnarray} \label{}
\frac{\dot{\Gamma}}{\partial_{\sigma}\Gamma} = \frac{U \pm \sqrt{P^2Q^2+U^2}}{Q^2} =  \frac{P^2}{-U \pm \sqrt{P^2Q^2+U^2}},
\end{eqnarray}
which we note is formally analogous to (\ref{Const_6.2*}) for appropriate choices of $P$, $Q$ and $U$. 
(These will be considered, in detail, shortly.) 
Next, adopting the definitions given in Eqs. (\ref{kappa_1*}) and (\ref{kappa_2*}), the components of $\mathcal{T}^{\mu}{}_{\nu}\sqrt{-g}$ that are analogous to those in Eqs. (\ref{EMT_3.2a})-(\ref{EMT_3.2c}) and (\ref{EMT_4.2a})-(\ref{EMT_4.2d}) can be written as
\begin{subequations}
\begin{align}
\mathcal{T}^{0}{}_{0}\sqrt{-g} &= 2\pi\eta^2|n|\frac{Q^2}{\sqrt{P^2Q^2+U^2}}\omega^{-2},  
\nonumber\\
\mathcal{T}^{\sigma}{}_{0}\sqrt{-g} &=  -2\pi\eta^2|n|\left[\frac{U}{\sqrt{P^2Q^2+U^2}}\omega^{-2} \mp \left(\frac{1-\omega^2}{\omega^2}\right)\right], \label{EMT_6.2a}\\
\mathcal{T}^{0}{}_{\sigma}\sqrt{-g} &=  2\pi\eta^2|n|\left[\frac{U}{\sqrt{P^2Q^2+U^2}}\omega^{-2} \pm \left(\frac{1-\omega^2}{\omega^2}\right)\right]\left(\frac{Q^2-(P^2Q^2+U^2)}{1-P^2}\right), 
\nonumber\\
\mathcal{T}^{\sigma}{}_{\sigma}\sqrt{-g} &=  2\pi\eta^2|n|\frac{P^2}{\sqrt{P^2Q^2+U^2}}\left[1-\left(\frac{1-\omega^2}{\omega^2}\right)\frac{(U \pm \sqrt{P^2Q^2+U^2})^2}{P^2Q^2}\right]\left(\frac{Q^2-(P^2Q^2+U^2)}{1-P^2}\right), \label{EMT_6.2b}\\
\mathcal{T}^{0}{}_{\theta}\sqrt{-g} &= \mp2\pi\eta^2|n| r_{c|n|}Q\frac{\sqrt{1-\omega^2}}{\omega}, \nonumber\\ \mathcal{T}^{\sigma}{}_{\theta}\sqrt{-g} &= \pm 2\pi\eta^2|n| r_{c|n|} \frac{U \pm \sqrt{P^2Q^2+U^2}}{Q} \frac{\sqrt{1-\omega^2}}{\omega}, \label{EMT_6.2c}
\end{align}
\end{subequations}
where we have defined $\omega^2(\sigma,t)$ as in Eq. (\ref{omega^2_2}) but with $Q^2=Q^2(\sigma,t)$. 
The relevant conservation equations are Eqs. (\ref{EOM_4.2a})-(\ref{EOM_4.2d}). 
Once more, we note that the condition which ensures the consistency of the field ansatz, in this case Eq. (\ref{Assume}), is precisely the condition required to ensure that the Lagrangian reduces to that of the Nielsen-Olesen string at the level of the energy-momentum tensor.
\\ \indent
For noncircular loops, the normalised components of the current density may be defined as
\begin{eqnarray} 
\mathcal{J}^{0}\sqrt{-g} &=& \frac{e}{2\pi}\frac{U(\sigma,t_i) \pm \sqrt{P^2(\sigma,t_i)Q^2(\sigma,t_i)+U^2(\sigma,t_i)}}{Q^2(\sigma,t_i)}
\frac{Q^2}{\sqrt{P^2Q^2+U^2}}\dot{\Gamma},
\nonumber\\
\mathcal{J}^{\sigma}\sqrt{-g} &=& \frac{e}{2\pi}\frac{U(\sigma,t_i) \pm \sqrt{P^2(\sigma,t_i)Q^2(\sigma,t_i)+U^2(\sigma,t_i)}}{Q^2(\sigma,t_i)}
\frac{\sqrt{P^2Q^2+U^2}}{Q^2}\partial_{\sigma}\Gamma,
\end{eqnarray}
by analogy with Eq. (\ref{J_compts**}), so that using the EOM in $\phi$ gives
\begin{eqnarray} \label{J_relnX}
&{}& \mathcal{J}^{0}\sqrt{-g} = \pm \frac{e}{2\pi}\frac{U(\sigma,t_i) \pm \sqrt{P^2(\sigma,t_i)Q^2(\sigma,t_i)+U^2(\sigma,t_i)}}{Q^2(\sigma,t_i)}\partial_{\sigma}\Gamma 
\nonumber\\
= &{}&-\mathcal{J}^{\sigma}\sqrt{-g} = \mp \frac{e}{2\pi}\frac{U(\sigma,t_i) \pm \sqrt{P^2(\sigma,t_i)Q^2(\sigma,t_i)+U^2(\sigma,t_i)}}{Q^2(\sigma,t_i)}\dot{\Gamma}.
\end{eqnarray}
by analogy with (\ref{J_reln**}). 
The components of the $4$-current may again be expressed in terms of $\mathcal{T}^{0}{}_{\theta}\sqrt{-g} $ and $\mathcal{T}^{\sigma}{}_{\theta}\sqrt{-g}$, now given by Eq. (\ref{EMT_6.2c}), and it is straightforward to show that the current conservation equation is then equivalent to Eq. (\ref{EOM_4.2d}).
\\ \indent
Although the full system of evolution equations, that is, the remaining Euler-Lagrange equations of the Abelian-Higgs model, (\ref{Kawa-scalar-imag}) and (\ref{Assume}), and the conservation equations, (\ref{EMT_6.2a})-(\ref{EMT_6.2c}), appear incredibly complicated, we will show that careful treatment allows us to demonstrate their consistency. 
Furthermore, this will allow us to write the general solution for $\Gamma(t,\sigma)$ in terms of $Q(t,\sigma)$, and its first and second derivatives in $t$ and $\sigma$, via the imposition of a suitable `generalised Finsler condition'.
\\ \indent
As a first step towards demonstrating the self-consistency of Eqs. (\ref{EOM_4.2a})-(\ref{EOM_4.2d}), using the components $\mathcal{T}^{\mu}{}_{\nu}\sqrt{-g}$ obtained in Eqs. (\ref{EMT_6.2a})-(\ref{EMT_6.2c}), we note that each contains terms in $\omega$, $\dot{\omega}$ and $\partial_{\sigma}\omega$. 
In the case of circular loops, we had $\partial_{\sigma}\omega = 0$, so that the EOM contained terms in only $\omega$ and $\dot{\omega}$. 
However, we were able to eliminate the latter via appropriate manipulation, leaving us with a single self-consistent EOM in $Q$ and $\omega$, in addition to the EOM in $\Gamma$. 
Even in this more general case, we still have three EOM and two quantities ($\dot{\omega}$ and $\partial_{\sigma}\omega$) we wish to eliminate, so this poses no problem. 
\\ \indent
We begin the elimination of $\dot{\omega}$ and $\partial_{\sigma}\omega$ by first writing each of the components of $\mathcal{T}^{\mu}{}_{\nu}\sqrt{-g}$ in Eqs. (\ref{EMT_6.2a})-(\ref{EMT_6.2b}) as the sum of two terms, one independent of the factor $(\omega^{-2}-1)$ and one directly proportional to $(\omega^{-2}-1)$. 
Thus, we set
\begin{subequations} 
\begin{align}
\mathcal{T}^{0}{}_{0}\sqrt{-g} = \beta + \beta(\omega^{-2}-1), \ \ \ \mathcal{T}^{\sigma}{}_{0}\sqrt{-g} = \gamma + \delta(\omega^{-2}-1), \label{GreekA}
\\
\mathcal{T}^{0}{}_{\sigma}\sqrt{-g} = \epsilon + \zeta(\omega^{-2}-1), \ \ \ \ \mathcal{T}^{\sigma}{}_{\sigma}\sqrt{-g} = \eta + \theta(\omega^{-2}-1), \label{GreekB} 
\end{align}
\end{subequations}
where the values of $\beta$, $\gamma$, $\delta$, $\epsilon$, $\zeta$, $\eta$ and $\theta$ are obtained by comparison of Eqs. (\ref{GreekA})-(\ref{GreekB}) with Eqs. (\ref{EMT_6.2a})-(\ref{EMT_6.2b}), giving
\begin{eqnarray} \label{beta}
\beta = \frac{Q^2}{\sqrt{P^2Q^2+U^2}},
\end{eqnarray}
\begin{eqnarray} \label{delta}
\gamma = \delta \mp 1, \ \ \ \delta = -\frac{U}{\sqrt{P^2Q^2+U^2}} \pm 1.
\end{eqnarray}
\begin{eqnarray} \label{zeta}
\epsilon = -\frac{Q^2-(P^2Q^2+U^2)}{1-P^2}(\delta \mp 1), \ \ \  \zeta = -\frac{Q^2-(P^2Q^2+U^2)}{1-P^2}\delta,
\end{eqnarray}
\begin{eqnarray} \label{theta}
\eta = \frac{Q^2-(P^2Q^2+U^2)}{\sqrt{P^2Q^2+U^2}},  \ \ \ \theta = -\frac{Q^2-(P^2Q^2+U^2)}{1-P^2}\frac{\delta^2}{\beta}.
\end{eqnarray}
Equations (\ref{GreekA})-(\ref{GreekB}) are formally equivalent to (\ref{GreekA*})-(\ref{GreekB*}) and the definitions of $\beta$, $\gamma$, $\delta$, $\epsilon$, $\zeta$, $\eta$ and $\theta$ (\ref{beta})-(\ref{theta}) are analogous to those in Eqs. (\ref{beta*})-(\ref{theta*}).
Substituting from Eqs. (\ref{GreekA})-(\ref{GreekB})
into Eqs. (\ref{EOM_4.2a})-(\ref{EOM_4.2b}), rearranging to make $\partial_{t}(\omega^{-2}-1)$ the subject, and equating the results, gives
\begin{eqnarray} \label{GreekEOM3}
[\beta(\dot{\zeta}+\partial_{\sigma}\theta) - \zeta(\dot{\beta}+\partial_{\sigma}\delta)](1-\omega^{2}) + [\beta(\dot{\epsilon}+\partial_{\sigma}\eta) - \zeta(\dot{\beta}+\partial_{\sigma}\gamma)]\omega^{2} = 0.
\end{eqnarray}
where we have used the fact that $\zeta\delta-\beta\theta = 0$, which may be explicitly verified by direct substitution. 
This equation involves only $\omega$ and is formally equivalent to (\ref{GreekEOM3*}).
\\ \indent
Next, we interpret the function $Q(\sigma,t)$ as an effective `local' radial coordinate (i.e. depending on both $t$ and $\sigma$), and make the coordinate transformation $\left\{t,r,\theta,\sigma\right\} \rightarrow \left\{t,x,Q,\sigma\right\}$. 
This corresponds to the change of line element
\begin{eqnarray} \label{Metric4}
{\rm d}s^2 &= P^2(\sigma,t){\rm d}t - {\rm d}r^2 - r^2{\rm d}\theta^2 - 2U(\sigma,t){\rm d}t{\rm d}\sigma - Q^2(\sigma,t){\rm d}\sigma^2 
\nonumber\\
&\rightarrow P^2(\sigma,t){\rm d}t - {\rm d}x^2 - {\rm d}Q^2(\sigma,t) - 2U(\sigma,t){\rm d}t{\rm d}\sigma - Q^2(\sigma,t){\rm d}\sigma^2.
\end{eqnarray}
We then impose the `generalised Finsler condition', which relates the metric components $g_{00} = P^2(\sigma,t)$, $g_{0\sigma} = -U(\sigma,t)$ and $g_{\sigma\sigma} = -Q^2(\sigma,t)$ to both the temporal and spatial derivates of the $Q-$coordinate, via the identifications
\begin{eqnarray} \label{P-Q-U}
P^2 = 1 - \dot{\rho}^2, \ \ \ Q^2 = \rho^2 + (\partial_{\sigma}\rho)^2, \ \ \ U = \dot{\rho}\partial_{\sigma}\rho,
\end{eqnarray}
where $\rho(\sigma,t)$ may be interpreted as the genuine local radius of the string. 
Together with the conditions (\ref{P-Q-U}), Eq. (\ref{Metric4}) implies that the metric is a generalised Finsler metric. 
This may be expressed as:
\begin{eqnarray} \label{Kawaguchi}
{\rm d}s^2 = \left(1-\left[\frac{\partial}{\partial t}\left(\frac{Q^2-(P^2Q^2+U^2)}{1-P^2}\right)^{1/2}\right]^2\right){\rm d}t - {\rm d}x^2 - {\rm d}Q^2 - Q^2{\rm d}\sigma^2,
\end{eqnarray}
where the behaviour of the functions $P(\sigma,t)$, $Q(\sigma,t)$ and $U(\sigma,t)$ must be reconstructed by solving the EOM for $\rho(\sigma,t)$. 
The method used from here on is precisely analogous to that used in Sec. \ref{Sect2} and we begin by rewriting the line element (\ref{Kawaguchi}) as:
\begin{eqnarray} \label{Kawaguchi*}
{\rm d}s^2 = (1-\dot{\rho}^2){\rm d}t^2 - {\rm d}x^2 - {\rm d}\rho^2 - (\rho^2+(\partial_{\sigma}\rho)^2){\rm d}\sigma^2.
\end{eqnarray}
\\ \indent
Next, we note that substituting for $(\dot{\epsilon}+\partial_{\sigma}\eta)$, $(\dot{\beta}+\partial_{\sigma}\gamma)$, $\beta$ and $\zeta$ from Eqs. (\ref{beta})-(\ref{theta}), the terms in the second set of square brackets in Eq. (\ref{GreekEOM3}) 
are proportional to the quantity
\begin{eqnarray} \label{chi}
\chi = (1-\dot{\rho}^2)\left(1-\frac{\partial^2_{\sigma}\rho}{\rho}\right) + \left(1+\frac{(\partial_{\sigma}\rho)^2}{\rho^2}\right)\rho\ddot{\rho} + 2\left[\frac{(\partial_{\sigma}\rho)^2}{\rho^2}-\frac{\dot{\rho}}{\rho}\partial_{\sigma}\rho\partial_{\sigma}\dot{\rho}\right],
\end{eqnarray}
which was defined previously in Eq. (\ref{chi*}). 
It may be verified that $\chi = 0$ is the unique EOM obtained from the conservation equations (\ref{EOM_4.2a})-(\ref{EOM_4.2d}) in the limit $\omega^2 \rightarrow 1$. 
In other words, this corresponds to the macroscopic evolution equation of either a Nambu-Goto or Nielsen-Olesen string, at least in the classical theories, ignoring gravitational effects. 
We then have
\begin{eqnarray}
\beta(\dot{\epsilon}+\partial_{\sigma}\eta) - \zeta(\dot{\beta}+\partial_{\sigma}\gamma) = \left[\beta \partial_{\sigma}\rho - \rho^2\dot{\rho}\delta\right] \times \frac{\rho^3}{[(1-\dot{\rho}^2)\rho^2 + (\partial_{\sigma}\rho)^2]^{\frac{3}{2}}}\chi,
\end{eqnarray}
\indent
The first set of square brackets in Eq. (\ref{GreekEOM3}) may again be written purely in terms of the quantities $\beta$ and $\delta$ (c.f. (\ref{needlater})) so that the final remaining independent EOM, involving only derivatives of $\rho$ and powers of $\omega$, but not $\dot{\omega}$ or $\partial_{\sigma}\omega$, may be written in a \emph{relatively} compact form as:
\begin{eqnarray} \label{relcompact}
\chi &+& \frac{\rho^3(\beta \partial_{\sigma}\rho - \rho^2\dot{\rho}\delta)}{[(1-\dot{\rho}^2)\rho^2 + (\partial_{\sigma}\rho)^2]^{\frac{3}{2}}}\left[\rho^2(\delta\dot{\beta}-\beta\dot{\delta}) - 2\delta(\rho\dot{\rho}\beta+\rho\partial_{\sigma}\rho\delta) + \frac{\rho^2\delta}{\beta}(\delta\partial_{\sigma}\beta-\beta\partial_{\sigma}\delta)\right]\left(\frac{1-\omega^2}{\omega^2}\right) 
\nonumber\\ &=& 0.
\end{eqnarray}
This is formally equivalent to (\ref{relcompact*}).
We see, therefore, that the term proportional to $(1-\omega^2)/\omega^2$ quantifies the contribution of the superconducting current. Specifically, this is expressed via the embedding function for the lines of constant phase, $\Gamma(\sigma,t)$, in terms of the associated parameter $\omega^2(\sigma,t)$. 
This, in turn, may be thought of as a kind of `current parameter', whereby $\omega^2 = 1$ implies $J^0=-J^z=0$, while $\omega^2 \rightarrow 0$ gives $J^0=-J^z \rightarrow \infty$. 
(As noted in Sec. \ref{Sect2}.)
It is straightforward to show that $\omega^2=1/2$ corresponds to the tensionless case, for which $\dot{\rho} = 0$, $\partial_{\sigma}\dot{\rho}=0$ and $\ddot{\rho}=0$ for all $t$, by directly substituting each of these conditions into Eq. (\ref{relcompact}). 
Substituting the generalised Finsler conditions (\ref{P-Q-U}) into Eq. (\ref{Assume}), we obtain
\begin{eqnarray} \label{Assume_Kawa}
(1-\dot{\rho}^2)(\partial_{\sigma}\Gamma)^2 - (\rho^2 + (\partial_{\sigma}\rho)^2)^2\dot{\Gamma}^2 + 2\dot{\rho}\partial_{\sigma}\rho\dot{\Gamma}\partial_{\sigma}\Gamma = 0,
\end{eqnarray}
and it is straightforward, but tedious, to demonstrate that, under these conditions, this is precisely equivalent to Eq. (\ref{Kawa-scalar-imag}) (see analogous results in Sec. \ref{Sect2}). 
Thus, Eqs. (\ref{relcompact}) and (\ref{Assume_Kawa}) alone are sufficient to completely specify the dynamics of the string.
\\ \indent
We now solve for $\Gamma(\sigma,t)$ in terms of $\rho(\sigma,t)$ and its derivatives. For the sake of notational simplicity, we first rewrite Eq. (\ref{relcompact}) as
\begin{eqnarray} \label{simplesX}
-Z\omega^2 + Y(1-\omega^2) = 0,
\end{eqnarray}
where the factors $Z$ and $Y$ are independent of $\partial_{\sigma}\Gamma$ or, equivalently
\begin{eqnarray} \label{simples+X}
\left(\frac{Z}{Y}\right) =  \left(\frac{1-\omega^2}{\omega^2}\right),
\end{eqnarray}
c.f. Eqs. (\ref{simples**})-(\ref{simples+*}).
Using the definition of $\omega^2(\sigma,t)$ (\ref{omega^2_2}), but with $Q(\sigma,t)$ given by Eq. (\ref{P-Q-U}), Eq. (\ref{simples+X}) is equivalent to
\begin{eqnarray} \label{simplesXX}
(\partial_{\sigma}\Gamma)^2 = \frac{1}{r_{c|n|}^2}(\rho^2+(\partial_{\sigma}\rho)^2)\left(\frac{1-\omega^2}{\omega^2}\right).
\end{eqnarray}
This expression has physical significance: as we we will see, it allows us to interpret the EOM in $\Gamma(t,\sigma)$ as dispersion relation governing longitudinal `waves' (i.e. compressions and rarefractions) in the helical phase lines within the string core, as in the case of long strings and circular loops. 
We then have 
\begin{eqnarray} \label{phi_dotX}
\dot{\Gamma}^2 = \frac{1}{r_{c|n|}^2}\left(\dot{\rho}\partial_{\sigma}\rho \pm \sqrt{(1-\dot{\rho}^2)\rho^2 +(\partial_{\sigma}\rho)^2}\right)^2\left(\frac{1-\omega^2}{\omega^2}\right),
\end{eqnarray}
which allows us to reconstruct the required form of $\Gamma(\sigma,t)$:
\begin{eqnarray} \label{phi_lastX}
\Gamma(\sigma,t) &=& \int \partial_{\sigma}\Gamma  {\rm d}\sigma + \int \dot{\Gamma}  {\rm d}t
\nonumber\\
&=& \frac{1}{r_{c|n|}} \int \sqrt{\rho^2+(\partial_{\sigma}\rho)^2}\frac{\sqrt{1-\omega^2}}{\omega}  {\rm d}\sigma \pm \frac{1}{r_{c|n|}} \int \left(\dot{\rho}\partial_{\sigma}\rho \pm \sqrt{(1-\dot{\rho}^2)\rho^2 +(\partial_{\sigma}\rho)^2}\right)\frac{\sqrt{1-\omega^2}}{\omega} {\rm d}t.
\nonumber
\end{eqnarray}
\begin{eqnarray}
{}
\end{eqnarray}
Clearly, Eqs. (\ref{simplesXX})-(\ref{phi_lastX}) are precisely analogous to (\ref{simples})-(\ref{phi_last*}).
\\ \indent
For circular strings ($\partial_{\sigma}\rho=0$), we recover 
\begin{eqnarray} \label{simples*}
n_{\sigma}^2 = \frac{\rho^2}{r_{c|n|}^2}\left(\frac{1-\dot{\rho}^2 + \rho\ddot{\rho}}{1-\dot{\rho}^2 - \rho\ddot{\rho}}\right) = \frac{\rho^2}{r_{c|n|}^2}\left(\frac{1-\omega^2}{\omega^2}\right),
\end{eqnarray}
which is equivalent to Eq. (\ref{EOM8_4.2}), using the expression for $\omega^2(t)$ from Eq. (\ref{omega^2_2}). 
This motivates the following definitions for noncircular loops:
\begin{eqnarray}\label{neff_last}
(n_{\sigma}^{\rm eff})^2 = \frac{(v_{\theta}^{\rm eff})^2(l_{\sigma}^{\rm eff})^2}{(v_{\sigma}^{eff})^2(2\pi)^2r_{c|n|}^2} = (\partial_{\sigma}\Gamma)^2,
\end{eqnarray}
where 
\begin{eqnarray}\label{leff_last}
(l_{\sigma}^{\rm eff})^2 = (2\pi)^2(\rho^2+(\partial_{\sigma}\rho)^2),
\end{eqnarray}
and
\begin{eqnarray}\label{enigma}
\frac{(v_{\theta}^{\rm eff})^2}{(v_{\sigma}^{\rm eff})^2} = \left(\frac{1-\omega^2}{\omega^2}\right).
\end{eqnarray}
Here $n_{\sigma}^{\rm eff}$ and $l_{\sigma}^{\rm eff}$ denote the effective local winding number and string radius, respectively, for given values of $\sigma$ and $t$, while $v_{\theta}^{\rm eff}$ and $v_{\sigma}^{\rm eff}$ denote the local velocities of the phase lines in the $\theta-$direction (perpendicular to the string core axis) and the $\sigma-$direction (parallel to the string core axis).
\\ \indent
Together with Eqs. (\ref{neff_last})-(\ref{enigma}), the following definitions then form a self-consistent set:
\begin{eqnarray}\label{1}
(l_{\sigma}^{\rm eff})^2 = (n_{\sigma}^{\rm eff})^2(\lambda_{\sigma}^{\rm eff})^2, \ \ \ 
(v_{\theta}^{\rm eff})^2(\lambda_{\sigma}^{\rm eff})^2 = (v_{\sigma}^{\rm eff})^2(2\pi)^2R^2,
\end{eqnarray}
\begin{eqnarray}\label{3}
(k_{\sigma}^{\rm eff})^2 = \frac{(2\pi)^2}{(\lambda_{\sigma}^{\rm eff})^2},
\end{eqnarray}
\begin{eqnarray}\label{4}
(\omega_{\sigma}^{\rm eff})^2 = (v_{\sigma}^{\rm eff})^2(k_{\sigma}^{\rm eff})^2 \equiv (\omega_{\theta}^{\rm eff})^2 =  \frac{(v_{\theta}^{\rm eff})^2}{r_{c|n|}^2} = \dot{\Gamma}^2,
\end{eqnarray}
which are compeletely analogous to their counterparts in the wound-string model. 
Combining Eqs. (\ref{enigma}) and (\ref{1}), we then have
\begin{eqnarray}\label{5}
(\lambda_{\sigma}^{\rm eff})^2 = (2\pi)^2r_{c|n|}^2\left(\frac{1-\omega^2}{\omega^2}\right).
\end{eqnarray}
where $\lambda_{\sigma}^{\rm eff}(\sigma,t)$ denotes the effective local wavelength of the phase line `twists'. Eq. (\ref{5}) reproduces the tensionless condition, $(\lambda_{\sigma}^{\rm eff})^2 = \lambda_{\sigma}= (2\pi)^2r_{c|n|}^2$, when $\omega^2(\sigma,t) = 1/2$ for all $\sigma, t$.
\\ \indent
We would like to be able to verify that the expressions for $(l_{\sigma}^{\rm eff})^2$, $(n_{\sigma}^{\rm eff})^2$, $(v_{\sigma}^{\rm eff})^2$, $(v_{\theta}^{\rm eff})^2$, $(\lambda_{\sigma}^{\rm eff})^2$ and 
$(\omega_{\sigma}^{\rm eff})^2 \equiv (\omega_{\theta}^{\rm eff})^2$ reduce to their equivalents for both circular loops and straight strings, in appropriate limits. 
However, from Eqs. (\ref{neff_last})-(\ref{5}), we do not have enough information to write down explicit expressions for $(\omega_{\sigma}^{\rm eff})^2$, $(v_{\sigma}^{\rm eff})^2$, or $(v_{\theta}^{\rm eff})^2$. 
By analogy with Sec. \ref{Sect2}, the missing piece comes from identifying $(\omega_{\sigma}^{\rm eff})^2$ with the integrand of the integral over ${\rm d}t$ in Eq. (\ref{phi_lastX}). 
Hence, we have that
\begin{eqnarray}\label{6X}
\Gamma(\sigma,t) = \int n_{\sigma}^{\rm eff} {\rm d}\sigma + \int \omega_{\sigma}^{\rm eff} {\rm d}t,
\end{eqnarray}
where
\begin{eqnarray}\label{7X}
(n_{\sigma}^{\rm eff})^2 = \frac{1}{r_{c|n|}^2}(\rho^2+(\partial_{\sigma}\rho)^2)\left(\frac{1-\omega^2}{\omega^2}\right),
\end{eqnarray}
\begin{eqnarray}\label{8X}
(\omega_{\sigma}^{\rm eff})^2 =  \frac{1}{r_{c|n|}^2}\left(\frac{(\dot{\rho}\partial_{\sigma}\rho \pm \sqrt{(1-\dot{\rho}^2)\rho^2 +(\partial_{\sigma}\rho)^2})^2}{\rho^2 +(\partial_{\sigma}\rho)^2}\right)\left(\frac{1-\omega^2}{\omega^2}\right),
\end{eqnarray}
and the sign of $\omega_{\sigma}^{\rm eff}$ relative to $n_{\sigma}^{\rm eff}$ depends on the sign of $\dot{\Gamma}$ relative to $\partial_{\sigma}\Gamma$. 
These are completely equivalent to their wound-string counterparts (\ref{6})-(\ref{8}), under the exchange $\rho \leftrightarrow a\rho$ and $r_{c|n|} \leftrightarrow R$.
\\ \indent
It may then be shown explicitly that the expressions for $(l_{\sigma}^{\rm eff})^2$, $(n_{\sigma}^{\rm eff})^2$, $(v_{\sigma}^{\rm eff})^2$, $(v_{\phi}^{\rm eff})^2$, $(\lambda_{\sigma}^{\rm eff})^2$ and 
$(\omega_{\sigma}^{\rm eff})^2 \equiv (\omega_{\phi}^{\rm eff})^2$, defined in Eqs. (\ref{neff_last})-(\ref{leff_last}) and (\ref{1})-(\ref{4}), reduce to their equivalents, for circular loops and long strings, in the limits 
$\partial_{\sigma}\rho \rightarrow 0$ and $(2\pi)^2\rho \rightarrow \Delta$, respectively. 
As in previous Sections, formally we may write down exact expressions for the constants of motion and integrated pressures and shears, but we are unable to evaluate them explicitly without further specifying the ansatz for $\rho(\sigma,t)$.
\\ \indent
We now note that, interpreting $\rho(t,\sigma)$ as the genuine radial coordinate for the loop, the associated conservation equation takes the form
\begin{eqnarray}\label{rho-cons}
\partial_{0}(T^{0}{}_{\rho}\sqrt{-g}) + \partial_{\sigma}(T^{\sigma}{}_{\rho}\sqrt{-g}) - \frac{1}{2}\partial_{\rho}g_{\sigma\sigma}T^{\sigma\sigma}\sqrt{-g}&=  0.
\end{eqnarray}
This, in turn, requires
\begin{subequations}
\begin{align}
\mathcal{T}^{0}{}_{\rho}\sqrt{-g} &=  -2\pi\eta^2|n|\frac{\rho^2\dot{\rho}}{\sqrt{(1-\dot{\rho}^2)\rho^2 + (\partial_{\sigma}\rho)^2)}}\omega^{-2} \mp 2\pi\eta^2|n|\partial_{\sigma}\rho\left(\frac{1-\omega^2}{\omega^2}\right), 
\\
\mathcal{T}^{\sigma}{}_{\rho}\sqrt{-g} &=  2\pi\eta^2|n|\frac{\partial_{\sigma}\rho}{\sqrt{(1-\dot{\rho}^2)\rho^2 + (\partial_{\sigma}\rho)^2)}}\left[1-\left(\frac{1-\omega^2}{\omega^2}\right)\frac{(\dot{\rho}\partial_{\sigma}\rho \pm \sqrt{(1-\dot{\rho}^2)\rho^2 +(\partial_{\sigma}\rho)^2}))^2}{(1-\dot{\rho}^2)(\rho^2+(\partial_{\sigma}\rho)^2)}\right]
\nonumber\\
&\pm  2\pi\eta^2|n|\dot{\rho}\left(\frac{1-\omega^2}{\omega^2}\right)\frac{(\dot{\rho}\partial_{\sigma}\rho \pm \sqrt{(1-\dot{\rho}^2)\rho^2 +(\partial_{\sigma}\rho)^2}))^2}{(1-\dot{\rho}^2)(\rho^2+(\partial_{\sigma}\rho)^2)}, 
\\
\mathcal{T}^{\sigma\sigma}\sqrt{-g} &=  -2\pi\eta^2|n|\frac{(1-\dot{\rho}^2)}{\sqrt{(1-\dot{\rho}^2)\rho^2 + (\partial_{\sigma}\rho)^2)}}\left[1-\left(\frac{1-\omega^2}{\omega^2}\right)\frac{(\dot{\rho}\partial_{\sigma}\rho \pm \sqrt{(1-\dot{\rho}^2)\rho^2 +(\partial_{\sigma}\rho)^2}))^2}{(1-\dot{\rho}^2)(\rho^2+(\partial_{\sigma}\rho)^2)}\right]. 
\end{align}
\end{subequations}
The expression for $\mathcal{T}^{\sigma\sigma}\sqrt{-g}$ follows directly from the definitions in Eqs. (\ref{U-T-mathcal{T}_3})-(\ref{kappa_2*}), whereas it may be shown that those for $\mathcal{T}^{0}{}_{\rho}\sqrt{-g}$ and $\mathcal{T}^{\sigma}{}_{\rho}\sqrt{-g}$ are necessary for consistency. 
Using these definitions, Eq. (\ref{rho-cons}) is equivalent to (\ref{relcompact}). 
Yet again, \emph{all} expressions obtained in this Section are equivalent to those obtained for arbitrary planar loops of wound-string in Sec. \ref{Sect2}, under the identifications $\Gamma(\sigma,t) \leftrightarrow \varphi(\sigma,t)$ and $r_{c|n|} \leftrightarrow R$, together with $\rho(\sigma,t) \leftrightarrow a \rho(\sigma,t)$, etc, where necessary in order to account for a warp factor of $a^2 <1$ in the string theory model.
\\ \indent
Finally we note that, though it is beyond the scope of the current work, it would be useful to define a parameter $\alpha(t)$, the generalization of the constant $\alpha$ defined for long straight strings, which quantifies the degree of nonlinearity in the phase / magnetic field-line `twists', at a given moment in time, for noncircular loops. 
By analogy with our previous results, we could then use this to determine the spatially-averaged values of the parameters defined in Eqs. (\ref{neff_last})-(\ref{leff_last}) and (\ref{1})-(\ref{4}), namely $\langle l_{\sigma}^2\rangle(t)$, $\langle n_{\sigma}^2\rangle(t)$, $\langle v_{\sigma}^2\rangle(t)$, $\langle v_{\phi}^2\rangle(t)$, $\langle \lambda_{\sigma}^2\rangle(t)$ and $\langle \omega_{\sigma}^2\rangle(t) \equiv \langle \omega_{\phi}^2\rangle(t)$, as well as a generalized expression for $\Omega^2(t)$, which is valid for arbitrary current-carrying loop configurations. 
As before, based on our previous findings, we expect the condition of zero net tension to correspond to $\Omega^2= 1/2$, for which $a^2\langle \lambda_{\sigma}^2\rangle  = (2\pi)^2r_{c|n|}^2$ and $\langle \omega_{\sigma}^2\rangle = 1/r_{c|n|}^2$ for all $t$. 
Such a definition would allow us to write the constants of motion and \emph{bulk} properties of noncircular superconducting strings in terms of spatially-averaged values of the parameters which determine the microscopic `internal' structure of the string core. 
This could be useful for describing the bulk properties of string networks and help yield further insights into the relation between their microscopic and macroscopic dynamics. 

\subsection{Summary of the defect-string model} \label{Sect5.4}
We have shown that, by adopting an appropriate ansatz for the Abelian-Higgs fields, and imposing an appropriate {\it Finsler condition} on the space-time metric such that 
\begin{eqnarray} \label{Finsler_Condition_Sec5}
&{}&{\rm d}s^2 = {\rm d}t^2 - {\rm d}r^2 - r^2{\rm d}\theta^2 - \rho^2(t){\rm d}\sigma^2
\nonumber\\
&\rightarrow& {\rm d}\tilde{s}^2 = (1-\dot{\rho}^2(t)){\rm d}t^2 - {\rm d}r^2 - r^2{\rm d}\theta^2 - \rho^2(t){\rm d}\sigma^2,
\end{eqnarray}
the evolution equations for a circular superconducting string loop can be derived, {\it exactly}, from the standard Abelian-Higgs field equations. 
The appropriate ansatz for scalar field describes rotating `twists' in the lines of constant phase, resulting in the existence of a second topological invariant, $N = \int \partial \vartheta/\partial \sigma|_{\theta = {\rm const.}}{\rm d}\sigma$ ($N \in \mathbb{Z}$), analogous to that found in the case of chiral strings \cite{Witten:1985,BlOlVi01}. 
\\ \indent
The time-dependent embedding of the lines of constant phase that are situated at the edge of the string core, $r = r_{c|n|}$, is {\it exactly} analogous to the embedding of the higher-dimensional windings of an $F$-string with {\it identical} macroscopic motion in the large space-time dimensions: the only difference is the replacement $R \leftrightarrow r_{c|n|}$, where $R$ is the radius of the compact internal space, and $a\rho(t) \rightarrow \rho(t)$, where necessary, due to the presence of warping caused by flux-compactifications. 
We conclude that the similarity (dissimilarity) of the embedding of the lines of constant phase, in the defect-string model, and the higher-dimensional windings, in the $F$-string model, is the key determinant of the similarity (dissimilarity) of the macroscopic string dynamics. 
Crucially, this similarity is {\it quantitative}: when the embedding of the phase-lines is exactly equivalent to the embedding of the windings, under the exchange of variables given above, then the macroscopic dynamics and {\it all} observable string parameters, including the $(3+1)$-dimensional energy and momentum densities, pressures, and shears, are {\it identical}. 
In this case, there is no observable difference, from a  $(3+1)$-dimensional perspective, between the species of topological defect string and its higher-dimensional $F$-string counterpart. 
\\ \indent
We conjecture that a similar equivalence holds in even more complicated scenarios, for example, when the modified Abelian-Higgs strings considered here are replaced by chiral strings, or vortons, etc. \cite{Test1,Test2,Carter&Martin(1993),Larsen(1993),Vortons(2013)}, and when simple $F$-strings are replaced by general $(p,q)$-strings, i.e., bound states of $p$ $F$-strings and $q$ $D$-strings \cite{hep-th/0505050v1,astro-ph/0410073v2,Sakellariadou:2009,Rajantie:2007}. 
For more complex defect string species, one must consider the embedding(s) of `twisted lines' in which the relevant order parameter(s) of the model remain constant. 
For more complex $F/D$-string bounds states, one must substitute the relevant effective string tensions and, where necessary, consider multiple embeddings, possibly corresponding to the presence of multiple order parameters in the defect string counterpart.  
\\ \indent
In the case of noncircular loops, we imposed an appropriate {\it generalised Finsler condition}, 
\begin{eqnarray} \label{gFinsler_Condition_Sec5}
&{}&{\rm d}s^2 = {\rm d}t^2 - {\rm d}r^2 - r^2{\rm d}\theta^2 - \rho^2(t,\sigma){\rm d}\sigma^2
\nonumber\\
&\rightarrow& {\rm d}\tilde{s}^2 = (1-\dot{\rho}^2(t,\sigma)){\rm d}t^2 - {\rm d}r^2 - r^2{\rm d}\theta^2 - \dot{\rho}(t,\sigma)\partial_{\sigma}\rho(t,\sigma){\rm d}t{\rm d}\sigma - (\rho^2(t,\sigma) + (\partial_{\sigma}\rho)^2(t,\sigma)){\rm d}\sigma^2,
\nonumber\\
\end{eqnarray}
and used this together with the previous ansatz for the microscopic structure of the Abelian-Higgs fields in the superconducting string. 
Imposing this condition, we again found that the fundamental Abelian-Higgs field equations reduce to those for the standard Nielsen-Olsen string, plus the EOM describing the macroscopic dynamics of an arbitrary planar wound-string loop. 
Yet again, the correspondence is {\it exact}: the key determinants of equivalent dynamics are the embeddings of the lines of constant phase (in the defect-string model) and the higher-dimensional windings (in the $F$-string case). 
Equivalent embeddings, under the replacements $R \leftrightarrow r_{c|n|}$ and $a\rho(t) \rightarrow \rho(t)$, give identical macroscopic dynamics, observable energy and momentum densities, pressure and shears. 
\\ \indent
Though tentative, these results suggest that the method we have outlined above is valid much more generally. 
If so, it may be extended to both arbitrary string configurations and arbitrary defect and $F/D$-string species in future works. 
This would allow us to determine, {\it exactly}, which string species are distinguishable, and which are not, via their observable cosmological and astrophysical signatures.   

\section{Conclusions} \label{Sect6}
We have outlined a detailed model of wound $F$-strings, in a physically intuitive coordinate system based on cylindrical polars in the large space-time dimensions, and have identified the key physical parameters which determine their macroscopic dynamics and physical observables. 
These include the effective $(3+1)$-dimensional energy and momentum densities, pressures, and shears, which may be expressed compactly in terms of the parameter $\omega^2(t,\sigma) \in [0,1]$. 
This represents the local fraction of the string length lying parallel to the macroscopic directions, so that $\omega^2 = 1$ represents an unwound string and the limit $\omega^2 \rightarrow 0$ represents a string contained entirely within the compact space. 
\\ \indent
Equivalently, for fixed winding radius $R$, the local fraction $\omega^2(t,\sigma)$ can be expressed in terms of an effective local winding number, $n_{\sigma}^{\rm eff}(t,\sigma)$, and wavelength, $\lambda_{\sigma}^{\rm eff}(t,\sigma)$, and the equation of motion for the angular embedding coordinate in the compact internal space takes the form of a canonical dispersion relation. 
This is, essentially, the dispersion relation for waves `in' the string, i.e., for compressions and rarefractions in the rotating higher-dimensional windings which, from a $(3+1)$-dimensional perspective, appear as superconducting currents. 
However, our treatment goes beyond the usual paradigm of dimensional reduction and we consider the full higher-dimensional string configuration, rather than integrating out the higher-dimensional variables to construct an approximate effective action \cite{CoHiTu87,Copeland:1987yv}. 
\\ \indent
This motivates us to consider a similarly detailed ansatz for superconducting defect strings, i.e., one that is capable of describing both the macroscopic `external' evolution of the string, as an extended object, and the evolution of the  `internal' variables that determine the microscopic structure of the string core. 
This is equivalent to describing the macroscopic string dynamics purely in terms of the evolution of fundamental fields, again without the use of an approximate effective action. 
The major advantage of this approach is that it allows us, in conjunction with our previous results for wound-strings, to determine {\it exactly} when two inequivalent string species are observationally {\it indistinguishable}, from a $(3+1)$-dimensional perspective. 
\\ \indent
We found two key factors that determine whether a given wound-string configuration is distinguishable from a superconducting defect string. 
These are (a) the embedding of the higher-dimensional windings (for $F$-strings) and (b) the embedding of the `twists' in the lines of constant phase along the string length (for defects). 
When the two embeddings are equivalent under the exchange $r_{c|n|} \leftrightarrow R$, where $r_{c|n|}$ is the radius of the defect-string core and $R$ is the radius of the windings in the compact space, the dispersion relation for the superconducting current in the former {\it exactly} matches the dispersion relation that governs the evolution of the windings in the latter. 
In this scenario, it is physically reasonable to expect the macroscopic dynamics of each string species to also be identical, thus rendering them indistinguishable to observers in $(3+1)$ dimensions: but how can such an exact equivalence be demonstrated? 
This is the question that was asked and answered in the final section of our paper, and its solution is the main achievement of this work.
\\ \indent
We have shown that, when equivalent embeddings for the phase-lines and higher-dimensional windings are imposed, the same equations of motion for the macroscopic string variables, e.g., the local effective loop radius $\rho(t,\sigma)$, can be obtained in each model. 
Remarkably, with the use of an appropriate ansatz, the Euler-Lagrange equations for the fundamental Abelian-Higgs fields reduce to the EOM of the standard Nielsen-Olesen string plus the EOM for the macroscopic string radius, $\rho(t,\sigma)$, obtained from the wound-string model. 
\\ \indent
However, there is a catch. 
In order to derive the correct EOM for $\rho(t,\sigma)$ from the Abelian-Higgs field equations, we must impose an appropriate {\it Finsler condition} on the space-time metric, for circular loops, and an appropriate {\it generalised Finsler condition} for non-circular loops. 
Mathematically, this trick is equivalent to modelling a time-evolving string, embedded in a fixed background geometry, as a fixed (comoving) string embedded in a time-evolving background. 
Interestingly, in the case of circular loops, the effective space-time background required is of Finsler type, with an effective {\it Finsler metric} of the form $\tilde{g}_{\mu\nu}(x,\partial_{\tau}x)$. 
For non-circular loops, an effective {\it generalised Finsler metric} of the form $\tilde{g}_{\mu\nu}(x,,\partial_{\tau}x,\partial_{\sigma}x)$ must be imposed. 
\\ \indent
These results demonstrate the potential importance of these novel geometric structures in modelling complex physical systems in astrophysics and cosmology, but also hint at a possible deeper connection between the mathematics of string theory, including higher-dimensional embeddings \cite{Kiritsis:2007zza,Avgoustidis:2004}, and the mathematics of Finsler and generalised Finsler spaces \cite{Vacaru:2002kp,Vacaru:1996ga}. 
In future work we will attempt to define, in more abstract and general terms, the conditions under which a given higher-dimensional configuration of a string, field, or other extended object (e.g. a $D$-brane), embedded in a physical pseudo-Riemannian space, can be described, from a $(3+1)$-dimensional viewpoint, in terms of an effective Finsler or generalised Finsler geometry. 

Finally, because the text of this paper is rather long and involved, we present a bullet point summary of the our main conclusions, below, for added clarity. 
The new and novel results presented in this work are:
\begin{enumerate}

\item The demonstration that the macroscopic EOM for current-carrying strings, or, equivalently, strings with higher-dimensional windings, can be obtained from a microscopic field theory ansatz, without the the need for the standard approximations used in constructing an effective action.

\item This provides a unified framework for studying various string-theoretic and field-theoretic string species in which both macroscopic ‘external’ and microscopic ‘internal’ variables are treated at the same level, i.e, , without integrating out the latter in order to obtain approximate dynamical equations for the former.

\item This, in turn, permits us to identify the precise conditions under which string-theoretic and field-theoretic string species become indistinguishable, observationally, to a macroscopic observer.

\item For Nambu-Goto strings with higher-dimensional windings and superconducting Abelian-Higgs strings this occurs when the helical embedding of the windings exactly matches the embedding of the lines of constant phase in the vortex-string core. However, a similar method of comparison can be used, in principle, to compare any two string-theoretic and field-theoretic string species.

\item Remarkably, the derivation of the macroscopic EOM for the field-theoretic string species requires us to model these in an effective Finsler or generalised Finsler geometry, where former applies to circular string loops and the latter to non-circular configurations.

\item This demonstrates the utility of these ‘exotic’ geometries in modelling complex physical systems in cosmology and astrophysics, but the exact mechanism behind this correspondence, i.e., why it works, requires further investigation.

\item Nonetheless, the correspondence discovered here, in a few test cases, hints at a deeper and as yet unexplored connection between Finsler-type geometries and string theory, perhaps along the lines considered decades ago by the mathematician Matsumoto \cite{Matsumoto-1986}. 
Though more work is needed to confirm this, it opens up an interesting new direction for research in both string theory and the theory of topological defects.

\end{enumerate}

\begin{center}
{\bf Acknowledgments}
\end{center}
My sincere thanks to Tiberiu Harko for help, advice, and interesting discussions, throughout the course of this project. This work was supported by the Natural Science Foundation of Guangdong Province, grant no. 008120251030.


%
\end{document}